\documentclass[journal]{IEEEtran}
\IEEEoverridecommandlockouts

\newtheorem{Remark}{Remark}
\usepackage{algorithm}
\usepackage{algorithmic}
\usepackage{graphicx}
\usepackage{textcomp}
\usepackage{stfloats}
\usepackage{subfigure}
\usepackage{amssymb,amsmath,bm}
\usepackage{cite}
\usepackage{times}
\usepackage{latexsym}
\usepackage{array}
\usepackage{fancyhdr}
\usepackage{psfrag}
\usepackage{multirow}
\usepackage{xcolor}
\usepackage{color}
\usepackage{makecell}
\usepackage{threeparttable}
\usepackage{hyperref} 
\usepackage[OT1]{fontenc}

\begin{document} 

\title{Multi-View Imaging in Networked Sensing Systems: A Covariance-based Approach\\}

 \author{\IEEEauthorblockN{Junyuan~Gao,
 Weifeng~Zhu, 
 Yanmo~Hu, 
 Shuowen~Zhang,
 Jiannong Cao,
 Yongpeng~Wu, 
 Giuseppe~Caire,
 and Liang~Liu }
 \thanks{This work was supported in part by the National Natural Science Foundation of China (NSFC) under Grant 62471421, Hong Kong RGC Collaborative Research Fund Young Collaborative Research Grant (CRF-YCRG) under Grant PolyU C5002-23Y, the Fundamental Research Funds for the Central Universities, the Yangtze River Delta Science and Technology Innovation Community Joint Research (Basic Research) Project under Grant BK20244006, 111 project BP0719010, STCSM 22DZ2229005, and the xG-RIC project as part of the research program Communication Systems ``Souver\"{a}n. Digital. Vernetzt.'' (16KIS2429K) of the German Federal Ministry of Research, Technology, and Space (BMFTR). 
\emph{(Corresponding author: Weifeng Zhu; Liang Liu)}}
 \thanks{J. Gao, W. Zhu, Y. Hu, S. Zhang, J. Cao, and L. Liu are with the Department of Electrical and Electronic Engineering, The Hong Kong Polytechnic University, Hong Kong SAR (e-mail: \{junyuan.gao, eee-wf.zhu, yanmo.hu, shuowen.zhang, jiannong.cao, liang-eie.liu\}@polyu.edu.hk).} 
 \thanks{Y. Wu is with the Department of Electronic Engineering, Shanghai Jiao Tong University, Minhang 200240, China (e-mail: yongpeng.wu@sjtu.edu.cn).}
 \thanks{G. Caire is with the Communications and Information Theory Group, Technische Universit{\"a}t Berlin, Berlin 10587, Germany (e-mail: caire@tu-berlin.de).} 
 }

\maketitle

\begin{abstract}  

This paper considers multi-view imaging in a sixth-generation (6G) integrated sensing and communication network, which consists of a transmit base-station (TBS), multiple receive base-stations (RBSs) connected to a central processing unit (CPU), and multiple extended targets. Our goal is to devise an effective multi-view imaging technique that can jointly leverage the echo signals at all the RBSs to precisely construct the image of these targets. To achieve this goal, we propose a two-phase framework. In Phase I, each RBS recovers an individual image of all the targets from its own view, which is obtained via utilizing its received signals' sample covariance matrix to detect the grids with non-zero effective scattering intensity in the region of interest. Moreover, the shape of each grid is adjusted to conform to target geometries. In Phase II, the CPU fuses the individual images of all the RBSs to construct a higher-quality image of all the targets. To this end, we first design an edge-preserving natural neighbor interpolation (EP-NNI) method and then formulate an optimization problem to fuse the interpolated results. Extensive numerical results show that the proposed scheme significantly enhances imaging performance, facilitating high-quality environment reconstruction for future 6G networks. 

\end{abstract}

\section{Introduction}\label{sec:intro} 

\subsection{Motivation}\label{sec:intro_motivation}  

Integrated sensing and communication (ISAC) emerges as one of the primary usage scenarios of future sixth-generation (6G) wireless systems. In 6G networks, sensing is performed by enabling devices to collect reflected signals from the targets and extract valuable information for decision-making, exploiting the same frequency band and hardware as those used for communication~\cite{ref:FLiu-ISAC2}. Due to this resource-sharing capability, 6G-oriented sensing is expected to play a crucial role in many emerging applications, such as smart transportation, augmented reality, and factory automation.

In most existing ISAC works, the sensing functionality is achieved based on the assumption of point targets. Under this assumption, the reflected signal of a target contains information on classical ``radar-type'' parameters such as delay (range), angle of arrival (AoA), and Doppler (velocity), which in turn yields information for localization and tracking, irrespective of the target's physical shape~\cite{ref:LLiu-ISAC,ref:ZhuWF,hu2024anchor}. However, it should be noted that the point target model remains valid only when the target size is much smaller than system resolution. In practical environments, there are a large number of targets, such as vehicles and industrial machinery, whose sizes substantially exceed resolution limits. In this case, the point target assumption does not hold, making it necessary to employ the extended target model. Compared to point targets, extended targets usually comprise numerous inseparable scattering points, rendering precise parameter estimation for individual scatterers impossible~\cite{ref:RCS2,ref:hou2006direct}. Thus, for extended targets, the core sensing objective shifts to imaging - constructing a high-resolution reflectivity map of the environment, from which the shape, size, localization, and structure of targets can be estimated. Environment imaging is regarded as a key sensing functionality in 6G networks, which is therefore the topic of this work.

Compared with radar systems, in which radars usually work independently due to their sparse deployment and the absence of a globally unified standard, networked sensing is a unique advantage of 6G-oriented sensing~\cite{ref:LLiu-ISAC,liu2025cooperative}. In typical communication systems, such as cloud radio access network and cell-free multiple-input multiple-output (MIMO) network, base-stations (BSs) are densely deployed and connected via the fronthaul links for cooperative communication. This inherently enables networked sensing, where the signals reflected by targets can be received by multiple BSs and these BSs can share their local information to enhance sensing performance. It is worth noting that such cooperation is indispensable for environment imaging. Specifically, due to the occlusion effect and weak scattering at some sites, each receiver only has a limited field of view about the target, and this field is usually unknown in advance~\cite{ref:PHDfusion,tian2023distributed,caire_multistatic,zhu2026joint}. Even for imaging within this field of view, the performance is fundamentally limited by the resolution of each receiver. Therefore, constructing a high-resolution image requires global environment awareness beyond the capabilities of individual receivers. Moreover, typical impairments that traditional synthetic aperture radar (SAR) imaging must cope with, such as synchronization errors between sensors and uncertain positions (in the case of fly-over SAR imaging from planes and drones) are not problems in 6G networks, since BSs are in fixed and precisely known locations, and virtually perfect synchronization is achieved with standard fronthaul technologies, which makes the wireless network well-suited for the task of networked sensing. Thus, in this paper, we present an investigation into the multi-view imaging problem within the 6G networked sensing framework.

\subsection{Related Works}
 
Environment imaging is recognized as an indispensable sensing functionality in future 6G wireless networks, attracting significant interests from researchers. Back projection (BP) and fast Fourier transform (FFT) / inverse FFT (IFFT) algorithms have been widely adopted for imaging~\cite{ref:BP,ref:FFT2,ref:huang2024fourier}. However, their reconstruction resolution is fundamentally constrained by physical limitations, such as antenna aperture and bandwidth. To enhance imaging performance, a number of super-resolution compressed sensing (CS) based methods have been proposed, where imaging is formulated as an inverse problem aiming at estimating instantaneous channel attenuation coefficients over a set of grid points by exploiting both angular and delay resolution. Specifically, the authors in \cite{ref:Tong_GAMP} developed a generalized approximate message passing based imaging algorithm, where the $\ell_1$-norm regularization term was induced to exploit the sparsity of attenuation coefficients. In \cite{ref:CAMP}, the complex approximated message passing method was extended to an $\ell_{(1/2)}$-norm based iterative thresholding scheme. Since discrete gradients of attenuation coefficients are sparse, i.e., discrete gradients only contain high values when the attenuation coefficients of adjacent grids are significantly different, \cite{ref:TV1} proposed to construct an image based on the total variation (TV) regularization. A sparse Bayesian learning based algorithm was proposed in \cite{ref:caire_SBL} to solve the CS problem in environment reconstruction.

The aforementioned algorithms are computationally expensive since they need to estimate high-dimensional instantaneous channels from the echo signal in each snapshot. In stationary scenarios, the shape, position, and structure of extended targets exhibit little variation across multiple snapshots, and thus the statistic scattering intensity remains invariant over time. This observation motivates us to reformulate the imaging task as a statistic scattering intensity estimation problem instead of relying on instantaneous channel estimation, thereby substantially reducing the number of parameters to be estimated. The advantage of estimating statistic information over instantaneous channels has been revealed from the information-theoretic view in \cite{ref:gaoIT1,ref:gaoIT2}. Inspired by this, our work aims to establish a covariance-based imaging framework that leverages statistical scattering properties to achieve precise imaging of extended targets.

Networked sensing substantially enhances sensing performance by enabling cooperation among multiple wireless nodes. Several works have explored networked sensing techniques for the localization of point targets. Specifically, the authors in \cite{ref:shiqin1} considered cell-free networks, where multiple BSs share their echo signals to jointly localize the targets. Effective scheme was designed to deal with the data association issue. The authors in \cite{ref:gaoISAC} proposed to integrate the massive communication and point target localization functionalities, where the uplink signals received at multiple BSs were leveraged to achieve joint activity detection, channel estimation, and target localization. Networked sensing has also been leveraged to improve the imaging performance of extended targets in existing literatures. Specifically, in \cite{ref:manzoni2024wavefield,ref:tagliaferri2024cooperative}, after obtaining single-view images at individual BSs via the BP technique, data fusion was performed by summing the estimates at all BSs together, under the assumption that BSs share the same field of view. However, in practice, receivers usually observe different scattering profiles of extended targets due to unknown occlusion effect and weak scattering at some sites, and thus their fields of view can be distinct and unknown in advance~\cite{tian2023distributed}. Under this condition, a clustering method was employed in \cite{ref:PHDfusion} to identify the subset of receivers having high-confidence estimates at a given grid point, subsequently performing weighted averaging over this subset, which, however, relies on the Gaussian mixture assumption on local estimates. For general local estimates without this assumption, the authors in \cite{caire_multistatic} proposed to fuse individual images applying the OR operation, where the existence of a scattering point was affirmed upon it was detected by one receiver. In \cite{ref:shiqin3}, a Gaussian-based filter was applied before summing the multi-view estimates. However, these approaches overlook the geometric characteristics of extended targets, thereby limiting their fusion performance. Thus, it is necessary to develop advanced fusion methods that incorporate target geometric characteristics while accommodating distinct and unknown fields of view at receivers.

\subsection{Main Contributions} 

In this paper, we consider a narrowband networked sensing system consisting of a transmit BS~(TBS), multiple receive BSs~(RBSs) that are connected to a CPU, and multiple extended targets that behave as dense clouds of scatterers. The transmitted signals are reflected by extended targets before arriving at RBSs, and thus the characteristics of targets, such as shape and size, are embedded in the received signals. Our goal is to devise an effective multi-view imaging technique that can leverage the echo signals at all RBSs to precisely construct the image of extended targets. The main contributions of this paper are summarized as follows.

\begin{itemize} 

    \item 
    We devise an advanced two-phase framework that exploits spatial resolution to achieve precise imaging of extended targets. Specifically, in Phase I, each RBS constructs an individual image of the extended targets from its own view by leveraging the spatial resolution provided by its own antenna array and that of TBS. In Phase II, the CPU fuses these individual results to construct a higher-quality image of all the targets, where the spatial diversity gain offered by multi-view observations at distributed RBSs is fully exploited.

    \item 
    In Phase I, we propose a novel covariance-based approach for single-view imaging of extended targets at each RBS. Specifically, we first discretize the region of interest into grids. Then, we formulate a penalized maximum likelihood (ML) problem to  alternatively estimate the effective scattering intensity for each grid point and update the grid positions, where both the sample covariance matrix of received signals and the property that extended targets typically occupy multiple adjacent grids are exploited. The targets can be recovered by the grids with high intensity values. Compared to FFT and CS based methods that need to estimate instantaneous channels, our approach significantly reduces the number of estimated variables. Moreover, by performing grid position optimization, grid shapes are dynamically adjusted to better align with target geometries. The advantage of employing dynamic grids over employing excessively dense and fixed grids for imaging is demonstrated.

    \item 
    In Phase II, we propose a novel multi-view fusion scheme, where cross-disciplinary concepts and tools from computer vision are introduced. The fusion task is challenging because grid topologies are heterogeneous and RBS fields of view are distinct and unknown. To this end, we design an edge-preserving natural neighbor interpolation (EP-NNI) method to map each single-view image onto a unified set of grids while preserving edge features. Then, we formulate an optimization problem to fuse interpolated results, which is solved by selecting the subset of RBSs that are ``informative'' on the scattering intensity corresponding to each grid, and updating the fused intensity based on this subset and target geometric features via the alternating optimization (AO) technique.

    \item 
    We provide extensive simulation results to verify the superiority of both the single-view imaging and data fusion stages, which collectively enable high-quality multi-view imaging of extended targets. In particular, it is shown that the proposed covariance-based imaging scheme consistently outperforms the FFT and CS based benchmark schemes, especially in the scenarios where resources, such as the number of antennas, transmit power, and pilot length, are limited. We interpret this performance gain as the exploitation of channel statistical properties and target geometric properties in our scheme, demonstrating the great potential of our scheme in supporting high-quality imaging in future 6G networks.
  
\end{itemize}

\emph{Organizations:} The rest of this paper is organized as follows. Section~\ref{Sec:model} describes the imaging system model. Section~\ref{Sec:Grid} formulates the imaging problem after grid discretization. In Section~\ref{Sec:image}, we devise the covariance-based single-view imaging scheme and theoretically analyze the advantages of adopting dynamic grids. Section~\ref{Sec:fusion} presents the multi-view fusion scheme for high-quality reconstruction based on local images. Numerical results are presented in Section~\ref{Sec:simulation}. Finally, Section~\ref{Sec:conclusion} concludes this paper.

  \emph{Notations:} Throughout this paper, uppercase and lowercase boldface letters denote matrices and vectors, respectively. The notation $\left[\mathbf{X} \right]_{m,:}$ denotes the $m$-th row of $\mathbf{X}$ and $\left[\mathbf{X} \right]_{:,n}$ denotes the $n$-th column of $\mathbf{X}$. The notation $\mathbf{I}_{n}$ denotes an $n\times n$ identity matrix. We use $\left(\cdot \right)^{T}$, $\left(\cdot \right)^{H}$, $\left\|\mathbf{x} \right\|_{p}$, and $\left\|\mathbf{X} \right\|_{F}$ to denote transpose,
  conjugate transpose, ${\ell}_p$-norm, and Frobenius norm, respectively. We use $\otimes$ and $\odot$ to denote Kronecker product and element-wise product, respectively. We use $\operatorname{diag} \left\{ \mathbf{x} \right\}$ to denote a diagonal matrix with $\mathbf{x}$ comprising its diagonal elements. The operation of the vectorization of a matrix is denoted as $\operatorname{vec}(\cdot)$. Given any complex variable, vector, or matrix, we use $\Re(\cdot)$ and $\Im(\cdot)$ to return its real and imaginary parts, respectively. We use $\cdot\backslash\cdot$ to denote set subtraction and $\left| \mathcal{A} \right|$ to denote the cardinality of a set $\mathcal{A}$. For an integer $k > 0$, let $[k] = \left\{1,\ldots,k \right\}$. We use $1[\mathcal{E}]$ to denote the indicator function of an event $\mathcal{E}$. The distribution of a circularly symmetric complex Gaussian~(CSCG) random vector $\mathbf{x}$ with mean $\bm{\mu}$ and covariance matrix $\bm{\Sigma}$ is denoted as $\mathcal{CN}(  \bm{\mu} , \bm{\Sigma} )$.

\section{System Model}\label{Sec:model}
 
  \begin{figure}
    \centering
    \includegraphics[width=0.95\linewidth]{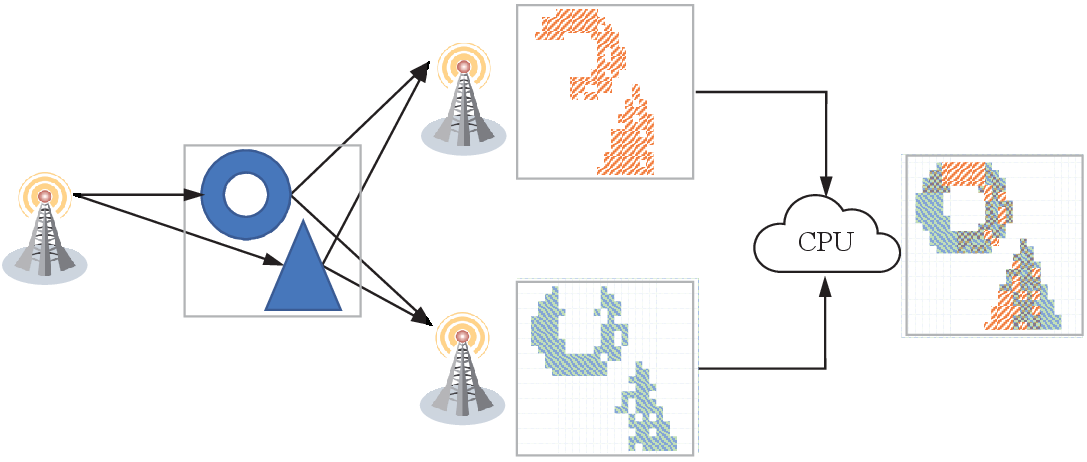}\\
    \caption{System model.}  
  \label{fig:SM}
  \end{figure}
    
We consider a multi-view imaging scenario in a 6G cellular network as illustrated in Fig.~\ref{fig:SM}, in which there are a TBS, $K$ RBSs that are connected to the CPU via fronthaul links, and $B$ extended targets. The TBS and each RBS are equipped with uniform linear arrays~(ULAs) consisting of $N_{\rm tx}$ and $N_{\rm rx}$ antennas spaced by half of the carrier wavelength, respectively. Under a two-dimensional~(2D) Cartesian coordinate system, the locations of the TBS and the $k$-th RBS are denoted as $\mathbf{p}_{{\rm tx}} = \left[ x_{{\rm tx}},y_{{\rm tx}}\right]^T \in \mathbb{R}^{2 \times 1}$ and $\mathbf{p}_{{\rm rx},k} = \left[ x_{{\rm rx},k},y_{{\rm rx},k} \right]^T \in \mathbb{R}^{2 \times 1}$ in meter, respectively, which are known in advance\footnote{As in~\cite{caire_multistatic,ref:CAMP,mandelli2022sampling,li2024radio,jiang20260}, this paper focuses on 2D imaging for the purpose of exposition. This can be extended to three-dimensional (3D) imaging by deploying uniform planar arrays at BSs, which enables the estimation of both azimuth and elevation angles and thus allows for detecting whether there exist scatterers in 3D space.}.

We consider the transmission over $M$ frames, each comprising $L$ symbols. In each frame, denote the transmitted pilot signal as $\mathbf{X} \in \mathbb{C}^{N_{\rm tx} \times L}$ satisfying $\|  [ \mathbf{X} ]_{n,:}  \|_2^2 = LP, \forall n\in [N_{\rm tx}]$. This signal reaches the receiver through propagation channels that include the scattering due to extended targets\footnote{The impact of clustered clutters can be effectively suppressed by using classical space-time adaptive processing (STAP) \cite{ref:STAP} to pre-filter these components in the angle-Doppler domain. Therefore, similar to \cite{ref:FFT2,ref:huang2024fourier,ref:CAMP,caire_multistatic,ref:tagliaferri2024cooperative}, we do not explicitly incorporate a clustered clutter model in this work.}. The signal received at the $k$-th RBS in the $m$-th frame is given by\footnote{Notice that \eqref{eq:receive_Y} is a narrowband model, corresponding for example to one subcarrier of an orthogonal frequency division multiplexing (OFDM) system. This means that in an ISAC framework, we can transmit imaging-oriented pilots as in \eqref{eq:receive_Y} on a single subcarrier, while using the remaining subcarriers for communication. Thus, the overhead of such form of imaging is very low, and can be easily plugged into any OFDM-based communication systems.} 
\begin{equation}\label{eq:receive_Y}
    \mathbf{Y}_{k,m} = \mathbf{H}_{k,m} \mathbf{X} + \mathbf{Z}_{k,m} \in \mathbb{C}^{N_{\rm rx} \times L}, \;\; \forall k,m, 
\end{equation} 
where $\mathbf{H}_{k,m}\in \mathbb{C}^{ N_{\rm rx} \times N_{\rm tx} }$ denotes the channel matrix between the TBS and the $k$-th RBS in frame $m$ and $\mathbf{Z}_{k,m} $ denotes the noise with i.i.d. $\mathcal{CN} \!\left( 0, \sigma^2_z\right)$ distributed elements.

According to the Born approximation~\cite{ref:Born} for the scattered field, the channel matrix $\mathbf{H}_{k,m}$ in \eqref{eq:receive_Y} can be expressed as  
\begin{equation}\label{eq:model_H} 
    \mathbf{H}_{k,m} = \iint_{ \mathcal{S}} c_{x,y,k,m} 
    \mathbf{b}\! \left(\theta_{x,y,k} \right)  \mathbf{a}^T\! \left(\phi_{x,y}\right) {\rm d} x {\rm d} y  ,  \;\; \forall k,m, 
\end{equation} 
where $\mathcal{S}$ denotes the spatial region occupied by extended targets, $[x,y]^T$ denotes the position of a general point on the target,
$\phi_{x,y}$ denotes the angle of departure (AoD) from the TBS to the scatterer located in $[x,y]^T$ given by
\begin{equation}\label{eq:AoD}
  \phi_{x,y} = \arctan  \left(  \tfrac{ y_{{\rm tx}} - y }{ x_{{\rm tx}} - x }  \right) + \pi 1 \left[ x_{{\rm tx}} < x \right], 
\end{equation} 
$\theta_{x,y,k}$ denotes the AoA from the point scatter 
to the $k$-th RBS given by
\begin{equation}\label{eq:AoA}
  \theta_{x,y,k} = \arctan  \left(  \tfrac{ y_{{\rm rx},k} - y }{ x_{{\rm rx},k} - x }  \right) + \pi 1 \left[ x_{{\rm rx},k} < x \right], 
\end{equation} 
$\mathbf{a}\! \left(\phi_{x,y}\right)$ and $\mathbf{b}\! \left(\theta_{x,y,k} \right)$ denote the transmit steering vector towards $\phi_{x,y}$ and the receive steering vector towards $\theta_{x,y,k}$, respectively, given by
\begin{equation}\label{eq:steer_a}
  \mathbf{a}\! \left(\phi_{x,y}\right) \!=\! \left[ 1, e^{-j\pi\sin( \phi_{x,y} )}, \ldots, e^{-j\pi(N_{\rm tx}-1)\sin( \phi_{x,y} )} \right]^T\!, 
\end{equation} 
\begin{equation}\label{eq:steer_b}
  \mathbf{b}\! \left(\theta_{x,y,k}\right) = \left[ 1, e^{-j\pi\sin( \theta_{x,y,k} )}, \ldots, e^{-j\pi(N_{\rm rx}-1)\sin( \theta_{x,y,k} )} \right]^T, 
\end{equation} 
and $c_{x,y,k,m} = \alpha_{x,y,k} \bar{c}_{x,y,k,m}$ denotes the effective attenuation coefficient of the path from the TBS to point $[x,y]^T$ and then to the $k$-th RBS in frame $m$. The binary variable $\alpha_{x,y,k}$ equals to $1$ if the signal reflected by point $[x,y]^T$ is received by the $k$-th RBS, and equals to $0$ if it is not received due to occlusion effects or weak scattering at some sites~\cite{ref:caire_SBL}. The attenuation coefficient $\bar{c}_{x,y,k,m}$ is contributed by path loss and radar cross section (RCS). Following the Swerling-II target model \cite{ref:RCS,ref:RCS2}, we assume $\bar{c}_{x,y,k,m} \sim \mathcal{CN}(0, \gamma_{x,y,k}), \forall x,y,k,m$, which stays constant during the transmission of $L$ symbols and changes i.i.d. in different frames. The variance $\gamma_{x,y,k}$ satisfies $\gamma_{x,y,k} = \bar{\gamma}_{r,x,y} \gamma_{\beta,x,y,k}$, where $\bar{\gamma}_{r,x,y}$ accounts for the scattering intensity and $\gamma_{\beta,x,y,k}$ denotes the path-loss.

The vectorized form of the received signal in \eqref{eq:receive_Y} is given~by
\begin{equation}\label{eq:receive_Y_Nc_vec}
    \mathbf{y}_{k,m} = \operatorname{vec}\!\left( \mathbf{Y}_{k,m} \right) 
    \!= \! \iint_{ \mathcal{S} } \!  \mathbf{v}_{x,y,k} c_{x,y,k,m}  {\rm d} x {\rm d} y    
    +   {\mathbf{z}}_{k,m} , \; \forall k,m, 
\end{equation} 
where $\mathbf{v}_{x,y,k} = \mathbf{X}^T {\mathbf{a}}\! \left(\phi_{x,y}\right)  \otimes \mathbf{b}\! \left(\theta_{x,y,k} \right)$ and $ {\mathbf{z}}_{k,m} = \operatorname{vec}\left(   {\mathbf{Z}}_{k,m} \right)$. The concatenation of $\mathbf{y}_{k,m}$ over $M$ frames is denoted as ${\mathbf{Y}}_k = \left[  {\mathbf{y}}_{k,1}, \ldots,  {\mathbf{y}}_{k,M} \right]$. Since both attenuation components and noises are i.i.d. Gaussian distributed over the frames, the columns of $\mathbf{Y}_k$ are i.i.d. distributed according to $\mathcal{CN}\left( \bm{0}, \bm{\Sigma}_{k,0} \right)$ with the covariance matrix $\bm{\Sigma}_{k,0}$ given by 
\begin{equation} \label{eq:receive_Y_sigmak0}
  {\bm{\Sigma}}_{k,0} \!=\! \iint_{ \mathcal{S}} \! \alpha_{x,y,k} \gamma_{x,y,k}   
  \mathbf{v}_{x,y,k}  \mathbf{v}_{x,y,k}^H   {\rm d} x {\rm d} y 
  +  {\sigma^2_z} \mathbf{I}_{ L N_{\rm rx}} , \;\;\!\! \forall k. 
\end{equation} 
Thus, the likelihood function of ${\mathbf{Y}}_k$ is given by
\begin{align}
  & p \left( {\mathbf{Y}}_k | \left\{ \alpha_{x,y,k} , \gamma_{x,y,k} : [x,y]^T \in \mathcal{S} \right\} \right)  \notag\\ 
  & = |\pi \bm{\Sigma}_{k,0} |^{-M} e^{ -\operatorname{tr} \left( M \bm{\Sigma}_{k,0}^{-1} \hat{\bm{\Sigma}}_k \right) }, \; \forall k ,  \label{eq:ML0}
\end{align}
where $\hat{\bm{\Sigma}}_k$ denotes the sample covariance matrix of the columns of the observation $ {\mathbf{Y}}_k$ given by
\begin{equation} \label{eq:receive_Y_sigmakhat}
    \hat{\bm{\Sigma}}_k = \frac{1}{M}  {\mathbf{Y}}_k  {\mathbf{Y}}_k^H, \;\; \forall k . 
\end{equation}
In this work, we aim to reconstruct the image of the extended targets based on this sample covariance matrix.

\section{Discrete Representation and Problem Statement} \label{Sec:Grid}

Different from point targets, extended targets usually comprise an infinite number of inseparable scattering points, rendering precise parameter estimation for individual scatterers impossible~\cite{ref:RCS2}. Consequently, the objective of imaging for extended targets is not to localize each single scatterer, but rather to create a high-resolution map of the reflectivity of the environment. To achieve this goal, we discretize the rectangular region of interest (RoI), denoted as $\mathcal{A} = [x_1, x_2] \times [y_1, y_2]$, into $Q$ grids for each RBS\footnote{In this work, we assume that the RoI $\mathcal{A}$ is known in advance. In practice, the RoI depends on applications. For example, if we want to image the targets in a room, we can set the room as RoI and perform our proposed algorithm in the room.}. The position of the $q$-th grid point at the $k$-th RBS is denoted as $\mathbf{p}_{k,q} = \left[ p_{x,k,q},p_{y,k,q} \right]^T$. The scattering of extended targets is approximated as that contributed by a set of scatterers located at these grid points~\cite{ref:caire_SBL}. Then, the channel matrix $\mathbf{H}_{k,m}$ in \eqref{eq:model_H} is approximated as $\tilde{\mathbf{H}}_{k,m} $ given~by 
\begin{equation}\label{eq:model_H_approx} 
    \tilde{\mathbf{H}}_{k,m} 
    = \sum_{q=1}^{Q}  c_{k,q,m} 
    \mathbf{b}\! \left(\theta_{k,q} \right) \mathbf{a}^T\! \left(\phi_{k,q}\right) ,  
\end{equation} 
where $\phi_{k,q}$ and $\theta_{k,q}$ denote the AoD and AoA corresponding to the $q$-th grid at the $k$-th RBS obtained from \eqref{eq:AoD} and \eqref{eq:AoA} by replacing $[x,y]^T$ to $\mathbf{p}_{k,q}$, respectively, and $c_{k,q,m} = \alpha_{k,q} \bar{c}_{k,q,m}$ denotes the effective attenuation coefficient of the path from the TBS to grid point $q$ (particular to the set of grids adopted by RBS $k$) and then to RBS $k$. Here, the binary variable $\alpha_{k,q}$ equals to $1$ if the $q$-th grid is in the field of view of RBS $k$, and equals to $0$ otherwise, $\forall k,q$. Due to the central limit theorem when approximating the scattering from infinitely many points in a grid cell as that from a single point, we model the attenuation coefficient $\bar{c}_{k,q,m}$ as a CSCG variable with mean $0$ and variance $\gamma_{k,q} = \bar{\gamma}_{r,k,q} \gamma_{\beta,k,q}$, $\forall k,q,m$, where $\bar{\gamma}_{r,k,q}$ accounts for the RCS part and $\gamma_{\beta,k,q}$ accounts for the path-loss. Since the positions of grid points and BSs are known, we assume that the path-loss vector $\bm{\gamma}_{\beta,k} = \left[ \gamma_{\beta,k,1} , \ldots , \gamma_{\beta,k,Q} \right]^T$ is known in advance, $\forall k$. For the $k$-th RBS, denote the effective scattering intensity corresponding to grid $q$ as $\gamma_{r,k,q} = \alpha_{k,q} \bar{\gamma}_{r,k,q}$. We denote $\bm{\gamma}_{r,k} = \left[ \gamma_{r,k,1} , \ldots , \gamma_{r,k,Q} \right]^T$, which is to be estimated for determining the single-view image.

The signal ${\mathbf{Y}}_{k,m}$ in \eqref{eq:receive_Y} is approximated as $\tilde{\mathbf{Y}}_{k,m}$ given by
\begin{equation}\label{eq:receive_Y_Nc_approx}
  \tilde{\mathbf{Y}}_{k,m} = \tilde{\mathbf{H}}_{k,m} \mathbf{X} +  {\mathbf{Z}}_{k,m} , \;\; \forall k,m. 
\end{equation} 
After vectorization, we have 
\begin{equation}\label{eq:receive_Y_Nc_approx_vec}
    \tilde{\mathbf{y}}_{k,m} = \operatorname{vec} ( \tilde{\mathbf{Y}}_{k,m}  ) 
    = \mathbf{V}_{k}\mathbf{c}_{k,m} +   {\mathbf{z}}_{k,m} , \;\; \forall k,m, 
\end{equation} 
where $\mathbf{c}_{k,m} = \left[ c_{k,1,m},  \ldots, c_{k,Q,m}\right]^T \in \mathbb{C}^{Q}$ and $\mathbf{V}_{k} = \left[ \mathbf{v}_{k,1}, \ldots, \mathbf{v}_{k,Q} \right] \in \mathbb{C}^{ L N_{\rm rx} \times Q} $ with 
\begin{equation}
  \mathbf{v}_{k,q} = \bar{\mathbf{a}}\! \left(\phi_{k,q}\right)  \otimes  \mathbf{b}\! \left(\theta_{k,q} \right), \;\; \forall k,q, 
\end{equation}
\begin{equation}
  \bar{\mathbf{a}}\! \left(\phi_{k,q}\right)
  = \mathbf{X}^T {\mathbf{a}}\! \left(\phi_{k,q}\right), \;\; \forall k,q . 
\end{equation} 
Then, the concatenation of $\tilde{\mathbf{y}}_{k,m}$ over $M$ frames is given by
\begin{equation}\label{eq:receive_Y_Nc_approx_vec_M}
   \tilde{\mathbf{Y}}_k = \mathbf{V}_k \mathbf{C}_k +  {\mathbf{Z}}_k , \;\; \forall k,  
\end{equation}
where $\tilde{\mathbf{Y}}_k = \left[ \tilde{\mathbf{y}}_{k,1}, \ldots, \tilde{\mathbf{y}}_{k,M} \right]$, $\mathbf{C}_k = \left[ \mathbf{c}_{k,1}, \ldots, \mathbf{c}_{k,M} \right]$, and $ {\mathbf{Z}}_k = \left[  {\mathbf{z}}_{k,1}, \ldots,  {\mathbf{z}}_{k,M} \right]$. Given the effective scattering intensity $\bm{\gamma}_{r,k}$ and grid position $\mathbf{P}_{k} = \left[\mathbf{p}_{k,1}, \ldots, \mathbf{p}_{k,Q}\right]$, the columns of $\tilde{\mathbf{Y}}_k$ are i.i.d. distributed according to $\mathcal{CN}\left( \bm{0}, \bm{\Sigma}_k \right)$, where the covariance matrix $\bm{\Sigma}_k$ is given by 
\begin{equation} \label{eq:receive_Y_sigmak}
    {\bm{\Sigma}}_k =  \mathbf{V}_k \bm{\Gamma}_{r,k} \bm{\Gamma}_{\beta,k} \mathbf{V}_k^H + \sigma^2_z  \mathbf{I}_{ L N_{\rm rx}} , \;\; \forall k, 
\end{equation}
with $\bm{\Gamma}_{r,k} = \operatorname{diag} \! \left\{  \bm{\gamma}_{r,k} \right\}$ and $\bm{\Gamma}_{\beta,k} = \operatorname{diag}\! \left\{  \bm{\gamma}_{\beta,k} \right\}$. By treating the quantized matrix ${\bm{\Sigma}}_k$ as the covariance matrix in the likelihood function of ${\mathbf{Y}}_k$ given in \eqref{eq:ML0}, we have 
\begin{equation}
  p( {\mathbf{Y}}_k | \bm{\gamma}_{r,k}, \mathbf{P}_{k} )  
  = |\pi \bm{\Sigma}_k|^{-M} e^{ -\operatorname{tr}  \left( M \bm{\Sigma}_k^{-1} \hat{\bm{\Sigma}}_k \right)  }, \; \forall k .   \label{eq:ML}
\end{equation}

Based on the discretized model established above, we propose a two-phase approach to address the environment reconstruction challenge. In Phase I, each RBS recovers an individual image of extended targets from its own view. Specifically, for the $k$-th RBS, we formulate the single-view imaging problem as a novel estimation problem of the effective scattering intensity $\bm{\gamma}_{r,k}$ and grid position $\mathbf{P}_{k}$, as will be introduced in Section~\ref{Sec:image}. Based on the estimated results, the target can be reconstructed by identifying grids with high intensity values. However, due to occlusion effects and weak RCS at some sites, each RBS has only a limited field of view about the target~\cite{ref:PHDfusion,tian2023distributed,caire_multistatic}, as shown in Fig.~\ref{fig:SM}. Even within its field of view, the performance is limited by the RBS's resolution. Thus, in Phase II, we fuse these individual results to construct a higher-quality image of all the extended targets, as will be introduced in Section~\ref{Sec:fusion}. Since the targets can be observed from different angular directions, significant diversity gain can be leveraged by fusing individual results.

\section{Phase I: Covariance-based Single-View Imaging} \label{Sec:image}

This section addresses the single-view imaging problem at each RBS. First, based on the ML criterion and target geometric characteristics, we formulate the imaging problem as an estimation problem of grid positions and effective scattering intensities. This problem is then solved applying the AO framework. Subsequently, we theoretically reveal the importance of grid optimization in achieving high-quality imaging of extended targets.

\subsection{Problem Formulation}

Conventional imaging methods, such as FFT and CS-based schemes, face inherent performance limitations due to their requirement to estimate high-dimensional instantaneous channel matrix $\mathbf{C}_k$ for determining the set of grids comprising scatterers. To overcome this challenge, we propose a covariance-based imaging scheme that aims to estimate the statistical effective scattering intensity $\bm{\gamma}_{r,k}$ instead of $\mathbf{C}_k$, thereby dramatically reducing the number of parameters to be estimated. Moreover, considering that the number of resolvable grid points cannot be excessively increased due to limited resolution, we propose to dynamically adjust grid points to conform to the geometry of targets. As a result, we formulate the single-view imaging problem at each RBS $k$ as a novel estimation problem of both scattering intensity $\bm{\gamma}_{r,k}$ and grid position $\mathbf{P}_{k}$. Building upon the grid representation introduced earlier, this problem admits an ML estimation framework by minimizing the approximate negative log-likelihood cost function given~by
\begin{equation}\label{eq:distributed_ML}
    - \frac{1}{M} \ln p( {\mathbf{Y}}_k | \bm{\gamma}_{r,k}, \mathbf{P}_{k} )  
    \propto  \ln \left| \bm{\Sigma}_{k} \right| + \operatorname{tr}\left( \bm{\Sigma}^{-1}_{k} \hat{\bm{\Sigma}}_{k} \right)  .
\end{equation} 
Moreover, the scatterers of extended targets typically exhibit clustering properties, forming spatially contiguous aggregations instead of isolated points. To leverage this structural prior, we first define the graph Laplacian matrix that encodes the grid adjacency relationship at the $k$-th RBS, whose element on the $q$-th row and the $q'$-th column is given as
\begin{equation} \label{eq:distributed_penalty_L}
    [\mathbf{L}_k]_{q,q'} =  
    \left\{ \!\!\!
             \begin{array}{ll}
             |\mathcal{N}_{k,q}| ,  &  \text{if}\;\; q=q',\!\! \\ 
             -1, &  \text{if}\;\; (q, q') \in \mathcal{N}_k ,\!\! \\
             0,  &  \text{otherwise} ,
             \end{array} 
             \right.  
\end{equation}
where the set $\mathcal{N}_k$ includes adjacent pairs after spatial discretization at the $k$-th RBS and the set $\mathcal{N}_{k,q}$ includes neighbor grids of the $q$-th grid observed at RBS $k$. 
Then, we introduce a penalty term as follows
\begin{align}
   & \left( \bm{\gamma}_{r,k} \odot \bm{\gamma}_{\beta,k} \right)^T \mathbf{L}_k \left( \bm{\gamma}_{r,k} \odot \bm{\gamma}_{\beta,k} \right) \notag\\
   & =  {\sum}_{(q, q') \in \mathcal{N}_k} \frac{1}{2} (\gamma_{r,k,q}\gamma_{\beta,k,q} - \gamma_{r,k,q'}\gamma_{\beta,k,q'} )^2.  \label{eq:distributed_penalty}
\end{align}  
By penalizing large energy differences between adjacent grids, it enforces spatial clusters in the reconstructed image while effectively suppressing isolated artifacts.

Thus, the penalized ML cost function is given by  
\begin{align}
    \mathcal{J}( \bm{\gamma}_{r,k}, \mathbf{P}_{k} ) & = \ln \left| \bm{\Sigma}_{k} \right| + \operatorname{tr}\left( \bm{\Sigma}^{-1}_{k} \hat{\bm{\Sigma}}_{k} \right) \notag\\
    & \;\;\; + \delta \left( \bm{\gamma}_{r,k} \odot \bm{\gamma}_{\beta,k} \right)^T \mathbf{L}_k \left( \bm{\gamma}_{r,k} \odot \bm{\gamma}_{\beta,k} \right), \label{eq:Problem1_obj}
\end{align}
where $\delta>0$ denotes the penalty parameter used to promote clusters. The optimization problem is then formulated as  
\begin{subequations} \label{eq:Problem1}
\begin{align}
  & \mathop{\min}\limits_{ \bm{\gamma}_{r,k}, \mathbf{P}_{k}  }  
& & \mathcal{J}( \bm{\gamma}_{r,k}, \mathbf{P}_{k} )  \\
&\;\;\;\; \text{s.t.} 
& & \gamma_{r,k,q} \geq 0,  \quad \forall q  \label{eq:constraint1}\\
&&& \mathbf{p}_{k,q} \in \mathcal{A},  \quad \forall q \label{eq:constraint2}\\ 
&&& \left\| \mathbf{p}_{k,q} - \mathbf{p}_{k,q}^{(0)} \right\|_2 \leq d_{\max},  \quad \forall q. \label{eq:constraint3}
\end{align} 
\end{subequations}
where the constraint \eqref{eq:constraint1} restricts energy is non-negative; the condition \eqref{eq:constraint2} restricts all grid points are within the RoI; and the constraint \eqref{eq:constraint3} not only prevents the grid index ambiguity problem (e.g., swaps between $(q,q')$ and $(q',q)$ coordinates) inherent to permutation-invariant formulations, but also guarantees the separability of adjacent grid points provided that initial positions $\mathbf{p}_{k,q}^{(0)}$'s are not too close. In the formulated covariance-based optimization problem, the sample covariance matrix $\hat{\bm{\Sigma}}_{k}$ of received signals provides sufficient information for imaging. In the following, we introduce an AO framework to iteratively refine $\bm{\gamma}_{r,k}$ and $\mathbf{P}_k$, with the outline of the proposed algorithm provided in Algorithm \ref{alg:image}.

\subsection{\texorpdfstring{Optimizing ${\bm{\gamma}}_{r,k}$ with Fixed ${\bm P}_k$}{Optimizing r with fixed P}} \label{Sec:opt1}

For fixed ${\bm P}_k$, the optimization with respect to ${\bm{\gamma}}_{r,k}$ is not convex considering that the objective function $\mathcal{J}( \bm{\gamma}_{r,k}, \mathbf{P}_{k} )$ is the sum of a concave log-determinant term and two convex terms. To address this problem, we apply the coordinate descent algorithm to update ${\bm{\gamma}}_{r,k}$. As shown in Algorithm \ref{alg:image}, at each step of one iteration, we optimize $\mathcal{J}( \bm{\gamma}_{r,k}, \mathbf{P}_{k} )$ with respect to one of the coordinates of $\bm{\gamma}_{r,k}$. Specifically, given ${\bm{\gamma}}_{r,k}$ and ${\mathbf{P}}_{k}$ obtained in the previous step, the coordinate optimization with respect to $\gamma_{r,k,q}$ is equivalent to the optimization of the increment $d$ in $\gamma_{r,k,q}$. The resulting optimization problem to determine $d$ can be expressed as 
\begin{equation}  \label{eq:Problem_r}
  \mathop{\min}\limits_{ -\gamma_{r,k,q} \leq d \leq 1-\gamma_{r,k,q} }  \mathcal{J}( \bm{\gamma}_{r,k} + d \mathbf{e}_q, \mathbf{P}_{k} ), 
\end{equation}
where $\mathbf{e}_{q}$ denotes the $q$-th canonical basis vector with a single $1$ at its $q$-th coordinate and zero elsewhere. The objective function in \eqref{eq:Problem_r} satisfies that 
\begin{align}
    & \mathcal{J}( \bm{\gamma}_{r,k} + d \mathbf{e}_q, \mathbf{P}_{k} ) \propto f_{k,q}(  d ) \label{eq:Problem_r1} \\
    & = \ln \!\left( 1 + d  \gamma_{\beta,k,q}  \mathbf{v}_{k,q}^H \bm{\Sigma}^{-1}_k \mathbf{v}_{k,q} \right)  
    - \tfrac{ d  \gamma_{\beta,k,q}  \mathbf{v}_{k,q}^H \bm{\Sigma}^{-1}_k \hat{\bm{\Sigma}}_k \bm{\Sigma}^{-1}_k \mathbf{v}_{k,q}  }{ 1+d \gamma_{\beta,k,q} \mathbf{v}_{k,q}^H \bm{\Sigma}^{-1}_k \mathbf{v}_{k,q}} \notag\\ 
    & \;\;   +\! \frac{\delta}{2} {\sum}_{q' \in \mathcal{N}_{k,q}} \! (d \gamma_{\beta,k,q} \!+\! \gamma_{r,k,q}\gamma_{\beta,k,q} \!-\! \gamma_{r,k,q'}\gamma_{\beta,k,q'} )^2 , \label{eq:Problem_r2}
\end{align}
where \eqref{eq:Problem_r2} is obtained applying the determinant identity
\begin{equation} \label{eq:Problem_r2_det}
    \left| \bm{\Sigma}_k + d \gamma_{\beta,k,q} \mathbf{v}_{k,q} \mathbf{v}_{k,q}^H  \right| 
    = \left( 1 + d \gamma_{\beta,k,q} \mathbf{v}_{k,q}^H \bm{\Sigma}^{-1}_k \mathbf{v}_{k,q} \right)  \left| \bm{\Sigma}_k \right| ,
\end{equation}
and the Sherman-Morrison rank-1 update identity~\cite{ref:rank1} 
\begin{equation} \label{eq:Problem_r2_rank1}
    \left( \bm{\Sigma}_k + d \gamma_{\beta,k,q} \mathbf{v}_{k,q} \mathbf{v}_{k,q}^H \right)^{-1} 
    = \bm{\Sigma}_k^{-1} - \tfrac{d \gamma_{\beta,k,q}  \bm{\Sigma}_k^{-1} \mathbf{v}_{k,q} \mathbf{v}_{k,q}^H \bm{\Sigma}_k^{-1}  }{ 1 + d \gamma_{\beta,k,q} \mathbf{v}_{k,q}^H \bm{\Sigma}^{-1}_k \mathbf{v}_{k,q} } . 
\end{equation}

The function $f_{k,q}(d)$ is well-defined only when $d > d_0 = -\tfrac{1}{  \gamma_{\beta,k,q} \mathbf{v}_{k,q}^H \bm{\Sigma}^{-1}_k \mathbf{v}_{k,q} }$. 
Taking the derivative of $f_{k,q}(d)$ yields
\begin{equation}
    f'_{k,q}(d ) = \tfrac{ c_3 d^3 + c_2 d^2 + c_1 d + c_0 }{ \left( 1+d  a_{k,q} \right)^2 } ,  \label{eq:gamma_cubic}
\end{equation} 
where
\begin{equation}
    c_3 = a_{k,q}^2 c_{k,q} , 
\end{equation}
\begin{equation}
    c_2 = 2 a_{k,q} c_{k,q} + a_{k,q}^2 e_{k,q} , 
\end{equation}
\begin{equation}
    c_1 = a_{k,q}^2 + c_{k,q} + 2 a_{k,q} e_{k,q}  , 
\end{equation}
\begin{equation}
    c_0 = a_{k,q} - b_{k,q} + e_{k,q} , 
\end{equation}
\begin{equation}
    a_{k,q} =  \gamma_{\beta,k,q} \mathbf{v}_{k,q}^H \bm{\Sigma}^{-1}_k \mathbf{v}_{k,q}, 
\end{equation}
\begin{equation}
    b_{k,q} =   \gamma_{\beta,k,q} \mathbf{v}_{k,q}^H \bm{\Sigma}^{-1}_k \hat{\bm{\Sigma}}_k \bm{\Sigma}^{-1}_k \mathbf{v}_{k,q} ,
\end{equation}
\begin{equation}
    c_{k,q} =   \delta |\mathcal{N}_{k,q}| {\gamma_{\beta,k,q}^2} , 
\end{equation}
\begin{equation}
    e_{k,q} =   \delta \gamma_{\beta,k,q} {\sum}_{q' \in \mathcal{N}_{k,q}} (\gamma_{r,k,q} \gamma_{\beta,k,q} -\gamma_{r,k,q'} \gamma_{\beta,k,q'} ). 
\end{equation}
The derivative $f'_{k,q}(d )$ involves a cubic function, which admits closed-form roots. Denote the set of real roots in the feasible region as
\begin{equation}\label{eq:gamma_rootset}
  \mathcal{D} = \left\{ d : \!f'_{k,q}(d ) = 0, \Im(d) = 0,   \Re(d)  \geq \max\!\left\{ -\gamma_{r,k,q} , d_0 \right\}  \right\}\!. 
\end{equation} 
The optimal increment $d$ is obtained by selecting $\max\{-\gamma_{r,k,q},d_0\}$ or root in the set $\mathcal{D}$ with the minimum objective function value. Then, $\gamma_{r,k,q}$ can be updated as $\gamma_{r,k,q} \leftarrow \gamma_{r,k,q} + d$. Applying the Sherman-Morrison rank-1 update identity, the covariance matrix is updated as 
\begin{equation}\label{eq:gamma_covariance}
  \bm{\Sigma}^{-1}_k \leftarrow \bm{\Sigma}^{-1}_k - d  \tfrac{  \gamma_{\beta,k,q}  \bm{\Sigma}^{-1}_k \mathbf{v}_{k,q} \mathbf{v}_{k,q}^H \bm{\Sigma}^{-1}_k }{ 1+d \gamma_{\beta,k,q} \mathbf{v}_{k,q}^H \bm{\Sigma}^{-1}_k \mathbf{v}_{k,q}} . 
\end{equation}
Following similar procedure, all coordinates of ${\bm{\gamma}}_{r,k}$ are updated once in each iteration, as shown in Algorithm~\ref{alg:image}.

\begin{algorithm}[t] 
\caption{Covariance-based Single-View Imaging}
\label{alg:image} 
\begin{algorithmic}[1]  
{ \small
\REQUIRE 
The sample covariance matrix $\hat{\bm{\Sigma}}_k$, 
RoI $\mathcal{A}$,
noise variance $\sigma_{z}^2$, 
tolerances $\epsilon_1$ and $\epsilon_2$,
and iterations ${\text{Iter}_{\rm max}}, {\text{Iter}_1}$, and $\text{Iter}_2$.

\ENSURE 
The resulting estimate $\bm{\gamma}_{r,k}$ and $\mathbf{P}_{k}$.

\STATE \textbf{Initialize:} 
$\bm{\Sigma}_k = \sigma_{z}^2 \mathbf{I}$, $\bm{\gamma}_{r,k} = \bm{0}$, and uniform grids $\mathbf{p}_{k,q}^{(0)}, \forall q$.

\FOR{$j = 1,2,\ldots, I_{\max}$}

\STATE \textbf{\% Update of $\bm{\gamma}_{r,k}$}
\FOR{$j_1 = 1,2,\ldots, I_1$}  
\STATE Select an index set $\mathcal{Q}$ from the random permutation of $[Q]$.

\FOR{$q \in \mathcal{Q}$} 
\STATE Solve the cubic function in \eqref{eq:gamma_cubic} and obtain $\mathcal{D}$ \eqref{eq:gamma_rootset}.  
\STATE Set $d = \arg \min_{d\in \mathcal{D} \cup  \max\{-\gamma_{r,k,q},d_0\}   } \! f_{k,q}(  d ) $.
\STATE Update $\gamma_{r,k,q} \leftarrow \gamma_{r,k,q} + d$.
\STATE Update $\Sigma_k^{-1}$ applying \eqref{eq:gamma_covariance}.
\ENDFOR
\ENDFOR

\STATE \textbf{\% Update of $\mathbf{P}_k$}
\FOR{$j_2 = 1,2,\ldots, I_2$}  
\STATE Calculate the gradient $\nabla_{\bm \psi} f_{k,x,y}({\bm {\psi}} )$ applying \eqref{eq:image_P_grad} - \eqref{eq:image_P_grad_end}.
\STATE Compute the stepsize $\xi$ by Armijo rule. 
\STATE Update $\bm \psi$ applying \eqref{eq:P_GD}. 
\STATE Apply the rectangle projection operator and the Euclidean ball projection operator in \eqref{eq:image_P_proj} for $\mathbf{P}_k$. 
\ENDFOR

\STATE If $ \tfrac{ \left\| \bm{\gamma}_{r,k}^{(j)} - \bm{\gamma}_{r,k}^{(j-1)} \right\|_2^2}{\left\|  \bm{\gamma}_{r,k}^{(j-1)} \right\|_2^2} \leq \epsilon_1$ and $\nabla_{\bm \psi} f_{k,x,y}({\bm {\psi}} ) \leq \epsilon_2$, stop.

\ENDFOR

}

\end{algorithmic}
\end{algorithm}

\subsection{\texorpdfstring{Optimizing ${\bm P}_k$ with Fixed ${\bm{\gamma}}_{r,k}$}{Optimizing P with fixed r}} \label{Sec:opt2} 

Given ${\bm{\gamma}}_{r,k}$ obtained in the previous step, we aim to optimize grid positions in this part. Denote the set of grid indexes with positive scattering intensity estimates as $\mathcal{Q}_{a,k} = \{q\in [Q]: \gamma_{r,k,q}>0\}$. Let $\bm{\psi} = \operatorname{vec}\left( {\mathbf{P}}_{a,k} \right)$, where ${\mathbf{P}}_{a,k}$ is a submatrix of ${\mathbf{P}}_{k}$ with the column indices lying in $\mathcal{Q}_{a,k}$. In the following, we aim to update $\bm{\psi}$, while keeping the grid parameters indexed by the set $[Q] \backslash \mathcal{Q}_{a,k}$ equal to those in the previous iteration. Specifically, given the effective scattering intensity ${\bm{\gamma}}_{r,k}$ and grid positions indexed by the set $q\in [Q] \backslash \mathcal{Q}_{a,k}$, the cost function with respect to $\bm{\psi}$ is expressed as
\begin{equation}\label{eq:cost_P}
    \mathcal{J}( \bm{\gamma}_{r,k}, \mathbf{P}_{k} )  \propto  f_{k,x,y} (\bm{\psi}) 
    =  \ln  \left| \bm{\Sigma}_{k } \right|  +  \operatorname{tr}  \left( \bm{\Sigma}_{k}^{-1}   \hat{\bm{\Sigma}}_{k} \right) . 
\end{equation}
It is challenging to find the optimal $\bm{\psi}$ that minimizes the cost function in \eqref{eq:cost_P} since it is non-convex with respect to $\bm{\psi}$. Thus, a gradient descent approach is employed. Specifically, the grid parameters in $\bm{\psi}$ are updated as   
\begin{equation} \label{eq:P_GD}
    \bm{\psi} \leftarrow \bm{\psi} - \xi  \nabla_{\bm \psi} f_{k,x,y}({\bm {\psi }})  , 
\end{equation}
where the step size $\xi$ is determined by the Armijo rule. The gradient $\nabla_{\bm \psi} f_{k,x,y}({\bm {\psi}} )$ is given by
\begin{align} \label{eq:image_P_grad}
  \nabla_{\bm \psi} f_{k,x,y}({\bm {\psi}} )  
  = & \left[ \tfrac{\partial f_{k,x,y}(\bm{\psi}) }{\partial p_{x,k, \mathcal{Q}_{a,k} (1) } } , \tfrac{\partial f_{k,x,y}(\bm{\psi}) }{\partial p_{y,k,\mathcal{Q}_{a,k} (1)} }, \ldots ,   \right.  \notag\\
  &  \;\;\left.   
  \tfrac{\partial f_{k,x,y}(\bm{\psi}) }{\partial p_{x,k, \mathcal{Q}_{a,k} ( |\!\mathcal{Q}_{a,k}\!| )  } }  , \tfrac{\partial f_{k,x,y}(\bm{\psi}) }{\partial p_{y,k, \mathcal{Q}_{a,k} ( |\!\mathcal{Q}_{a,k}\!| ) } }  \right]^T \!,
\end{align}
where for each $q\in  \mathcal{Q}_{a,k} $, we have
\begin{align} 
    & \tfrac{\partial f_{k,x,y}( {\bm {\psi}} )  }{\partial p_{x,k,q}}  
    = \tfrac{ 2 \gamma_{r,k,q} \gamma_{\beta,k,q} \Re\left\{  \tilde{\mathbf{v}}_{k,q,x}^H  \bm{\Sigma}^{-1}_{k,\backslash q} 
    \left( \mathbf{I}_{L N_{\rm rx}} - \hat{\bm{\Sigma}}_k 
    \bm{\Sigma}^{-1}_{k,\backslash q}   \right) {\mathbf{v}}_{k,q}   \right\} }{1 + \gamma_{k,q}  {\mathbf{v}}^H_{k,q}  \bm{\Sigma}^{-1}_{k,\backslash q}   {\mathbf{v}}_{k,q} } \notag\\
    & \;\;\;\;\;   + \! \tfrac{ \!2 \gamma_{r,k,q}^2 \gamma_{\beta,k,q}^2 {\mathbf{v}}^H_{k,q} 
    \bm{\Sigma}^{-1}_{k,\!\backslash q} 
    \hat{\bm{\Sigma}}_k 
    \bm{\Sigma}^{-1}_{k,\!\backslash q} 
     {\mathbf{v}}_{k,q} 
    \Re  \left\{ \! \tilde{\mathbf{v}}_{k,q,x}^H   
    \bm{\Sigma}^{-1}_{k,\!\backslash q} 
     {\mathbf{v}}_{k,q}\! \right\}
     }{ \left( 1 + \gamma_{k,q}  {\mathbf{v}}^H_{k,q}  \bm{\Sigma}^{-1}_{k,\backslash q}   {\mathbf{v}}_{k,q} \right)^2 } , \label{eq:image_P_gradx}
\end{align}
\begin{align}
    & \tfrac{\partial f_{k,x,y}( {\bm {\psi}} )  }{\partial p_{y,k,q}} = \tfrac{ 2 \gamma_{r,k,q} \gamma_{\beta,k,q} \Re\left\{ \tilde{\mathbf{v}}_{k,q,y}^H  \bm{\Sigma}^{-1}_{k,\backslash q} 
    \left( \mathbf{I}_{L N_{\rm rx} } - \hat{\bm{\Sigma}}_k 
    \bm{\Sigma}^{-1}_{k,\backslash q}   \right) {\mathbf{v}}_{k,q}   \right\} }{1 + \gamma_{k,q}  {\mathbf{v}}^H_{k,q}  \bm{\Sigma}^{-1}_{k,\!\backslash q}   {\mathbf{v}}_{k,q} } \notag\\
    & \;\;\;\; + \!\tfrac{\! 2 \gamma_{r,k,q}^2 \gamma_{\beta,k,q}^2 {\mathbf{v}}^H_{k,q} 
    \bm{\Sigma}^{-1}_{k,\!\backslash q} 
    \hat{\bm{\Sigma}}_k 
    \bm{\Sigma}^{-1}_{k,\!\backslash q} 
     {\mathbf{v}}_{k,q} 
    \Re \left\{ \!  \tilde{\mathbf{v}}_{k,q,y}^H  
    \bm{\Sigma}^{-1}_{k,\!\backslash q} 
     {\mathbf{v}}_{k,q} \!\right\}
     }{ \left( 1 + \gamma_{k,q}  {\mathbf{v}}^H_{k,q}  \bm{\Sigma}^{-1}_{k,\backslash q}   {\mathbf{v}}_{k,q} \right)^2 } , \label{eq:image_P_grady}
\end{align} 
\begin{align} 
  \bm{\Sigma}^{-1}_{k,\backslash q} &= \left( \bm{\Sigma}_k - \gamma_{k,q} {\mathbf{v}}_{k,q} {\mathbf{v}}_{k,q}^H \right)^{-1} \\
  & =\bm{\Sigma}_k^{-1} + \tfrac{ \gamma_{k,q}  \bm{\Sigma}_k^{-1} \mathbf{v}_{k,q} \mathbf{v}_{k,q}^H \bm{\Sigma}_k^{-1}  }{ 1 - \gamma_{k,q} \mathbf{v}_{k,q}^H \bm{\Sigma}^{-1}_k \mathbf{v}_{k,q} }, \label{eq:image_P_grad_inv}
\end{align}
\begin{equation} 
   \tilde{\mathbf{v}}_{k,q,x}  = (\mathbf{X}^T \otimes \mathbf{I}_{N_{\rm rx} } ) \bar{\bm{\Lambda}}_{k,q,x} \left( {\mathbf{a}}\! \left(\phi_{k,q}\right) \otimes \mathbf{b}\! \left(\theta_{k,q} \right)  \right) ,
\end{equation}
\begin{equation} 
   \tilde{\mathbf{v}}_{k,q,y}  = (\mathbf{X}^T \otimes \mathbf{I}_{N_{\rm rx} } ) \bar{\bm{\Lambda}}_{k,q,y}  \left( {\mathbf{a}}\! \left(\phi_{k,q}\right) \otimes \mathbf{b}\! \left(\theta_{k,q} \right)   \right)  ,
\end{equation} 
\begin{equation} 
  \bar{{\bm \Lambda}}_{k,q,x} = 
  a_{ k,q,x}   {\bm{\Lambda}}_{N_{\rm tx}}  \otimes  \mathbf{I}_{N_{\rm rx} } 
  + b_{ k,q,x} \mathbf{I}_{N_{\rm tx} }  \otimes  {\bm{\Lambda}}_{N_{\rm rx}} ,
\end{equation} 
\begin{equation} 
  \bar{{\bm \Lambda}}_{k,q,y} = 
  a_{ k,q,y}   {\bm{\Lambda}}_{N_{\rm tx}}  \otimes  \mathbf{I}_{N_{\rm rx} } 
  + b_{ k,q,y} \mathbf{I}_{N_{\rm tx} }  \otimes  {\bm{\Lambda}}_{N_{\rm rx}} ,
\end{equation} 
\begin{equation}
    \bm{\Lambda}_{N_{\rm tx}} = -j\pi\operatorname{diag} \left\{ 0 , 1 , \ldots , N_{\rm tx}-1  \right\},
\end{equation} 
\begin{equation}
    \bm{\Lambda}_{N_{\rm rx}} = -j\pi\operatorname{diag} \left\{ 0 , 1 , \ldots , N_{\rm rx}-1  \right\},
\end{equation}  
\begin{equation}
  a_{ k,q,x} = \tfrac{ \left( x_{\rm tx} - p_{x,k,q} \right) \left( y_{\rm tx} - p_{y,k,q} \right) }{ \left(  \left( x_{\rm tx} - p_{x,k,q} \right)^2 + \left( y_{\rm tx} - p_{y,k,q} \right)^2  \right)^{\frac{3}{2}} }  ,
\end{equation}
\begin{equation}
   b_{ k,q,x}  = \tfrac{ \left( x_{{\rm rx},k} - p_{x,k,q} \right) \left( y_{{\rm rx},k} - p_{y,k,q} \right) }{ \left(  \left( x_{{\rm rx},k} - p_{x,k,q} \right)^2 + \left( y_{{\rm rx},k} - p_{y,k,q} \right)^2  \right)^{\frac{3}{2}} }  ,
\end{equation} 
\begin{equation}
  a_{k,q,y} = \tfrac{ -\left( x_{{\rm tx} } - p_{x,k,q} \right)^2 }{ \left(  \left( x_{{\rm tx},k} - p_{x,k,q} \right)^2 + \left( y_{{\rm rx},k} - p_{y,k,q} \right)^2  \right)^{\frac{3}{2}} } ,
\end{equation} 
\begin{equation}\label{eq:image_P_grad_end}
   b_{k,q,y}  = \tfrac{ -\left( x_{{\rm rx},k} - p_{x,k,q} \right)^2 }{ \left(  \left( x_{{\rm rx},k} - p_{x,k,q} \right)^2 + \left( y_{{\rm rx},k} - p_{y,k,q} \right)^2  \right)^{\frac{3}{2}} }  .
\end{equation}

After obtaining $\bm{\psi}$, we project grid positions onto the feasible set defined by the constraints \eqref{eq:constraint2} and \eqref{eq:constraint3}. We adopt the alternating projection method since both the rectangular region $\mathcal{A}$ and the Euclidean ball are convex. Specifically, we first project $\bm{p}_{k,q}$ onto the rectangle by performing $p_{k,q,x} \leftarrow \min( \max( p_{k,q,x} , x_1 ) , x_2) $ and $p_{k,q,y} \leftarrow \min( \max( p_{k,q,y} , y_1 ) , y_2)$, and then project it onto the ball via
\begin{equation}
    \bm{p}_{k,q}  \!\leftarrow \!\! 
    \left\{ \!\!\!
             \begin{array}{ll}
             \bm{p}_{k,q}, \!\!\!\!\!&  \left\| \bm{p}_{k,q} \!-\! \bm{p}_{k,q}^{(0)} \right\|_2 \!\leq\! d_{\max} ,\!\! \\ 
             \bm{p}_{k,q}^{(0)} \!+\! \frac{d_{\max}  ( \bm{p}_{k,q} \!-  \bm{p}_{k,q}^{(0)}  )  }{  \|  \bm{p}_{k,q} - \bm{p}_{q}^{(0)} \|_2  } , \!\!\!\!\!&  \left\| \bm{p}_{k,q} \!-\! \bm{p}_{k,q}^{(0)} \right\|_2 \!>\! d_{\max} .\!\!
             \end{array} 
             \right. \label{eq:image_P_proj}
\end{equation}
The above projections iterate for a few times until convergence to ensure both constraints are satisfied. We choose uniform grids in $\mathcal{A}$ as initial grids. By ensuring that each optimized grid point is in an Euclidean ball centered by the corresponding initial grid, we can prevent the grid index ambiguity problem while making sure adjacent points are not too close.

\subsection{Role of Dynamic Grids}  \label{Sec:theory}

In this part, we establish the theoretical foundation of adopting dynamic grids for achieving high-quality imaging. 
The cost function in \eqref{eq:cost_P} can be rewritten as 
\begin{align}
  \ln \! \left| \bm{\Sigma}_{k } \right|  \!+\!  \operatorname{tr} \! \left( \bm{\Sigma}_{k}^{-1}   \hat{\bm{\Sigma}}_{k} \right) \! 
  & = \!\ln  \left| \mathbf{T}_k \right|  +  \operatorname{tr}  \left( \mathbf{T}_k^{-1} \right)  \!+\! \ln \! \left| \hat{\bm{\Sigma}}_{k } \right|  \\
  & = \!{\sum}_{n=1}^{N} \!\!\left(  \ln \tau_n \! + \!\frac{1}{\tau_n}  \right)
  \!+\! \ln \! \left| \hat{\bm{\Sigma}}_{k } \right|\! ,  \label{eq:cost_P2}
\end{align}
where $N = L N_{\rm rx}$ and $\tau_n$ denotes the eigenvalue of $\mathbf{T}_k = \hat{\bm{\Sigma}}_{k }^{-\frac{1}{2}} \bm{\Sigma}_{k } \hat{\bm{\Sigma}}_{k }^{-\frac{1}{2}} $. 
The cost function \eqref{eq:cost_P2} is minimized when $\tau_n \to 1, \forall n$. 
The deviation from optimality is bounded by
\begin{align}
  & {\sum}_{n=1}^{N} \left(  \tau_n  - 1  \right)^2   \notag\\  
  & = \big\| \hat{\bm{\Sigma}}_{k }^{-\frac{1}{2}} \big( \bm{\Sigma}_{k }  -  \hat{\bm{\Sigma}}_{k }  \big) \hat{\bm{\Sigma}}_{k }^{-\frac{1}{2}}    \big\|_F^2 \label{eq:theory_norm1}\\
  & \leq \big\| \hat{\bm{\Sigma}}_{k }^{-\frac{1}{2}} \big\|_{F}^4  \big\|  \bm{\Sigma}_{k }  -  \hat{\bm{\Sigma}}_{k }  \big\|_F^2 \label{eq:theory_norm2} \\
  & \leq \big\| \hat{\bm{\Sigma}}_{k }^{-\frac{1}{2}} \big\|_{F}^4  \big( \big\|  \bm{\Sigma}_{k,0}  -  \hat{\bm{\Sigma}}_{k }  \big\|_F
   + \left\|  \bm{\Sigma}_{k }  -  \bm{\Sigma}_{k,0}  \right\|_F 
    \big)^2. \label{eq:theory_norm}
\end{align} 
where \eqref{eq:theory_norm2} follows from the sub-multiplicative property of Frobenius norm and \eqref{eq:theory_norm} follows from the triangle inequality. The term $\big\|  \bm{\Sigma}_{k,0}  -  \hat{\bm{\Sigma}}_{k }  \big\|_F$ quantifies the gap between the sample covariance matrix and the true one, which can be reduced by increasing the number $M$ of frames based on the law of large numbers. The term $\left\|  \bm{\Sigma}_{k }  -  \bm{\Sigma}_{k,0}  \right\|_F$ quantifies discretization error. Minimizing this term requires careful grid design.

It is evident that we can reduce $\left\|  \bm{\Sigma}_{k }  -  \bm{\Sigma}_{k,0}  \right\|_F$ by increasing the number $Q$ of grids, but this introduces a critical limitation in the distinguishability of adjacent grids. Specifically, the matrix coherence of \(\mathbf{V}_k\), defined as the maximum inner product between its normalized columns, is given by
\begin{align}
  \mu \left( \mathbf{V}_k \right) 
  & \triangleq 
  \max_{q' \neq q} \tfrac{|\mathbf{v}_{k,q'}^H \mathbf{v}_{k,q}|}{\|\mathbf{v}_{k,q'}\|_2 \|\mathbf{v}_{k,q}\|_2}  \\ 
  & = \max_{q' \neq q} \tfrac{| \bar{\mathbf{a}}^H \!\left(\phi_{k,q'}\right) \bar{\mathbf{a}} \left(\phi_{k,q}\right) |}{ L } 
    \tfrac{ | \mathbf{b}^H \!\left(\theta_{k,q'}\right) \mathbf{b}\left(\theta_{k,q}\right) |}{ N_{\rm rx} } . 
\end{align} 
Generally speaking, smaller coherence $\mu \left( \mathbf{V}_k \right)$ provides stronger recovery guarantees. However, as $Q$ increases, $\mu \left( \mathbf{V}_k \right)$ approaches $1$, thereby violating the distinguishability of adjacent grids. Taking $\frac{| \mathbf{b}^H\! \left(\theta_{k,q'}\right) \mathbf{b}\left(\theta_{k,q}\right) |}{ N_{\rm rx} }$ as an example, we have
\begin{equation}
  \tfrac{| \mathbf{b}^H \!\left(\theta_{k,q'}\right) \mathbf{b}\left(\theta_{k,q}\right) |}{ N_{\rm rx} }  
  \!\!=\! \tfrac{ \big| 1 - e^{j\pi N_{\rm rx}  \Delta_{\theta,q,q'}  } \big|}{ N_{\rm rx} \big| 1 - e^{j\pi  \Delta_{\theta,q,q'} } \big|} 
  \!\!=\! \left|\!\tfrac{\sin\left(\!\frac{\pi N_{\rm rx} }{2} \Delta_{\theta,q,q'} \right)}{N_{\rm rx} \sin\left(\frac{\pi  }{2} \Delta_{\theta,q,q'} \right)}\!\right|\!,   
\end{equation}
where $\Delta_{\theta,q,q'} =  \sin \theta_{k,q'} - \sin \theta_{k,q}$. In the case of $ \left| \Delta_{\theta,q,q'} \right| \leq \frac{1}{N_{\rm rx}} $, it is evident that $\frac{| \mathbf{b}^H\! \left(\theta_{k,q'}\right) \mathbf{b}\left(\theta_{k,q}\right) |}{ N_{\rm rx} }$ increases as $|\Delta_{\theta,q,q'}|$ decreases. Once $|\Delta_{\theta,q,q'}| \ll \frac{1}{N_{\rm tx}} $, we have $\frac{| \mathbf{b}^H\! \left(\theta_{k,q'}\right) \mathbf{b}\left(\theta_{k,q}\right) |}{ N_{\rm rx} } \to 1$. Similar convergence can be observed for the DoA term. Thus, for fixed numbers of transmit and receive antennas, we have $\mu \left( \mathbf{V}_k \right) \to 1$ as the number $Q$ of grids increases, thereby violating the incoherence requirements for stable recovery.

To avoid this issue, we propose to use dynamic grids to reduce the discretization error, instead of excessively increasing the number of grids. Specifically, we have 
\begin{align} 
  & \left\|  \bm{\Sigma}_{k }  -  \bm{\Sigma}_{k,0}  \right\|_F^2  \notag\\
  & = \left\| \bm{\Psi}_k - \bm{\Psi}_{k,0} \right\|_F^2 \\
  & = \big\| \bm{\Psi}_k - \mathcal{P}_{\bar{\mathbf{V}}_k} \bm{\Psi}_{k,0} \mathcal{P}_{\bar{\mathbf{V}}_k} 
  - \mathcal{P}_{\bar{\mathbf{V}}_k} \bm{\Psi}_{k,0} \mathcal{P}_{\bar{\mathbf{V}}_k}^{\bot}  \notag\\
  & \;\;\;\;\;   - \mathcal{P}_{\bar{\mathbf{V}}_k}^{\bot} \bm{\Psi}_{k,0} \mathcal{P}_{\bar{\mathbf{V}}_k} 
  - \mathcal{P}_{\bar{\mathbf{V}}_k}^{\bot} \bm{\Psi}_{k,0} \mathcal{P}_{\bar{\mathbf{V}}_k}^{\bot}  \big\|_F^2 \\
  & = \left\| \bm{\Psi}_k -  \mathcal{P}_{\bar{\mathbf{V}}_k} \bm{\Psi}_{k,0} \mathcal{P}_{\bar{\mathbf{V}}_k} \right\|_F^2 
  +  \big\| \mathcal{P}_{\bar{\mathbf{V}}_k} \bm{\Psi}_{k,0} \mathcal{P}_{\bar{\mathbf{V}}_k}^{\bot}   \big\|_F^2 \notag\\
  & \;\;\;\;\; + \big\| \mathcal{P}_{\bar{\mathbf{V}}_k}^{\bot} \bm{\Psi}_{k,0} \mathcal{P}_{\bar{\mathbf{V}}_k}  \big\|_F^2
  +  \big\| \mathcal{P}_{\bar{\mathbf{V}}_k}^{\bot} \bm{\Psi}_{k,0} \mathcal{P}_{\bar{\mathbf{V}}_k}^{\bot}  \big\|_F^2 \label{eq:theory_grid0} \\
  & \leq  \left\| \bm{\Psi}_k -  \mathcal{P}_{\bar{\mathbf{V}}_k} \bm{\Psi}_{k,0} \mathcal{P}_{\bar{\mathbf{V}}_k} \right\|_F^2 
  + 2  \big\| \mathcal{P}_{\bar{\mathbf{V}}_k}^{\bot} \bm{\Psi}_{k,0}   \big\|_F^2 , \label{eq:theory_grid}
\end{align}
where $\bm{\Psi}_{k,0} = \iint_{ \mathcal{S} }  \alpha_{x,y,k} \gamma_{x,y,k}   \mathbf{v}_{x,y,k}  \mathbf{v}_{x,y,k}^H   {\rm d} x {\rm d} y$, $\bm{\Psi}_k = \mathbf{V}_k \bm{\Gamma}_{r,k} \bm{\Gamma}_{\beta,k} \mathbf{V}_k^H$, $\bar{\mathbf{V}}_k$ is a submatrix of $\mathbf{V}_k$ with the column indices lying in the set $\mathcal{R}_{k} = \{q\in[Q]: {\gamma}_{r,k,q} > 0\}$, $\mathcal{P}_{\bar{\mathbf{V}}_k}$ denotes the projection matrix onto the subspace spanned by the columns of $\bar{\mathbf{V}}_k$, $\mathcal{P}_{\bar{\mathbf{V}}_k}^{\bot} = \mathbf{I} - \mathcal{P}_{\bar{\mathbf{V}}_k}$, \eqref{eq:theory_grid0} follows because $\left\|\mathbf{A}+\mathbf{B}\right\|_F^2 = \left\|\mathbf{A}\right\|_F^2 + \left\|\mathbf{B}\right\|_F^2$ in the case of $\Re(\operatorname{tr}(\mathbf{A}^H\mathbf{B})) = 0$, and \eqref{eq:theory_grid} follows because 
\begin{align}
  & \big\| \mathcal{P}_{\bar{\mathbf{V}}_k}^{\bot} \bm{\Psi}_{k,0} \mathcal{P}_{\bar{\mathbf{V}}_k}  \big\|_F^2
  + \big\| \mathcal{P}_{\bar{\mathbf{V}}_k}^{\bot} \bm{\Psi}_{k,0} \mathcal{P}_{\bar{\mathbf{V}}_k}^{\bot}  \big\|_F^2 \notag \\ 
  & = \operatorname{tr}\big( \mathcal{P}_{\bar{\mathbf{V}}_k}^{\bot} \bm{\Psi}_{k,0} \mathcal{P}_{\bar{\mathbf{V}}_k} \bm{\Psi}_{k,0} \big) + \operatorname{tr}\big( \mathcal{P}_{\bar{\mathbf{V}}_k}^{\bot} \bm{\Psi}_{k,0} \mathcal{P}_{\bar{\mathbf{V}}_k}^{\bot}  \bm{\Psi}_{k,0}  \big) \\
  & = \operatorname{tr}\big( \bm{\Psi}_{k,0} \mathcal{P}_{\bar{\mathbf{V}}_k}^{\bot} \bm{\Psi}_{k,0}   \big) \\
  & = \big\| \mathcal{P}_{\bar{\mathbf{V}}_k}^{\bot} \bm{\Psi}_{k,0}   \big\|_F^2, 
\end{align}
and $\big\| \mathcal{P}_{\bar{\mathbf{V}}_k} \bm{\Psi}_{k,0} \mathcal{P}_{\bar{\mathbf{V}}_k}^{\bot}  \big\|_F^2 + \big\| \mathcal{P}_{\bar{\mathbf{V}}_k}^{\bot} \bm{\Psi}_{k,0} \mathcal{P}_{\bar{\mathbf{V}}_k}  \big\|_F^2 = \big\| \mathcal{P}_{\bar{\mathbf{V}}_k}^{\bot} \bm{\Psi}_{k,0}  \big\|_F^2$ obtained in a similar way.

The upper bound in \eqref{eq:theory_grid} consists of two terms. The first term $\left\| \bm{\Psi}_k -  \mathcal{P}_{\bar{\mathbf{V}}_k} \bm{\Psi}_{k,0} \mathcal{P}_{\bar{\mathbf{V}}_k} \right\|_F^2$ quantifies how well the discretized matrix $\bm{\Psi}_k$ approximates the projection of the true matrix $\bm{\Psi}_{k,0}$ onto the subspace spanned by the columns of $\bar{\mathbf{V}}_k$. For fixed grids, this term can be reduced by optimizing the effective scattering intensity $\bm{\gamma}_{r,k}$. The second term $2 \big\| \mathcal{P}_{\bar{\mathbf{V}}_k}^{\bot} \bm{\Psi}_{k,0} \big\|_F^2$ measures the energy of $\bm{\Psi}_{k,0}$ that is orthogonal to the subspace spanned by the columns of $\bar{\mathbf{V}}_k$. It arises due to subspace mismatch. For fixed scattering intensity, we can use grid optimization to align $\bar{\mathbf{V}}_k$ with the dominant eigenvectors of $\bm{\Psi}_{k,0}$, thereby reducing the subspace mismatch effect. This process helps to generate a coordinate system intrinsically aligned with the target, where grid positions are adjusted to conform to target geometries, leading to better imaging performance than optimizing the scattering intensity alone.

\subsection{Complexity Analysis} \label{Sec:IV_complexity}

The complexity of Algorithm 1 is primarily determined by the calculation of the sample covariance matrix $\hat{\bm{\Sigma}}_k$ and the AO of the scattering intensity $\bm{\gamma}_{r,k}$ and the grid position $\mathbf{P}_k$. Specifically, the initial calculation of $\hat{\bm{\Sigma}}_k$ in \eqref{eq:receive_Y_sigmakhat} incurs a complexity of $\mathcal{O}( M(L N_{\rm rx})^2 )$. In the update of $\bm{\gamma}_{r,k}$ with fixed $\mathbf{P}_k$, the complexity is dominated by the calculation of the objective function $f_{k,q}(d)$ in \eqref{eq:Problem_r2} and the update of the inverse covariance matrix $\bm{\Sigma}_k^{-1}$ in \eqref{eq:gamma_covariance} for each grid point $q$, both of which have complexity of $\mathcal{O}( (L N_{\rm rx})^2 )$. In the subsequent update of $\mathbf{P}_k$ with fixed $\bm{\gamma}_{r,k}$, the complexity of calculating each element in the gradient $\nabla_{\bm \psi} f_{k,x,y}({\bm {\psi}} )$ is $\mathcal{O}( (L N_{\rm rx})^2 + LN_{\rm rx} N_{\rm tx} )$. Consequently, to obtain single-view imaging at RBS $k$, the overall complexity of Algorithm 1 is $\mathcal{O}( M(L N_{\rm rx})^2 + I_{\rm max} I_1 Q (L N_{\rm rx})^2 + I_{\rm max} I_2 |\mathcal{Q}_{a,k}| ( (L N_{\rm rx})^2 + LN_{\rm rx} N_{\rm tx}) )$, where $I_{\rm max}$, $I_1$, and $I_2$ represent the number of iterations for the outer AO loop, the inner $\bm{\gamma}_{r,k}$-update loop, and the inner $\mathbf{P}_k$-update loop, respectively.

\section{Phase II: Multi-View Fusion}  \label{Sec:fusion} 

Each RBS obtains individual images applying the method proposed in Section~\ref{Sec:image}. This section addresses the challenge of fusing individual images to produce high-quality environment reconstruction.

\subsection{Edge-Preserving Interpolation} \label{Sec:fusion1}

The number of resolvable grid points is constrained by available resources, such as the numbers of transmit and receive antennas. This limited number of grids will inevitably result in non-smooth contours. Moreover, the RBSs operate on heterogeneous grids tailored to their respective spatial positions. Such topological distinction makes it challenging for fusion. To address these issues, we introduce some cross-disciplinary concepts and approaches from computer vision to the field of imaging for the first time. Based on these techniques, we interpolate each RBS's coarse image onto a set of common and finer grids, thereby establishing a unified foundation for the subsequent fusion process.

While traditional space-invariant interpolation methods, such as nearest-neighbor, bi-linear, bi-cubic, and Gaussian techniques, are computationally efficient, their fundamental limitation lies in determining weights solely by geometric distance to the interpolation point. These methods overlook local image structures by employing isotropic kernels to forcibly average intensities across physical boundaries, thereby inevitably suffering from edge blurring. This motivates us to explore space-variant interpolation, in which interpolation weights are dynamically adjusted to exploit image structural features. However, some existing space-variant interpolation methods, such as new edge-directed interpolation~\cite{ref:NEDI}, are constrained to regular grids, failing to address the irregular grid positions inherent to our imaging problem. To this end, we propose an EP-NNI method that simultaneously preserves edge features and accommodates irregular grid points.

Specifically, we aim to interpolate $\bm{\gamma}_{r,k}$ corresponding to $Q$ irregular points to ${\bm{\gamma}}'_{r,k}$ corresponding to $Q' = Q'_1 \times Q'_2$ regular grids shared by $K$ RBSs while preserving edge features. The core idea of EP-NNI lies in integrating gradient-based geometric features into natural neighbor interpolation (NNI) weights, thereby effectively inhibiting smoothing across edges while maintaining smoothness along edges. The implementation of this idea proceeds as follows. First, we estimate the gradient at each irregular point using local plane fitting, which operates based on the assumption that within a small neighborhood around a point $q\in[Q]$, the scattering intensity can be modeled as a linear function ${\gamma}_{r,k,j} \approx  c_{k,q,1} p_{x,k,j} + c_{k,q,2} p_{y,k,j} + c_{k,q,3}$, where $j\in\mathcal{N}_{k,q}$ and the set $\mathcal{N}_{k,q}\subset[Q]$ includes irregular points in the neighborhood of point $q$~\cite{fan2018local}. Based on the least square criterion, $\mathbf{c}_{k,q} = \left[c_{k,q,1},c_{k,q,2},c_{k,q,3}\right]^T$ is estimated as
\begin{align} \label{eq:fusion_gradq}
  \mathbf{c}_{k,q} 
  & = \arg\min_{\mathbf{c}_{k,q}} \left\| \mathbf{r}_{k,q} - \mathbf{A}_{k,q} \mathbf{c}_{k,q} \right\|_2^2 \\
  & = (\mathbf{A}_{k,q}^T \mathbf{A}_{k,q} )^{-1} \mathbf{A}_{k,q}^T\mathbf{r}_{k,q}, 
\end{align} 
where $\mathbf{r}_{k,q} \in \mathbb{R}^{|\mathcal{N}_{k,q}| \times 1}$ is formed by stacking the elements $\gamma_{r,k,j}, \forall j\in\mathcal{N}_{k,q}$ into a vector, and the matrix $\mathbf{A}_{k,q} \in \mathbb{R}^{|\mathcal{N}_{k,q}| \times 3}$ is constructed by stacking $p_{x,k,j}, \forall j\in\mathcal{N}_{k,q}$ in its first column, stacking $p_{y,k,j}, \forall j\in\mathcal{N}_{k,q}$ in its second column, and placing an all-one vector in its third column. Then, the gradient at the irregular grid $q\in[Q]$ can be approximated as $\nabla {\gamma}_{r,k,q} = \left[ \frac{\partial {\gamma}_{r,k,q}}{\partial p_{x,k,q}}  \;\; \frac{\partial {\gamma}_{r,k,q}}{\partial p_{y,k,q}} \right]^T  \approx \left[ c_{k,q,1} \;\; c_{k,q,2} \right]^T$.

Next, based on the gradients at irregular grids for each $k\in[K]$, we compute the structure tensor at each of the $Q'$ regular grid points to guide the anisotropic interpolation. As a powerful concept from computer vision, the structure tensor characterizes the local geometric structure by forming a smoothed second-moment matrix of the gradients~\cite{weickert1998anisotropic}. For each $k\in[K]$ and $q' \in [Q']$, the structure tensor is given by 
\begin{align}
\mathbf{J}_{k,q'} & = {\sum}_{q\in\mathcal{N}'_{k,q'}} w_{k,q,q'}^{\text{sib}} \nabla {\gamma}_{r,k,q} (\nabla {\gamma}_{r,k,q})^T \\
& \approx {\sum}_{q\in\mathcal{N}'_{k,q'}} w_{k,q,q'}^{\text{sib}}
\begin{bmatrix}
c_{k,q,1}^2 & c_{k,q,1} c_{k,q,2} \\
c_{k,q,1} c_{k,q,2} & c_{k,q,2}^2
\end{bmatrix},  \label{eq:fusion_gradqprime}
\end{align}
where the set $\mathcal{N}'_{k,q'} \subset [Q]$ includes irregular grid points near the regular grid point $q'$, and $ w_{k,q,q'}^{\text{sib}} $ denotes the Sibson natural neighbor weight determined by two Voronoi diagrams \cite{sibson1981brief}. In the first diagram, the RoI $\mathcal{A}$ is partitioned into $Q$ cells centered by the original $Q$ irregular grid points of RBS $k$, with the Voronoi cell $\text{Vor}_k(q)$ containing all points in the plane $\mathcal{A}$ closer to the $q$-th grid than to any other grids. The second diagram is constructed based on the target $Q'$ regular grid points, with $\text{Vor}(q')$ representing the Voronoi cell corresponding to the grid $q' \in [Q']$. Applying Sibson's method, the weight $ w_{k,q,q'}^{\text{sib}} $ is proportional to the area overlap between the Voronoi cells $\text{Vor}(q')$ and $\text{Vor}_k(q)$, given by \cite{sibson1981brief} 
\begin{equation} \label{eq:fusion_weight_Sibson}
   w_{k,q,q'}^{\text{sib}} = \tfrac{\text{Area}(\text{Vor}_k(q) \cap \text{Vor}(q'))}{\text{Area}(\text{Vor}(q'))} .
\end{equation}

Based on the structure tensor in \eqref{eq:fusion_gradqprime} and the steering kernel in \cite{takeda2007kernel}, we design an edge-preserving weight to penalize cross-edge interpolation, given by 
\begin{equation} \label{eq:fusion_weight}
\tilde{w}_{k,q,q'} = w_{k,q,q'}^{\text{sib}}  \exp \Big\{-\tfrac{ \mathbf{d}_{k,q,q'}^T  \mathbf{J}_{k,q'}  \mathbf{d}_{k,q,q'} }{ 2\sigma^2_{\rm EP} } \Big\}, 
\end{equation}
where $\mathbf{d}_{k,q,q'} = \mathbf{p}_{k,q} - \mathbf{p}_{k,q'} $ and $\sigma^2_{\rm EP}$ serves as an edge-sensitive decay parameter. Applying singular value decomposition (SVD), the structure tensor satisfies $\mathbf{J}_{k,q'} = \mathbf{F}_{k,q'} \mathbf{\Lambda}_{k,q'} \mathbf{F}^T_{k,q'}$, where $\mathbf{\Lambda}_{k,q'} = \operatorname{diag} \{ \lambda_{1,k,q'} , \lambda_{2,k,q'} \}$ with $\lambda_{1,k,q'} \geq \lambda_{2,k,q'}$ and $\mathbf{F}_{k,q'} = \left[ \mathbf{f}_{1,k,q'} \;\; \mathbf{f}_{2,k,q'} \right]$. Here, $\mathbf{f}_{1,k,q'}$ represents the dominant edge normal direction corresponding to the maximum intensity variation, while its orthogonal counterpart $\mathbf{f}_{2,k,q'}$ indicates the edge tangent direction along minimal variation~\cite{weickert1998anisotropic}. Then, the quadratic term in the exponent of \eqref{eq:fusion_weight} can be expressed as $\mathbf{d}_{k,q,q'}^T \mathbf{J}_{k,q'} \mathbf{d}_{k,q,q'} = \lambda_{1,k,q'} (\mathbf{d}_{k,q,q'}^T \mathbf{f}_{1,k,q'})^2 + \lambda_{2,k,q'} (\mathbf{d}_{k,q,q'}^T \mathbf{f}_{2,k,q'})^2$. As we can see, the anisotropic weight $\tilde{w}_{k,q,q'}$ imposes maximal penalty when $\mathbf{d}_{k,q,q'}$ is parallel to $\mathbf{f}_{1,k,q'}$ and minimal penalty when orthogonal to $\mathbf{f}_{1,k,q'}$, thereby preserving sharp edge structures.

By adaptively weighting contributions from neighbor observation points, the scattering intensity at the $q'$-th interpolation point is given by
   \begin{equation} \label{eq:fusion_EPNNI}
   \gamma'_{r,k,q'} = {\sum}_{q \in \mathcal{N}'_{k,q'}} \tfrac{\tilde{w}_{k,q,q'}}{\sum_{q''\in \mathcal{N}'_{k,q'}} \tilde{w}_{k,q'',q'}} \gamma_{r,k,q}. 
   \end{equation}
Then, for each RBS $k$, the scattering intensity corresponding to $Q'$ interpolation points is denoted as $\bm{\gamma}'_{r,k} = [\gamma'_{r,k,1}, \ldots, \gamma'_{r,k,Q'} ]^T \in \mathbb{R}^{Q'}$.


\subsection{Fusion} \label{Sec:fusion2}

In this section, we aim to aggregate the interpolated scattering intensities at $K$ RBSs to obtain high-accuracy environment reconstruction. We formulate this task as a novel optimization problem that integrates a weighted least squares term, a sparsity-promoting term, and a cluster-promoting term. This problem is given by
\begin{subequations} \label{eq:Problem_fusion}
\begin{align}
  & \mathop{\min}\limits_{ \bm{\gamma}'_r , \bm{\Lambda} }  
& & f_{\text{WLS}}(\bm{\gamma}'_r, \bm{\Lambda})  + \mu  f_{\text{sparse}}(\bm{\gamma}'_r)    + \eta  f_{\text{TV}}(\bm{\gamma}'_r) \label{eq:fusion_obj}  \\
&\;\;  \text{s.t.} 
& &  \gamma'_{r,q} \geq 0,  \quad \forall q  \label{eq:fusion_constraint1}\\
&&& \lambda_{k,q} \in\{0,1\},  \quad \forall k, q , \label{eq:fusion_constraint2} 
\end{align} 
\end{subequations} 
where $\mu>0$ and $\eta>0$ are penalty parameters. As we can see, the inputs of this problem are the interpolated scattering intensities $\{\bm{\gamma}'_{r,k}: k\in[K]\}$ over a common set of $Q'$ grids from $K$ RBSs obtained in the previous subsection. The optimization variables are the fused scattering intensity $\bm{\gamma}'_r = \left[{\gamma}'_{r,1}, \ldots,  {\gamma}'_{r,Q'} \right]^T \in \mathbb{R}^{Q'}$ and the binary matrix $\bm{\Lambda} \in \{0,1\}^{K\times Q'}$ with its $(k,q)$-th element $\lambda_{k,q}$ indicating whether grid $q$ lies within the field of view of RBS $k$.

In this work, we propose to employ the following three functions to constitute the objective function in \eqref{eq:fusion_obj}. Specifically, the weighted least squares term in \eqref{eq:fusion_obj} is given by
\begin{align} \label{eq:Problem_fusion_WLS}
  f_{\text{WLS}}(\bm{\gamma}'_r, \bm{\Lambda}) 
  & = {\sum}_{k=1}^K {\sum}_{q=1}^{Q'} \left( \lambda_{k,q}  {\gamma}_{\beta,k,q} (  {\gamma}'_{r,k,q} - {\gamma}'_{r,q} )^2  \right. \notag\\
  & \;\;\;\;\; \;\;\;\;\; \;\;\;\;\; \;\; \left. +   (1-\lambda_{k,q}) {\gamma}_{\beta,k,q} ({\gamma}'_{r,k,q})^2 \right) . 
\end{align}
The large-scale fading coefficient $\gamma_{\beta,k,q}$ serves as a weighting factor. Larger values indicate better channel conditions and thus higher confidence in the corresponding estimate $\gamma'_{r,k,q}$. Moreover, as introduced before, each RBS only has a limited field of view about the target, and this field is unknown in advance. The binary variable $\lambda_{k,q}$ acts as an adaptive selector, excluding RBSs that receive weak scattering signals from grid $q$ and only fusing estimates from RBSs whose fields of view likely contains this grid. Specifically, when $\lambda_{k,q} = 0$, grid \(q\) is presumed to lie outside the field of view of RBS $k$, and thus we only penalize ${\gamma}_{\beta,k,q} ({\gamma}'_{r,k,q})^2 $ because ${\gamma}'_{r,k,q}$ is supposed to approach $0$ in this case. For valid estimate $\gamma'_{r,k,q}$ with $\lambda_{k,q}=1, \forall k$, \eqref{eq:Problem_fusion_WLS} penalizes their squared deviation from the fusion target $\gamma'_{r,q}$, ensuring the final fused value represents a channel-weighted consensus of trustworthy estimates. Thus, the criterion in \eqref{eq:Problem_fusion_WLS} allows us to estimate effective observation zones without prior knowledge of RBSs' fields of view, and ensures only the estimates from RBSs with strong scattering intensity are aggregated. Moreover, we propose to introduce two $\ell_1$-norm based regularization terms into the objective function. Particularly, to exploit the fact that targets occupy a limited number of grids, we employ the following sparsity-promoting term 
\begin{equation}\label{eq:Problem_fusion_sparse}
  f_{\text{sparse}}(\bm{\gamma}'_r) = \left\| {\bm{\gamma}}'_{r} \right\|_1 .  
\end{equation} 
This term promotes a solution where only a small number of grids exhibit nonzero scattering intensities. Considering that extended targets usually occupy multiple adjacent grids, we introduce the $l_1$-based anisotropic total variation regularization term, which is defined as~\cite{ref:TVl1}
\begin{equation}\label{eq:Problem_fusion_TV}
f_{\text{TV}}(\bm{\gamma}'_r) 
=  \left\| \mathbf{D} \bm{\gamma}'_r \right\|_1 ,  
\end{equation}
where $\mathbf{D} = \left[ \mathbf{D}_x^T, \mathbf{D}_y^T \right]^T$ with $\mathbf{D}_x$ and $\mathbf{D}_y$ denoting the discrete gradient operator along $x$- and $y$-axis. Specifically, we have $\mathbf{D}_x = \mathbf{I}_{Q'_1} \otimes \mathbf{D}_{x,1}$ and $\mathbf{D}_y = \mathbf{D}_{y,1} \otimes \mathbf{I}_{Q'_2}$, where $\mathbf{D}_{x,1}$ is given by 
\begin{equation} 
  \mathbf{D}_{x,1} = \begin{bmatrix}
-1 & 1 & 0 & \cdots & 0 \\
0 & -1 & 1 & \cdots & 0 \\
\vdots & \ddots & \ddots & \ddots & \vdots \\ 
1 & 0 & \cdots & 0 & -1
\end{bmatrix} \in \mathbb{R}^{Q'_2 \times Q'_2}
\end{equation}
and $\mathbf{D}_{y,1}$ follows an identical structure to $\mathbf{D}_{x,1}$ but with dimension $Q'_1 \times Q'_1$. It is evident that the discrete gradient exhibits high values when scattering intensities of adjacent grids are significantly different, and approaches zero in regions where the scattering intensity remains approximately constant. For large extended targets, transitions between empty grids and scatterer-occupied grids are sparse. By promoting gradient sparsity, one can effectively preserve the cluster structure of target scatterers in the constructed image~\cite{ref:TV1,ref:TVAL3}.

Based on the functions in \eqref{eq:Problem_fusion_WLS}, \eqref{eq:Problem_fusion_sparse}, and \eqref{eq:Problem_fusion_TV}, the constraint in \eqref{eq:fusion_constraint1} can be removed while still maintaining the non-negativity of $\gamma'_{r,q}$'s after optimization. This is because the input interpolated scattering intensities $\{\bm{\gamma}'_{r,k}: k\in[K]\}$ are non-negative. In this case, for any solution where some $\gamma'_{r,q}$ become negative, there always exists a corresponding non-negative assignment that achieves a lower objective value, making negative values suboptimal within the optimization framework. After removing \eqref{eq:fusion_constraint1}, the problem in \eqref{eq:Problem_fusion} is still challenging since $\bm{\gamma}'_r$ and $\bm{\Lambda}$ are coupled. To this end, we adopt the AO framework to iteratively update $\bm{\gamma}'_r$ and $\bm{\Lambda}$.

When $\bm{\Lambda}$ is fixed, the optimization problem is convex with respect to $\bm{\gamma}'_r$. Although this problem can be directly solved using generic solvers (e.g., CVX based on interior-point methods), such approaches typically exhibit a computational complexity of $\mathcal{O}((Q')^{3.5})$, which is prohibitive in the setting with a large number $Q'$ of grids. To address this issue, we employ the alternating direction method of multipliers (ADMM) technique. Specifically, by introducing the auxiliary variable $\mathbf{z}_r = \mathbf{D} \bm{\gamma}'_r$, we have
\begin{subequations} \label{eq:Problem_fusion2}
\begin{align}
  & \mathop{\min}\limits_{ \bm{\gamma}'_r, \mathbf{z}_r }  
& & f_{\text{WLS}}(\bm{\gamma}'_r, \bm{\Lambda})   + \mu  \left\| {\bm{\gamma}}'_{r} \right\|_1   + \eta  \left\| \mathbf{z}_r \right\|_1 \label{eq:Problem_fusion2_obj}  \\
&\;\;  \text{s.t.}
& &\mathbf{z}_r = \mathbf{D} \bm{\gamma}'_r.   
\end{align} 
\end{subequations}
The augmented Lagrangian is expressed as
   \begin{align}
     \mathcal{L}(\bm{\gamma}'_r, \bm{\Lambda} ,\mathbf{z}_r , \mathbf{u}_r )  
     & = f_{\text{WLS}}(\bm{\gamma}'_r, \bm{\Lambda})     + \mu  \left\| {\bm{\gamma}}'_{r} \right\|_1  + \eta \left\| \mathbf{z}_r \right\|_{1}  
     \notag\\
     & \;\;\;\; + \frac{\rho}{2} \| \mathbf{z}_r - \mathbf{D} \bm{\gamma}'_r + \mathbf{u}_r\|_2^2 , 
   \end{align}
where $\rho$ denotes the penalty factor and $\mathbf{u}_r$ denotes the scaled dual variable. Then, we update $ \bm{\gamma}'_r,  \mathbf{z}_r $, and $\mathbf{u}_r$ in an alternating manner. Specifically, the scattering intensity $\bm{\gamma}'_r$ is updated by 
    \begin{align}  
    \bm{\gamma}'_r \leftarrow  &  \arg \min_{ \bm{\gamma}'_r } \; {\sum}_{k=1}^K {\sum}_{q=1}^{Q'} 
     \lambda_{k,q}  \gamma_{\beta,k,q} (  {\gamma}'_{r,k,q} - {\gamma}'_{r,q} )^2   
     \notag\\
     & \;\;\;\;\; \;\;\;\;\; \;\;\; + \mu {\sum}_{q=1}^{Q'} {\gamma}'_{r,q}   + \frac{\rho}{2} \| \mathbf{z}_r  - \mathbf{D} \bm{\gamma}'_r + \mathbf{u}_r \|_2^2 .  \label{eq:Problem_fusion_r}
   \end{align} 
By setting the gradient of \eqref{eq:Problem_fusion_r} to be zero, we have
   \begin{align} 
     & \left( 2 {\sum}_{k=1}^K  \mathbf{B}_k \bm{\Lambda}_k  + \rho \mathbf{D}^T \mathbf{D}  \right) \bm{\gamma}'_r \notag\\
     & = 2 {\sum}_{k=1}^K  \mathbf{B}_k \bm{\Lambda}_k \bm{\gamma}_{r,k} 
     + \rho  \mathbf{D}^T \left(  \mathbf{z}_r + \mathbf{u}_r  \right)
     - \mu \bm{1}_{Q'} ,  \label{eq:fusion_ADMM_r_condition}
   \end{align}
where $\mathbf{B}_k = \operatorname{diag}\left\{ {\gamma}_{\beta,k,q}: q\in[Q'] \right\}$, $ \bm{\Lambda}_k = \operatorname{diag}\left\{ {\lambda}_{k,q}: q\in[Q'] \right\}$, and $\bm{1}_{Q'}$ denotes an all-one vector with dimension $Q'$. To avoid the computational burden of obtaining the inverse of the matrix $2 \sum_{k=1}^K  \mathbf{B}_k \bm{\Lambda}_k  + \rho \mathbf{D}^T \mathbf{D}$, we adopt the conjugate gradient method to find the solution $\bm{\gamma}'_r$. The subproblem corresponding to $\mathbf{z}_r$ is given by   
    \begin{equation}  
      \min_{ \mathbf{z}_r } \; \eta  \| \mathbf{z}_r \|_1 + \frac{\rho}{2} \| \mathbf{z}_r  - \mathbf{D} \bm{\gamma}'_r + \mathbf{u}_r \|_2^2 .  \label{eq:Problem_fusion_z}
   \end{equation}
Apply proximal operator for $\ell_1$-norm, we have 
     \begin{align} \label{eq:fusion_ADMM_z}
       \mathbf{z}_r & \leftarrow  \text{prox}_{ \eta / \rho  \|\cdot\|_{1}} \left( \mathbf{D} \bm{\gamma}'_r - \mathbf{u}_r   \right) \\
       & = \operatorname{sgn} \left( \mathbf{D} \bm{\gamma}'_r - \mathbf{u}_r \right)   \max\left( \left| \mathbf{D} \bm{\gamma}'_r - \mathbf{u}_r \right|  - \tfrac{\eta}{\rho}  , 0 \right), 
     \end{align}    
where $\operatorname{sgn}\left( \cdot \right)$ denotes the sign function, defined as $\operatorname{sgn}\left( s \right) = -1$ for $s\leq 0$ and $\operatorname{sgn}\left( s \right) = 1$ otherwise. The dual variable \(\mathbf{u}_r \) is updated in the following way: 
   \begin{equation}  \label{eq:fusion_ADMM_u}
     \mathbf{u}_r \leftarrow \mathbf{u}_r + \left( \mathbf{z}_r - \mathbf{D}\bm{\gamma}'_r  \right) .
   \end{equation}

When $\bm{\gamma}'_r$ is fixed, the optimal $\lambda_{k,q}, \forall k,q$ admits a closed-form as follows
  \begin{align}  
    \lambda_{k,q} & \leftarrow \arg\min_{ \lambda_{k,q} } \;  \lambda_{k,q}  \gamma_{\beta,k,q} (  {\gamma}'_{r,k,q} - {\gamma}'_{r,q} )^2   \notag\\
    & \;\;\;\;\; \;\;\;\;\;\;\;\;\;\; \;\;\; +   (1-\lambda_{k,q}) \gamma_{\beta,k,q}  ({\gamma}'_{r,k,q})^2    \\
    & = \left\{
     \begin{array}{ll}
        0, &  (\gamma'_{r,q})^{2} - 2 \gamma'_{r,q} \gamma'_{r,k,q}  > 0, \\
        1, &  (\gamma'_{r,q})^{2} - 2 \gamma'_{r,q} \gamma'_{r,k,q} \leq 0 .
     \end{array}
     \right. \label{eq:fusion_ADMM_lambda}
  \end{align}  
This criterion ensures that $\gamma'_{r,k,q}$ is incorporated into fusion when it exhibits sufficient strength relative to $\gamma'_{r,q}$; otherwise, it is excluded to prevent invalid data from distorting the fusion.

The overall multi-view fusion algorithm is shown in Algorithm \ref{alg:fusion}. At each iteration of the AO algorithm, for the update of $\bm{\gamma}'_r$ with fixed $\bm{\Lambda}$, the objective \eqref{eq:Problem_fusion2_obj} approaches its minimum value applying ADMM \cite{Boyd_ADMM}. For the optimization of $\bm{\Lambda}$ with fixed $\bm{\gamma}'_r$, the objective is minimized via the closed-form solution in \eqref{eq:fusion_ADMM_lambda}. This AO scheme ensures that the objective in \eqref{eq:fusion_obj} is non-increasing over iterations. Meanwhile, \eqref{eq:fusion_obj} is lower-bounded by zero due to the non-negativity of all terms. Thus, the convergence of the overall algorithm is guaranteed.

\subsection{Complexity Analysis} \label{Sec:V_complexity}

Algorithm 2 consists of two main steps: interpolating each RBS's coarse image onto a common set of finer grids, and aggregating the interpolated scattering intensities from all RBSs. The complexity of EP-NNI is dominated by calculating the Sibson natural neighbor weights in \eqref{eq:fusion_weight_Sibson}, incurring a complexity of $KQ' \log Q$ \cite{ref:Voronoi_complexity}. For the aggregation step, the AO approach is adopted. When $\bm{\Lambda}$ is fixed, the computational complexity of the ADMM-based approach is $\mathcal{O}(I'_1I'_2Q')$, where $I'_1$ and $I'_2$ denote the iteration number of ADMM and that of inner conjugate gradient for each ADMM iteration, respectively. This linear scaling is achieved since the matrix $\mathbf{D}$ is sparse, with only two non-zero elements in each row. When $\bm{\gamma}'_r$ is fixed, the computational complexity for updating $\bm{\Lambda}$ is $\mathcal{O}(KQ')$. Consequently, the overall complexity of Algorithm 2 is $\mathcal{O}( KQ' \log Q + I'_{\rm max}(I'_1I'_2Q'+KQ'))$, where $I'_{\rm max}$ denotes the iteration number of AO.

\begin{algorithm}[t] 
\caption{Multi-View Fusion}
\label{alg:fusion} 
\begin{algorithmic}[1]  
{ \small
\REQUIRE 
The estimates $\bm{\gamma}_{r,k}, \forall k$ at $K$ RBSs, 
grid positions $\mathbf{P}_{k}, \forall k$,  
tolerances $\epsilon_1$ and $\epsilon_2$,
and iterations $I'_{\rm max}, I'_1$, and $I'_2$.

\ENSURE  
The aggregated result $\bm{\gamma}'_{r}$. 

\STATE $\forall k$: Compute the gradient at grid $q \in [Q]$ based on \eqref{eq:fusion_gradq}. 
\STATE $\forall k$: Compute the structure tensor at interpolation grid $q' \in [Q']$ based on \eqref{eq:fusion_gradqprime} and perform SVD. 
\STATE $\forall k$: Compute the edge-preserving weight in \eqref{eq:fusion_weight}. 
\STATE $\forall k$: Obtain EP-NNI result $\bm{\gamma}'_{r,k}$ applying \eqref{eq:fusion_EPNNI}.

\STATE Initialize $\bm{\gamma}'_{r} = \sum_k \bm{\gamma}'_{r,k}, \mathbf{z}_r = \mathbf{D}\bm{\gamma}'_{r}$, and $\mathbf{u}_r = \bm{0}$. 

\FOR{$j = 1,2,\ldots, I'_{\max}$} 

\STATE Update $\bm{\Lambda}$ according to \eqref{eq:fusion_ADMM_lambda}. 

\FOR{$j_1 = 1,2,\ldots, I'_1$} 
\STATE Update $\!\bm{\gamma}'_{r}\!$ by solving \eqref{eq:fusion_ADMM_r_condition} via conjugate gradient method with iteration number $I'_2$.
\STATE Update $\mathbf{z}_{r}$ according to \eqref{eq:fusion_ADMM_z}.
\STATE Update $\mathbf{u}_{r}$ according to \eqref{eq:fusion_ADMM_u}.
\ENDFOR

\STATE If $\tfrac{\left\|  {\bm{\gamma}'_{r}}^{(j)} - {\bm{\gamma}'_{r}}^{(j-1)} \right\|_2^2}{ \left\| {\bm{\gamma}'_{r}}^{(j-1)} \right\|_2^2 } \leq \epsilon_1$   and $\tfrac{\left\| \bm{\Lambda}^{(j)} - \bm{\Lambda}^{(j-1)} \right\|_F^2}{ \left\| \bm{\Lambda}^{(j-1)} \right\|_F^2 } \leq \epsilon_2$, stop. 

\ENDFOR

}

\end{algorithmic}
\end{algorithm}

\begin{Remark} \label{remark:asy} 
  The commonly used clock synchronization technologies in wireless networks include the global navigation satellite system (GNSS) and generalized precision time protocol (gPTP/IEEE 802.1AS), which provide a synchronization accuracy at the nanosecond level \cite{ref:GPS,ref:PTP}. Moreover, as mentioned before, our imaging framework is designed to be compatible with OFDM-based systems by utilizing a single subcarrier for transmitting imaging-oriented pilots. In such systems, the cyclic prefix (CP) length is typically on the order of microseconds. Since the synchronization accuracy is significantly smaller than the CP length, the timing offset $\tau$ does not cause inter-symbol interference but manifests as a phase shift $e^{-j2\pi f \tau}$ on the channel matrix, where $f$ denotes the frequency of this subcarrier. It should be noted that our covariance-based approach in Phase I is inherently robust to this phase shift, since any common phase shift is mathematically canceled out when calculating the sample covariance matrix in \eqref{eq:receive_Y_sigmakhat}. In Phase II, a fusion operation is triggered only when the CPU has successfully collected the scattering intensities and grid coordinates from $K$ RBSs. Due to the nanosecond-level synchronization accuracy, the timing jitter between different RBSs is extremely small. Thus, the observation window only needs to be slightly longer than the fronthaul transmission delay to ensure that all $K$ packets are correctly received, thereby enabling robust multi-view fusion.   
\end{Remark}

\section{Simulation Results}    \label{Sec:simulation}

In this section, we provide numerical results to verify the effectiveness of the proposed scheme. The system setup is as follows. The TBS is located at $[-3,7.5]^T$ in meter. There are $K=3$ RBSs, located at $[18,7.5]^T, [7.5,18]^T,$ and $[7.5,-3]^T$ in meter, respectively. Extended targets are located in the RoI with $x_1=y_1=0$ and $x_2=y_2=15$. The power spectral density of the AWGN at RBSs is $-169$~dBm/Hz and the channel bandwidth is $1$~MHz. The path-loss of the path from TBS to the scatterer located in $[x,y]^T$ and then to the $k$-th RBS is given by $\gamma_{\beta,x,y,k} = \beta_0^2 d_{{\rm tx},x,y}^{-2} d_{{\rm rx},k,x,y}^{-2}$, where $d_{{\rm tx},x,y}$ and $d_{{\rm rx},k,x,y}$ denote the distance between the TBS to scatterer and that between the scatterer and the $k$-th RBS, respectively, and $\beta_0 = -35$~dB measures the path-loss at the reference~distance. The penalty parameters in \eqref{eq:Problem1_obj} and \eqref{eq:fusion_obj} are empirically determined by conducting extensive pre-simulations across a variety of sensing scenarios and selecting the values that consistently yield excellent imaging performance \cite{zhang2011sparse,lu2008uncorrelated}. These parameters are kept constant throughout the iterations of Algorithm 1 and Algorithm 2 to ensure the stability of the framework.

\begin{figure*}
    \centering 
    \hspace{-3mm}
    \subfigure[]{\includegraphics[width=0.16\linewidth]{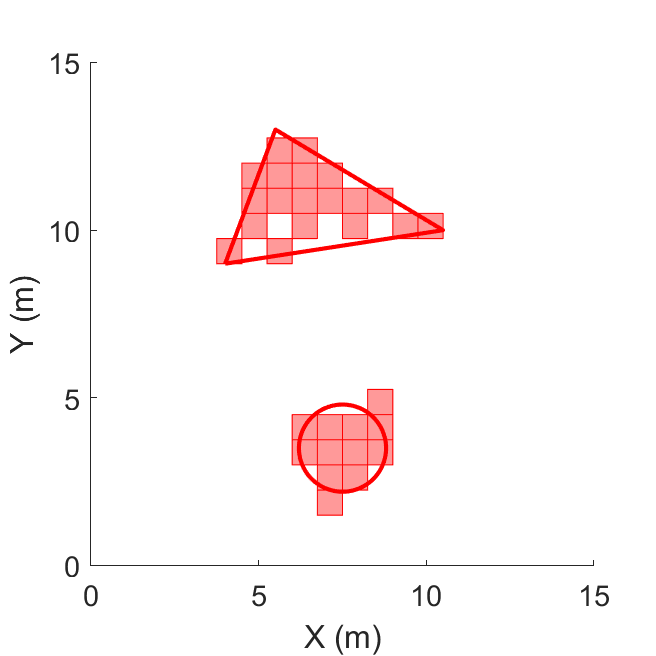}   \label{fig:grideta_a}}   \hspace{-4mm}
    \subfigure[]{\includegraphics[width=0.16\linewidth]{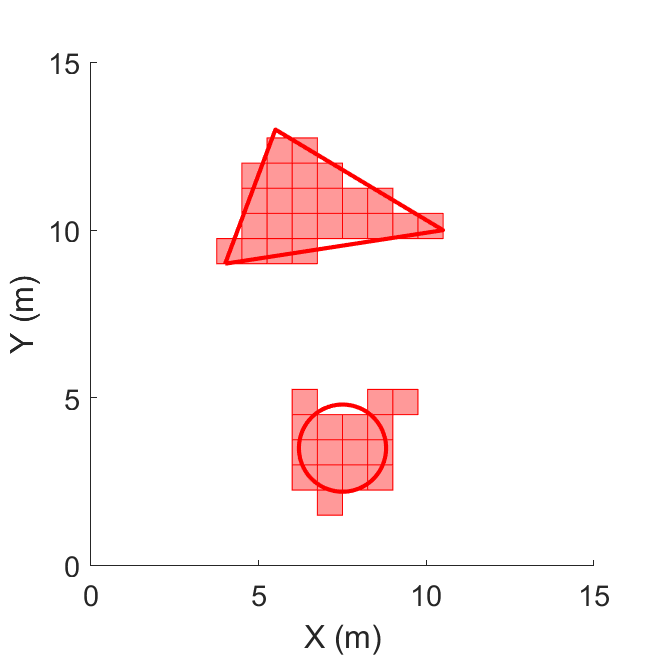}  \label{fig:grideta_b}}  \hspace{-4mm}   
    \subfigure[]{\includegraphics[width=0.16\linewidth]{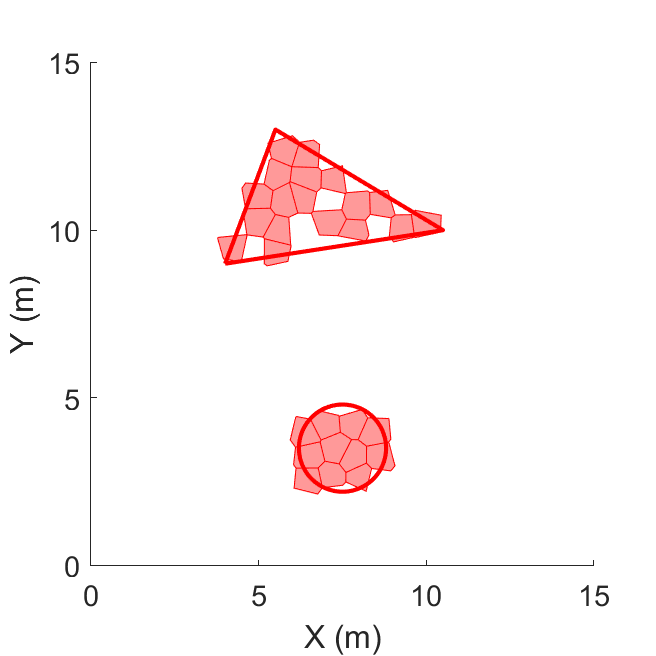}   \label{fig:grideta_c}}  \hspace{-4mm} 
    \subfigure[]{\includegraphics[width=0.16\linewidth]{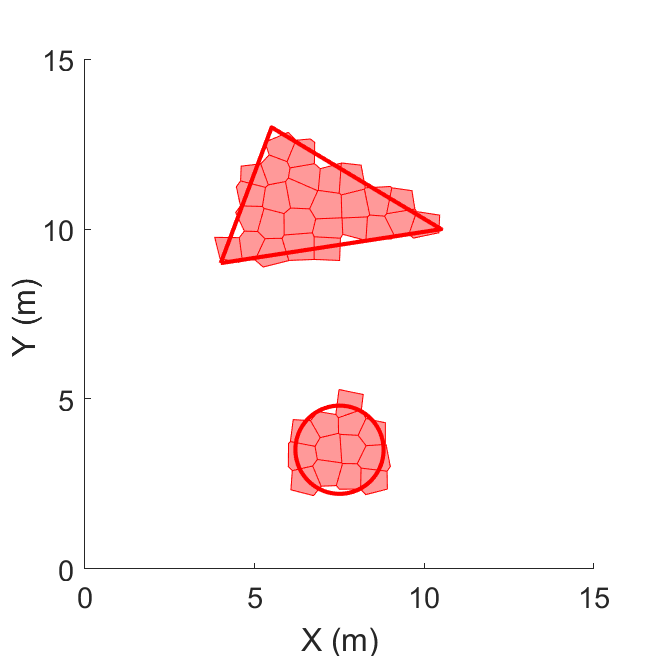}  \label{fig:grideta_d}}  \hspace{-4mm}   
    \subfigure[]{\includegraphics[width=0.16\linewidth]{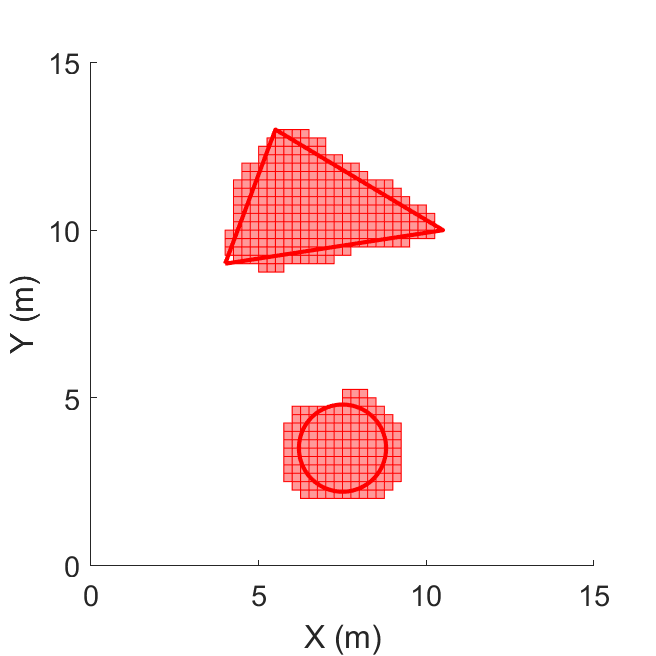}   \label{fig:grideta_e}}  \hspace{-4mm}
    \subfigure[]{\includegraphics[width=0.16\linewidth]{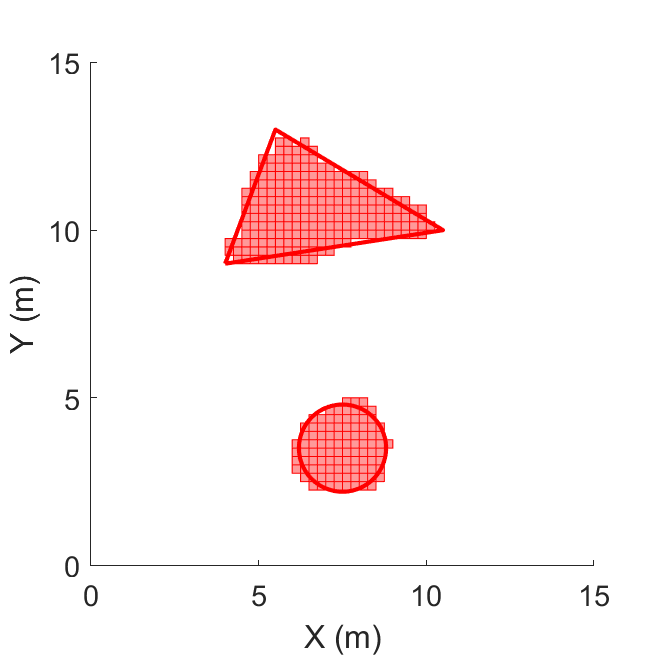}  \label{fig:grideta_f}}  \hspace{-3mm}     
	\caption{Covariance-based single-view images at RBS $1$ with $N_{\rm tx} = 16, N_{\rm rx} = 16, 
L = 16, M = 20, P = 10$ dBm, and $Q = 400$: (a) Fixed grid without term \eqref{eq:distributed_penalty}; (b) Fixed grid with term \eqref{eq:distributed_penalty}; (c) Optimized grid without term \eqref{eq:distributed_penalty}; (d) Optimized grid with term \eqref{eq:distributed_penalty}; (e) NNI; (f) EP-NNI.} \label{fig:grideta} 
\end{figure*}

In Fig.~\ref{fig:grideta}, we present the single-view images at the first RBS (located at $[18,7.5]^T$) applying the proposed covariance-based scheme, in the cases with and without the cluster-promoting term, grid optimization, and edge-preserving weight. We assume that both the TBS and RBS are equipped with $16$ antennas, the length of orthogonal pilot is $L=16$, the number of frames is $M=20$, and the transmit power is $P = 10$~dBm. We choose a triangle and a circle as extended targets with uniform scattering intensity across all scatterers. The contours of extended targets are plotted with red solid lines. We assume that the RoI is within the field of view of the RBS to verify the single-view imaging performance. In Fig.~\ref{fig:grideta}, the detected grids with high scattering intensity are displayed, which are selected by sorting all detected grids according to their intensity and retaining the highest-intensity ones until their cumulative intensity reaches $95\%$ of the total recovered intensity. In Fig.~\ref{fig:grideta_a}, we divide the RoI into $Q=400$ uniform grids, and optimize the effective scattering intensity $\bm{\gamma}_{r,k}$ with fixed grids based on the ML criterion \eqref{eq:distributed_ML}. In Fig.~\ref{fig:grideta_b}, the cluster-promoting term \eqref{eq:distributed_penalty} is taken into consideration for the optimization of $\bm{\gamma}_{r,k}$. In Fig.~\ref{fig:grideta_c} and Fig.~\ref{fig:grideta_d}, we optimize both grid positions and scattering intensity, using only the ML criterion and adding the cluster-promoting term, respectively. It is evident that by introducing the cluster-promoting term, the recovered extended targets exhibit enhanced continuity and reduced internal voids. Moreover, grid optimization sharply concentrates the detected energy within the targets, significantly reducing sidelobes and yielding boundaries that closely align with the true shape. Fig.~\ref{fig:grideta_e} and Fig.~\ref{fig:grideta_f} are obtained by interpolating the imaging result in Fig.~\ref{fig:grideta_d} onto $Q' = 60 \times 60$ grids via NNI and EP-NNI techniques, respectively. As we can see, the proposed EP-NNI scheme demonstrates superior performance over conventional NNI in preserving edge features and reducing blurring effects. The comparison between Fig.~\ref{fig:grideta_d} and Fig.~\ref{fig:grideta_f} shows that EP-NNI achieves smoother target boundaries by mitigating staircase artifacts inherent to discrete grid representations.

\begin{figure}
    \centering 
    \subfigure[]{\includegraphics[width=0.38\linewidth]{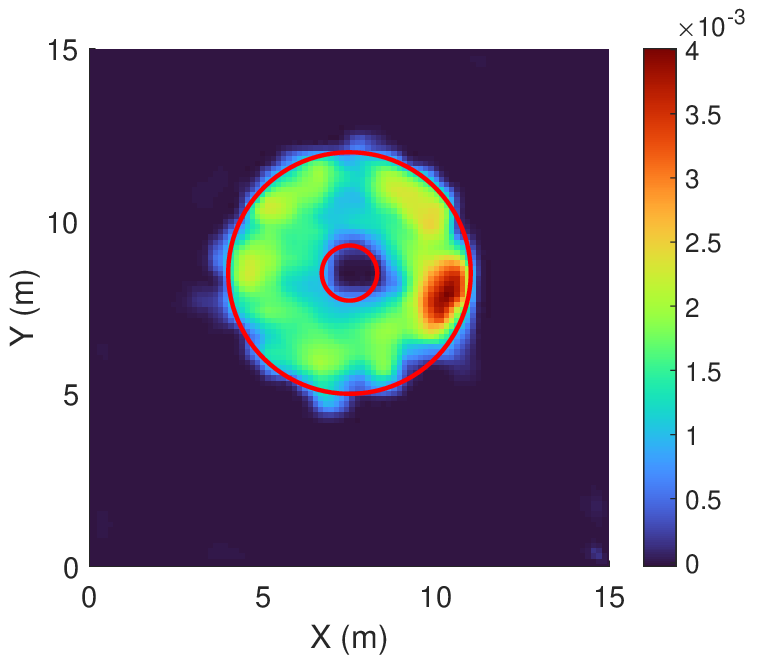} \label{fig:gridnuma}}  
    \subfigure[]{\includegraphics[width=0.38\linewidth]{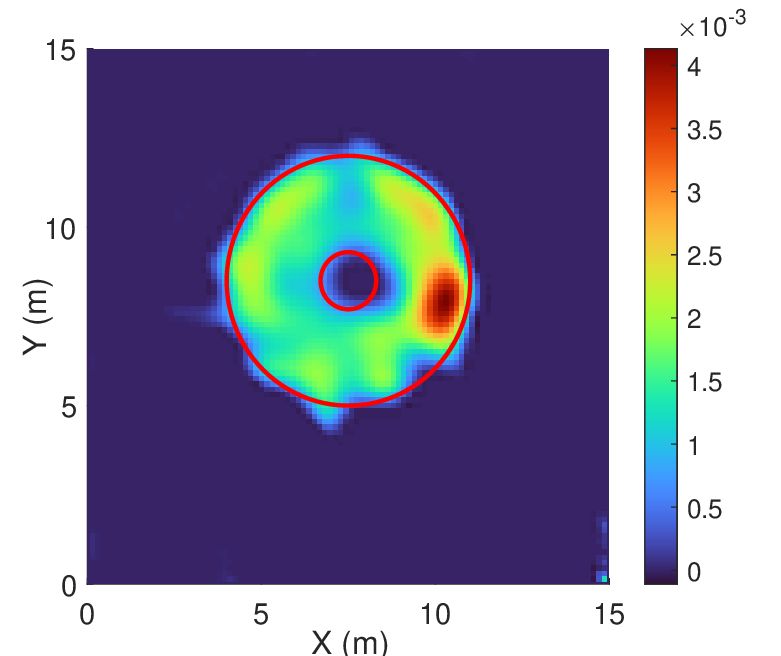} \label{fig:gridnumb} }   
    \subfigure[]{\includegraphics[width=0.38\linewidth]{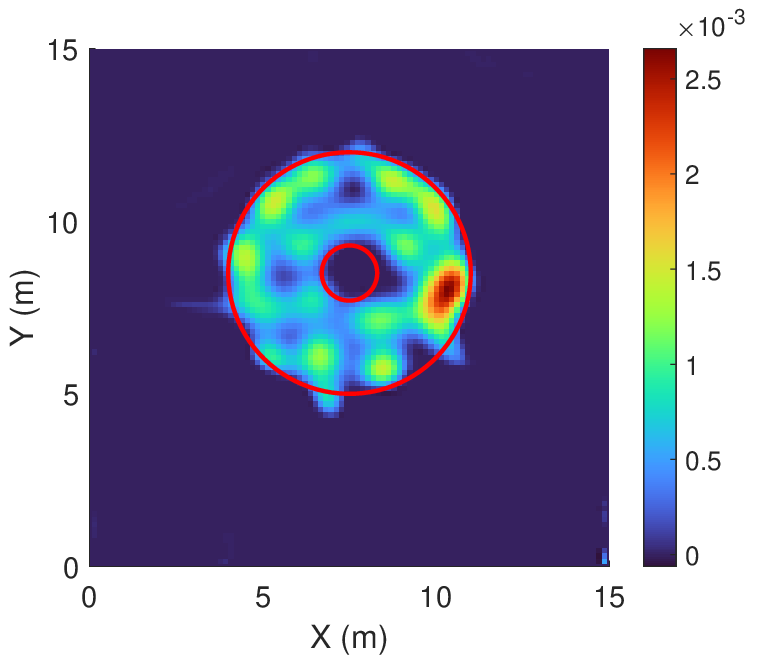}  \label{fig:gridnumc}} 
    \subfigure[]{\includegraphics[width=0.38\linewidth]{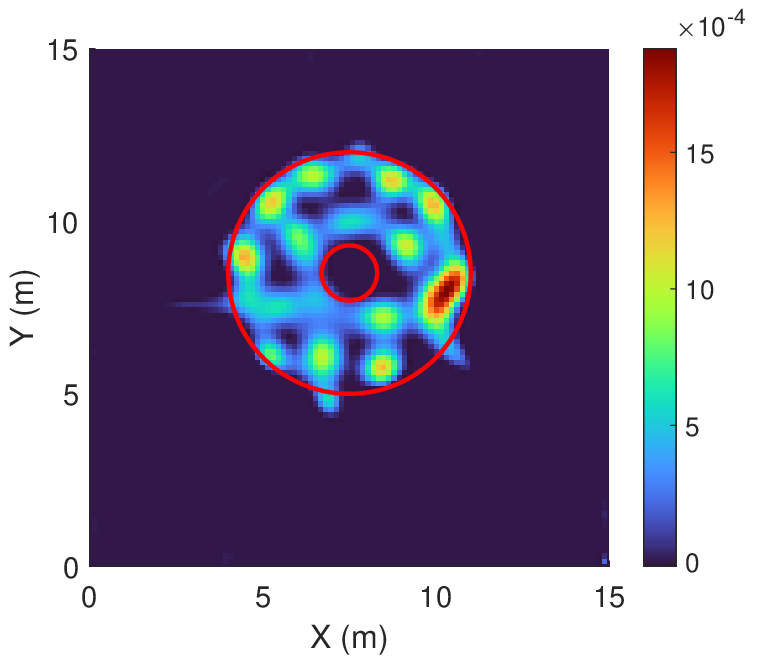}  \label{fig:gridnumd}}   
	\caption{Covariance-based single-view images at RBS $2$ with $N_{\rm tx} = 8, N_{\rm rx} = 8, M = 20, L = 8$, and $P = 10$~dBm: (a) Grid optimization with $Q = 900$ and interpolation to uniform grids with $Q' = 60\times 60$; (b) Fixed uniform grids with $Q = 60\times 60$; (c) Fixed uniform grids with $Q = 90\times 90$; (d) Fixed uniform grids with $Q = 120\times 120$.} \label{fig:gridnum}  
\end{figure}

In Fig.~\ref{fig:gridnum}, we present the single-view images at RBS $2$ (located at $[7.5,18]^T$) with $N_{\rm tx}=N_{\rm rx}=L=8, M = 20$, and $P = 10$~dBm. We choose an annulus as the extended target. Fig.~\ref{fig:gridnuma} is obtained by jointly estimating effective scattering intensity and grid positions with $Q=900$ based on the criterion in \eqref{eq:Problem1}, followed by EP-NNI onto finer uniform grids with $Q' = 60 \times 60$. In comparison, Figs.~\ref{fig:gridnumb}, \ref{fig:gridnumc}, and \ref{fig:gridnumd} present the images using fixed uniform grids with $Q = 60\times60$, $Q = 90\times90$, and $Q = 120\times120$, respectively, where only the scattering intensity is optimized. As we can see, Figs.~\ref{fig:gridnuma} and \ref{fig:gridnumb} present similar performance, i.e., the approach with dynamic but a small number of grids achieves comparable performance to fixed-grid reconstruction with a slightly higher grid density, both clearly recovering the annular target. However, considering that the distance between grid points that can be distinguished is constrained by the number of antennas, when the number of grids is further increased to $Q = 90\times90$ and $Q = 120\times120$, the reconstructed images exhibit multiple hollow regions beyond the true inner ring, making it impossible to identify the structure of target. This demonstrates the significance of adopting dynamic grids instead of arbitrarily increasing the grid density.

\begin{figure}
    \centering 
    \subfigure[]{\includegraphics[width=0.38\linewidth]{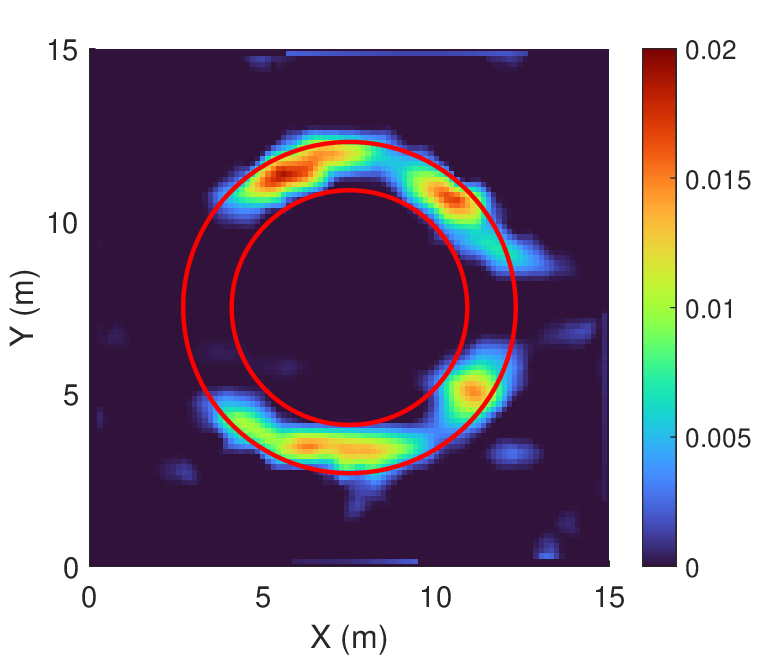} \label{fig:fusea} }  
    \subfigure[]{\includegraphics[width=0.38\linewidth]{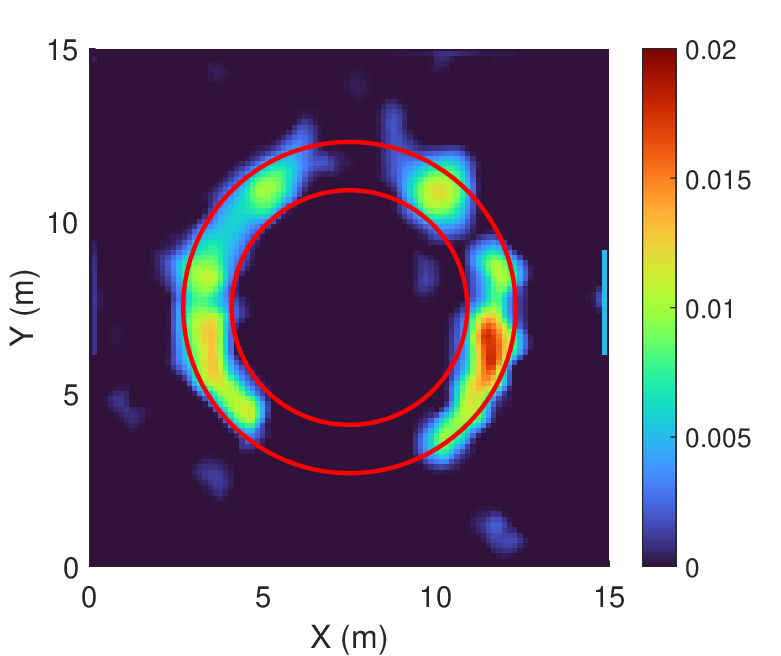} \label{fig:fuseb} }  
    \subfigure[]{\includegraphics[width=0.38\linewidth]{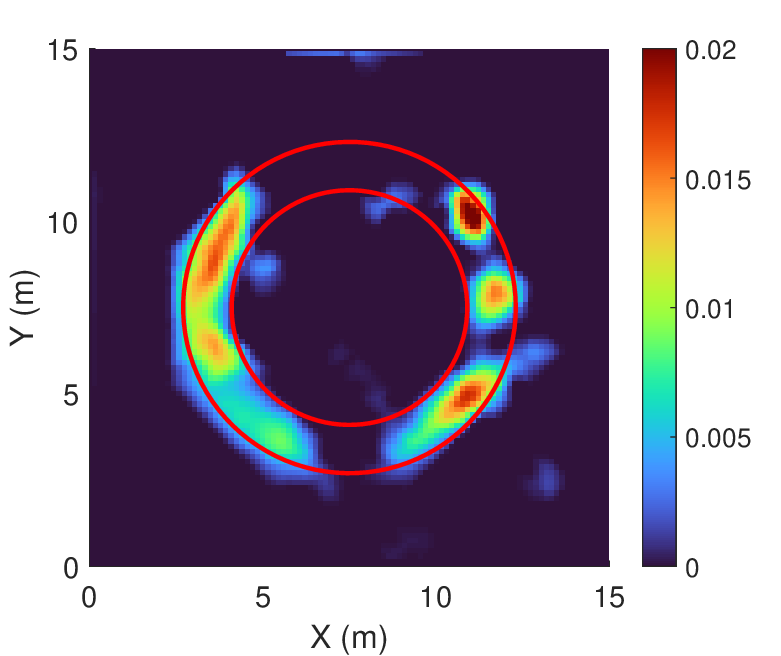}  \label{fig:fusec} }  
    \subfigure[]{\includegraphics[width=0.38\linewidth]{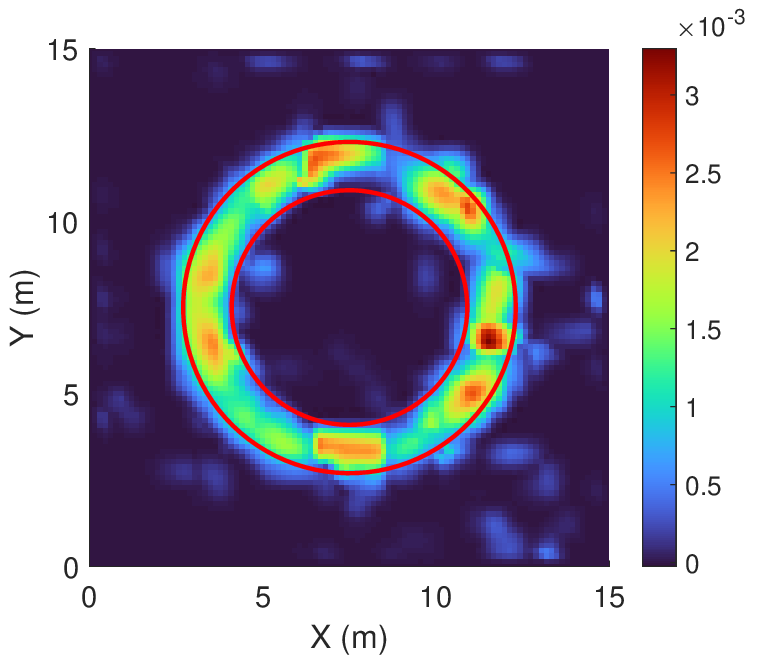}  \label{fig:fused} }  
    \subfigure[]{\includegraphics[width=0.38\linewidth]{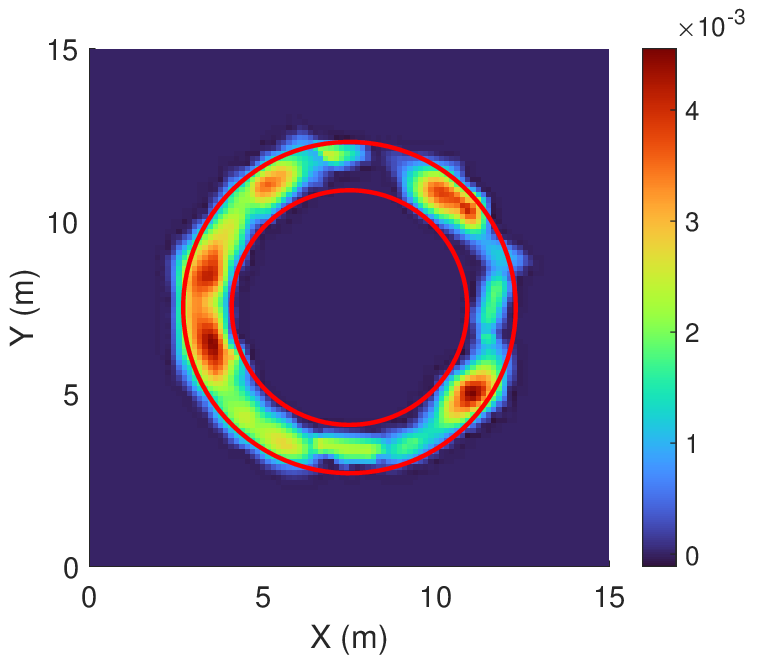}  \label{fig:fusee} }  
    \subfigure[]{\includegraphics[width=0.38\linewidth]{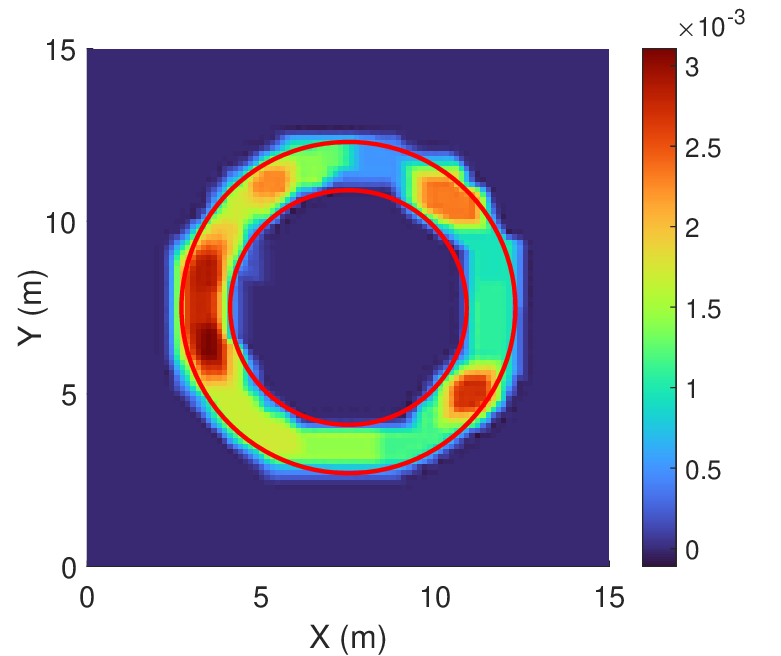}  \label{fig:fusef} }   
	\caption{Covariance-based single-view images and multi-view fusion results with $N_{\rm tx} = N_{\rm rx} = L = 8, M = 20,  P = 0$ dBm, $Q = 900$, and $Q' = 60 \times 60$: (a) Image at RBS $1$; (b) Image at RBS $2$; (c) Image at RBS $3$; (d) Multi-view fusion with WLS term; (e) Multi-view fusion with WLS and sparse terms; (f) Multi-view fusion with WLS, sparse, and TV terms.} \label{fig:fuse}  
\end{figure}

In Fig.~\ref{fig:fuse}, we present the single-view images at three RBSs and the fused images at the CPU, in the scenario with $N_{\rm tx}=N_{\rm rx}=L=8, M=20$, and $P = 0$~dBm. We assume that each RBS has a limited field of view, with a $\pi/8$ obstacle-induced angular blind sector centered on the boresight direction of the array. The covariance-based single-view images at three RBSs are provided in Figs.~\ref{fig:fusea}, \ref{fig:fuseb}, and \ref{fig:fusec}, respectively, each of which is obtained applying Algorithm \ref{alg:image}. We can observe that each RBS captures only a part of the extended target with poor quality. Fig.~\ref{fig:fused} presents the fused image based on the WLS criterion in \eqref{eq:Problem_fusion_WLS}, exhibiting serious spurious detections outside the extended target. Thus, we introduce the sparsity-promoting term~\eqref{eq:Problem_fusion_sparse}, which effectively suppresses false alarms, as shown in Fig.~\ref{fig:fusee}. However, this comes at the cost of inducing structural fragmentation in the reconstruction. To address this problem, we additionally incorporate the TV in~\eqref{eq:Problem_fusion_TV} to enhance the structural continuity by promoting gradient sparsity. As we can see from Fig.~\ref{fig:fusef}, this combined cost function yields substantially improved image quality. By leveraging diversity gain from multi-view observations, data fusion contributes to superior reconstruction performance compared to single-view images.

\begin{figure*}
    \centering  
    \subfigure[]{\includegraphics[width=0.165\linewidth]{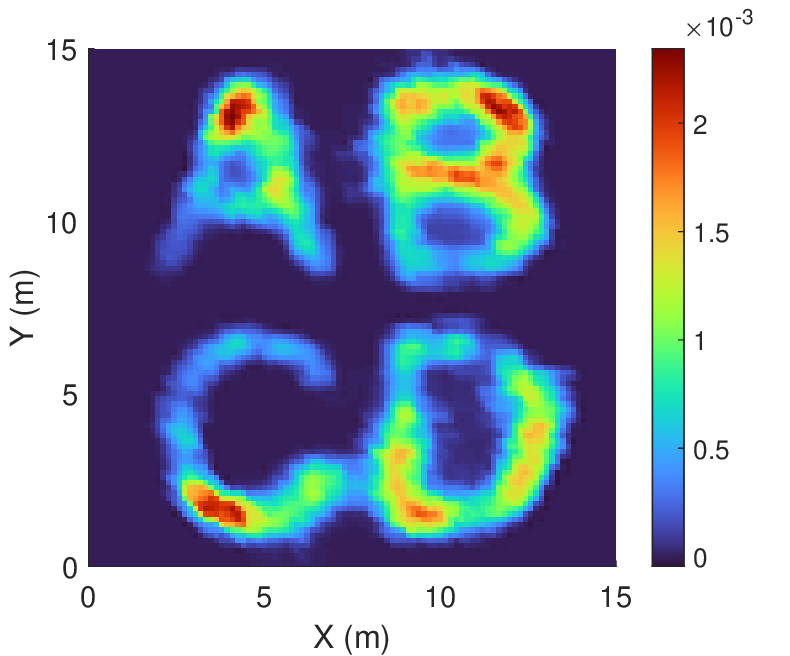} }  \hspace{-3.5mm}
    \subfigure[]{\includegraphics[width=0.165\linewidth]{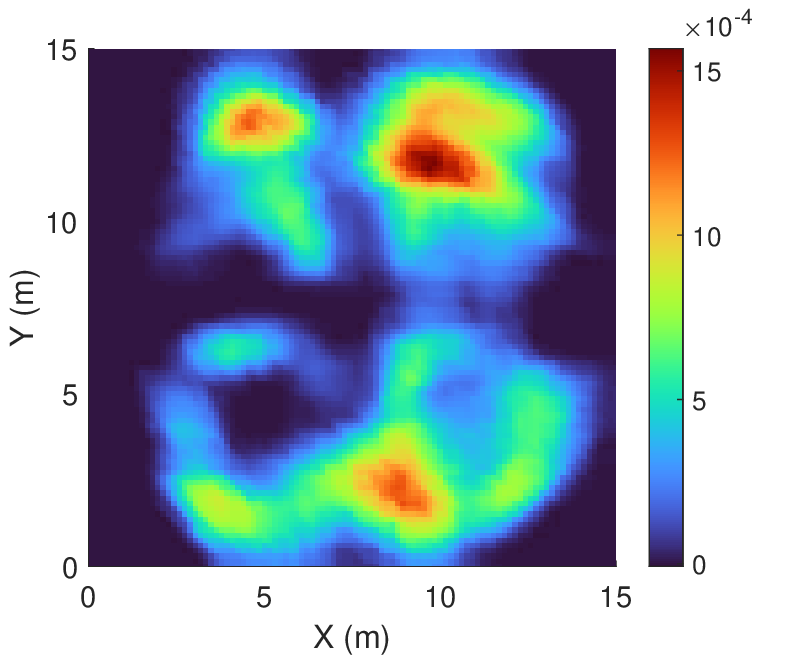} }  \hspace{-3.5mm}
    \subfigure[]{\includegraphics[width=0.165\linewidth]{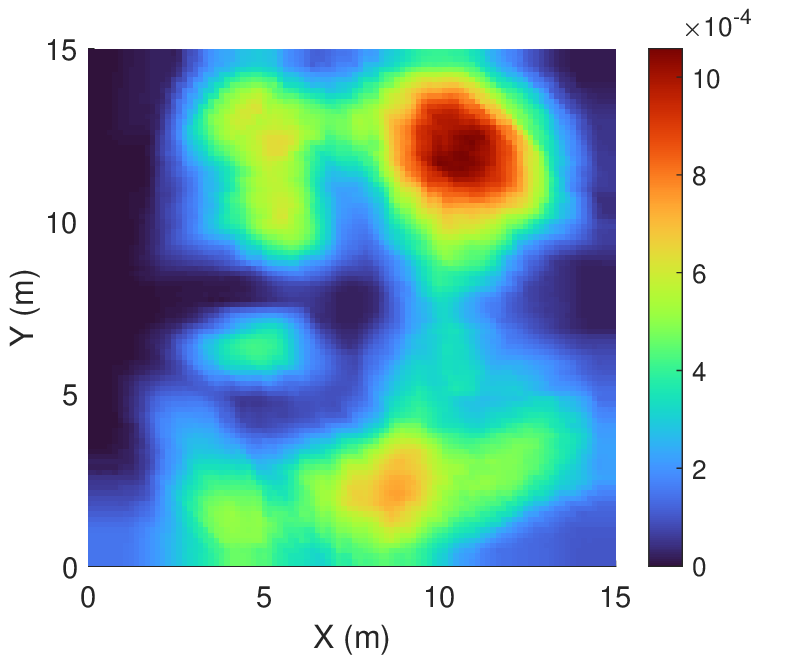} }  \hspace{-3.5mm}
    \subfigure[]{\includegraphics[width=0.165\linewidth]{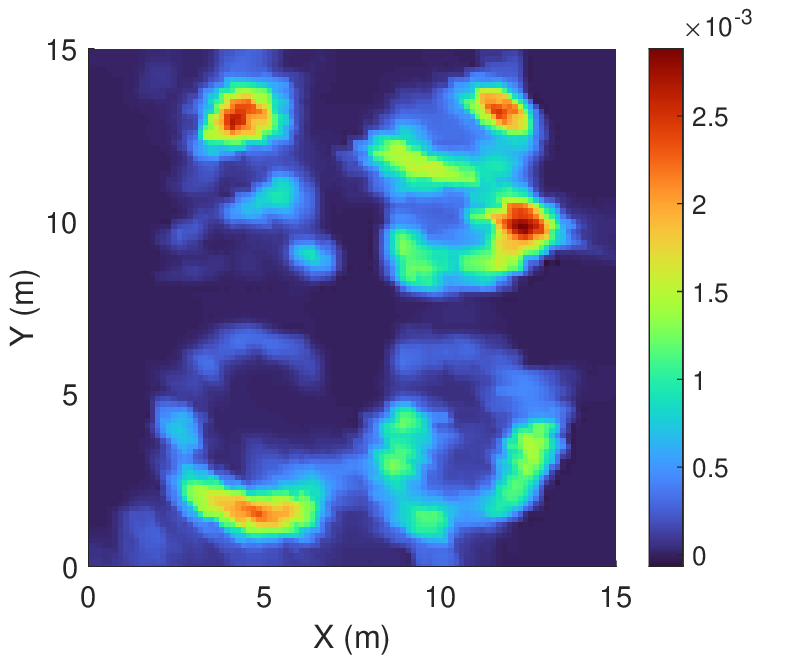}  } \hspace{-3.5mm}
    \subfigure[]{\includegraphics[width=0.165\linewidth]{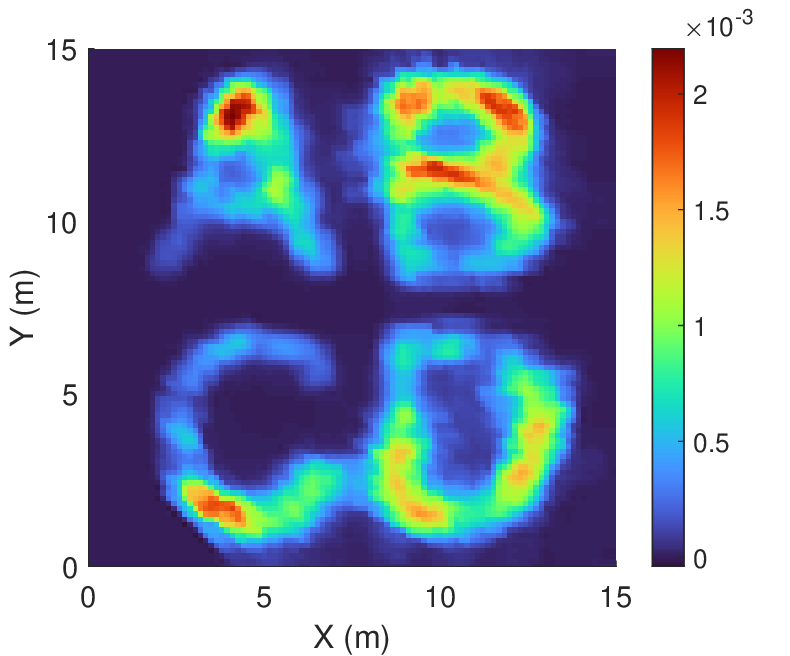}  } \hspace{-3.5mm}
    \subfigure[]{\includegraphics[width=0.165\linewidth]{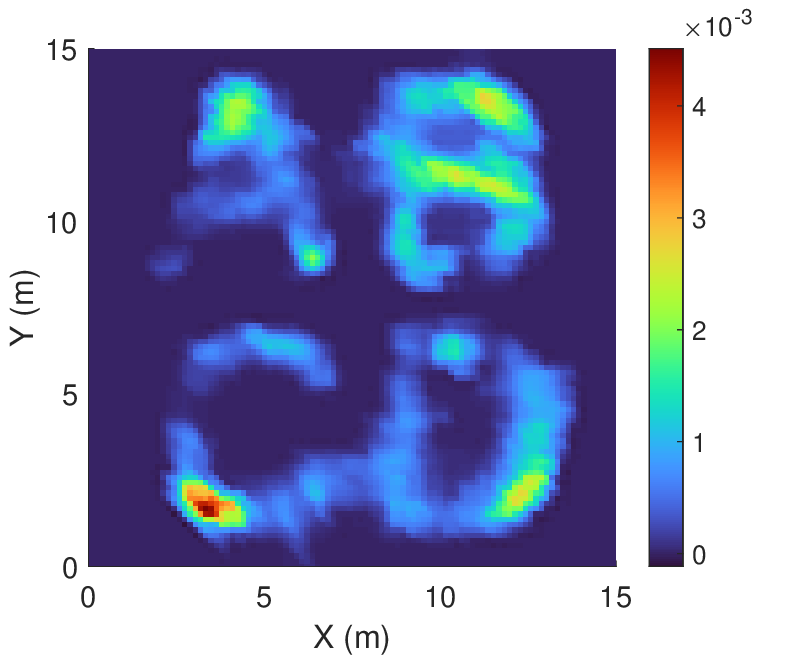}  }  
    \subfigure[]{\includegraphics[width=0.165\linewidth]{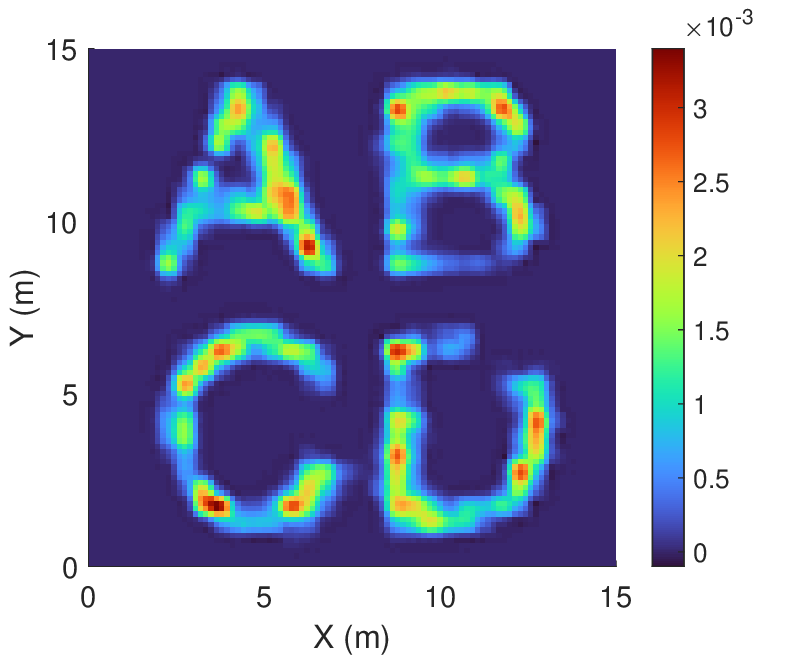}  } \hspace{-3.5mm}
    \subfigure[]{\includegraphics[width=0.165\linewidth]{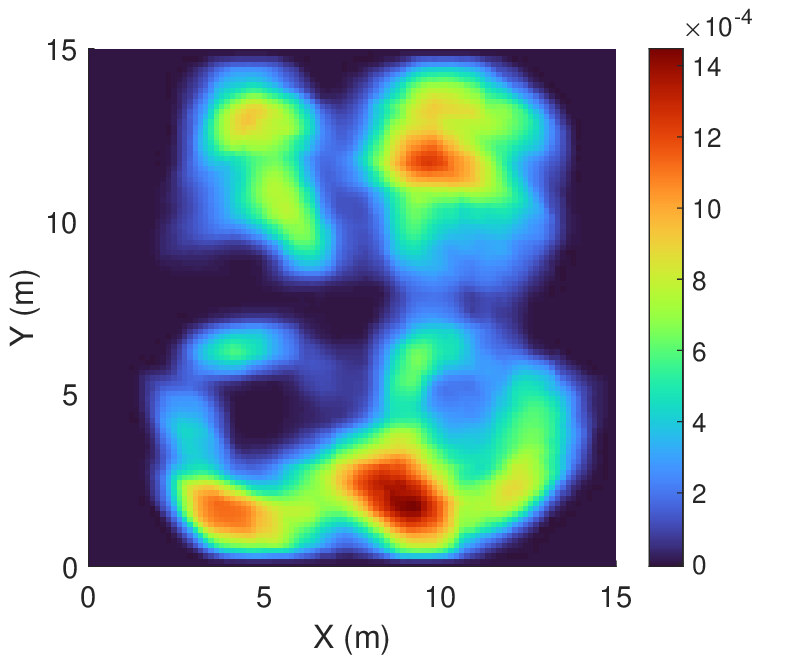}  } \hspace{-3.5mm}
    \subfigure[]{\includegraphics[width=0.165\linewidth]{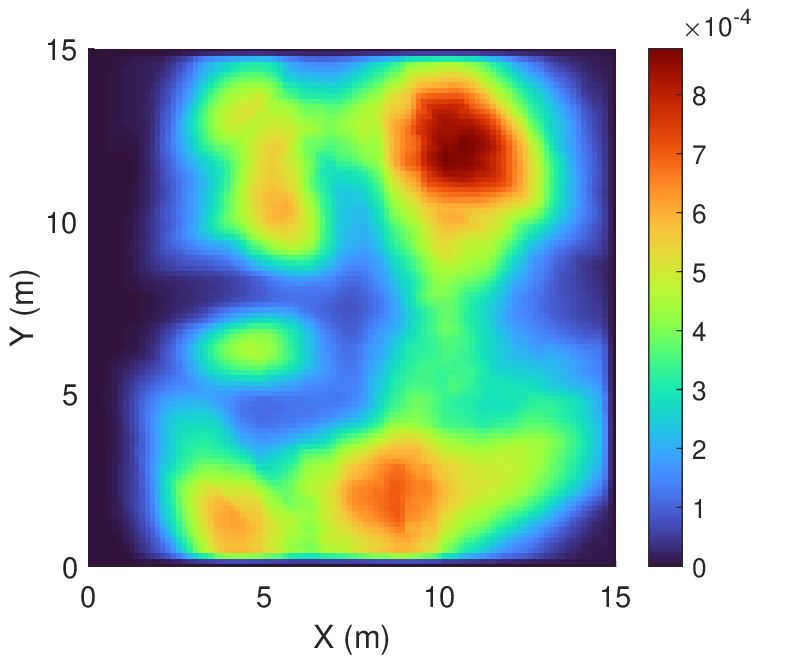}  } \hspace{-3.5mm}
    \subfigure[]{\includegraphics[width=0.165\linewidth]{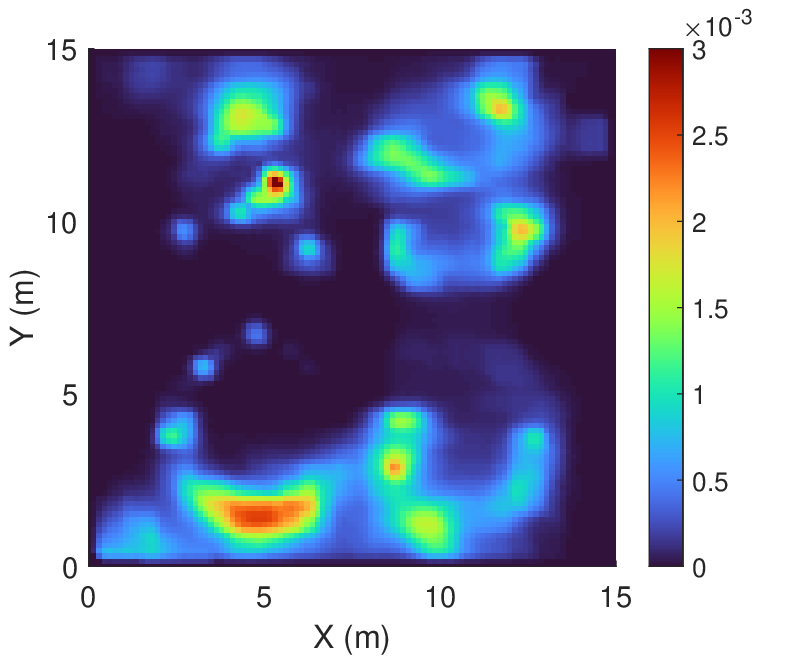}  }  \hspace{-3.5mm}
    \subfigure[]{\includegraphics[width=0.165\linewidth]{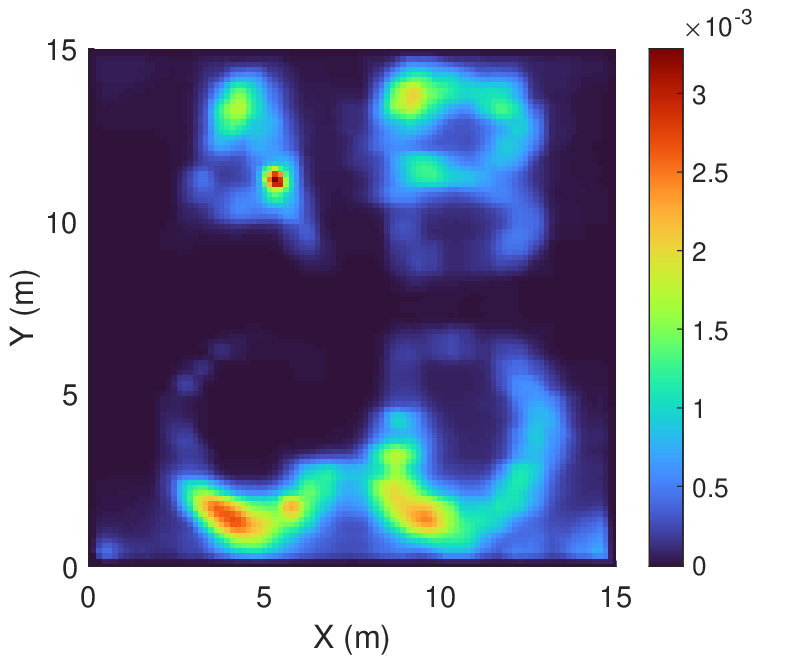}  } \hspace{-3.5mm}
    \subfigure[]{\includegraphics[width=0.165\linewidth]{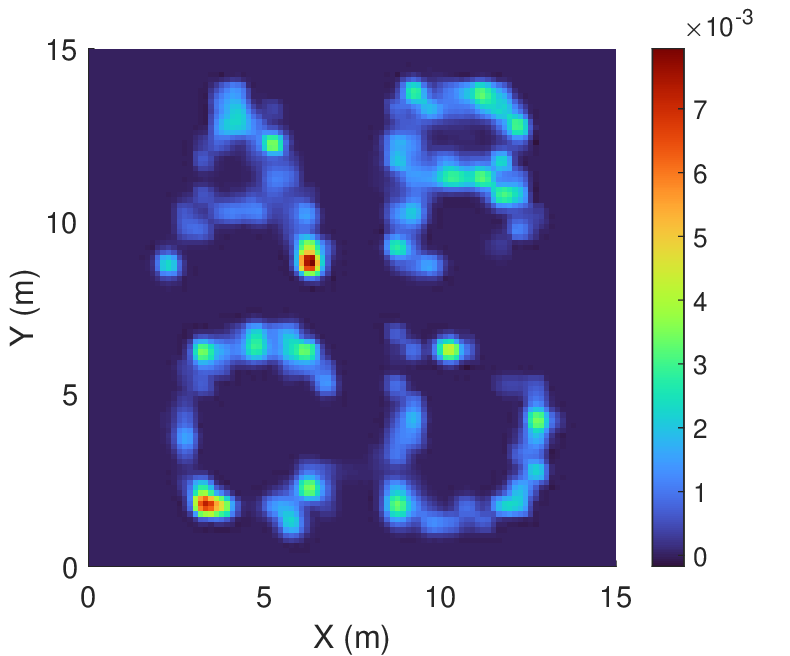}  }  \hspace{-1mm}
    \subfigure[]{\includegraphics[width=0.165\linewidth]{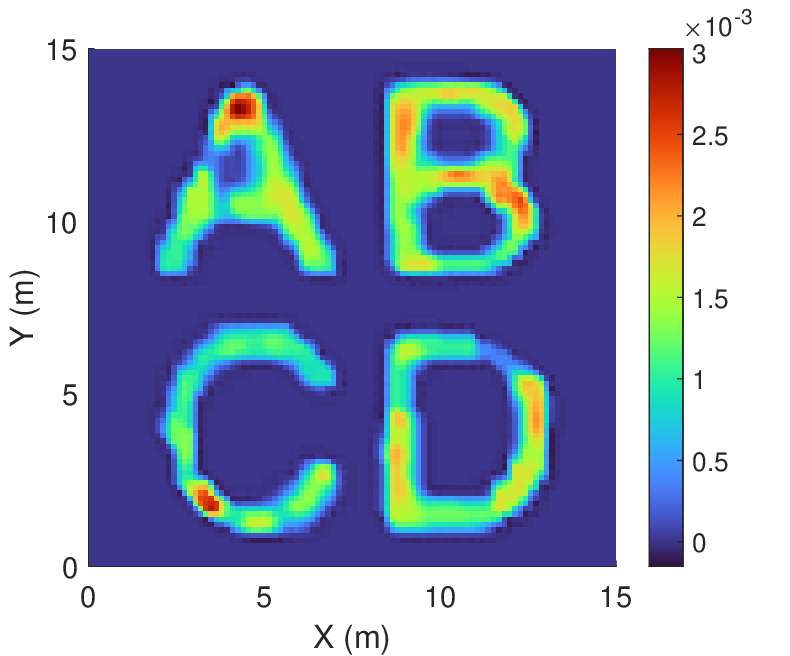}  }  \hspace{-3.5mm}
    \subfigure[]{\includegraphics[width=0.165\linewidth]{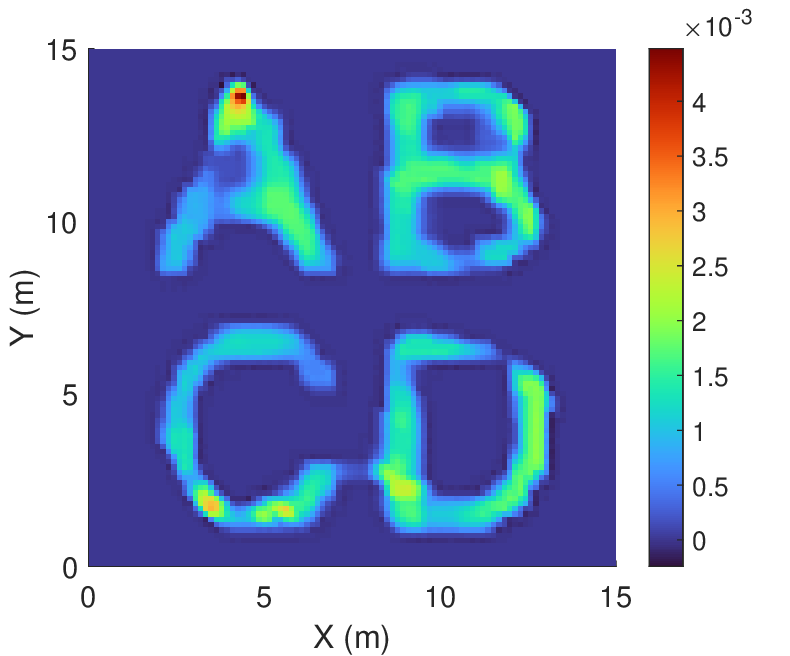}  } \hspace{-3.5mm}
    \subfigure[]{\includegraphics[width=0.165\linewidth]{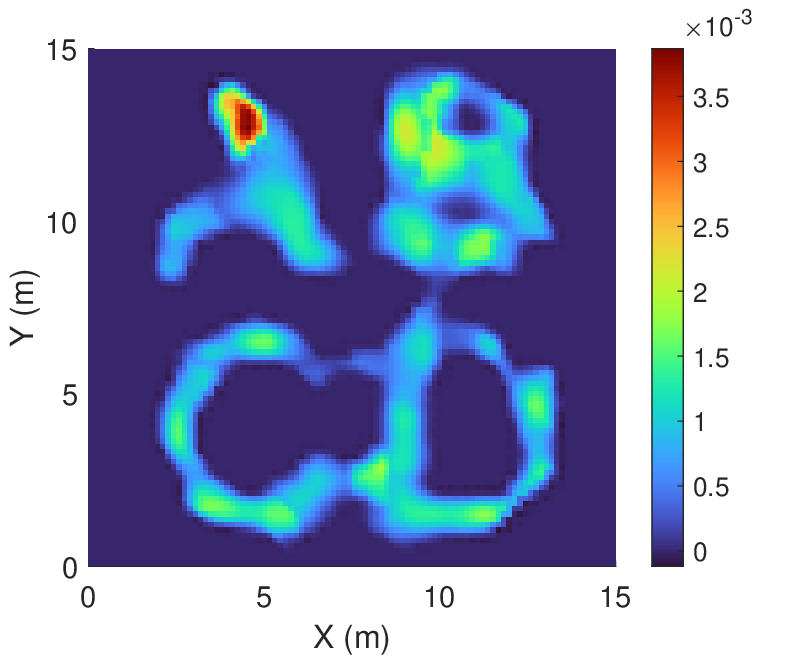}  } \hspace{-3.5mm}
    \subfigure[]{\includegraphics[width=0.165\linewidth]{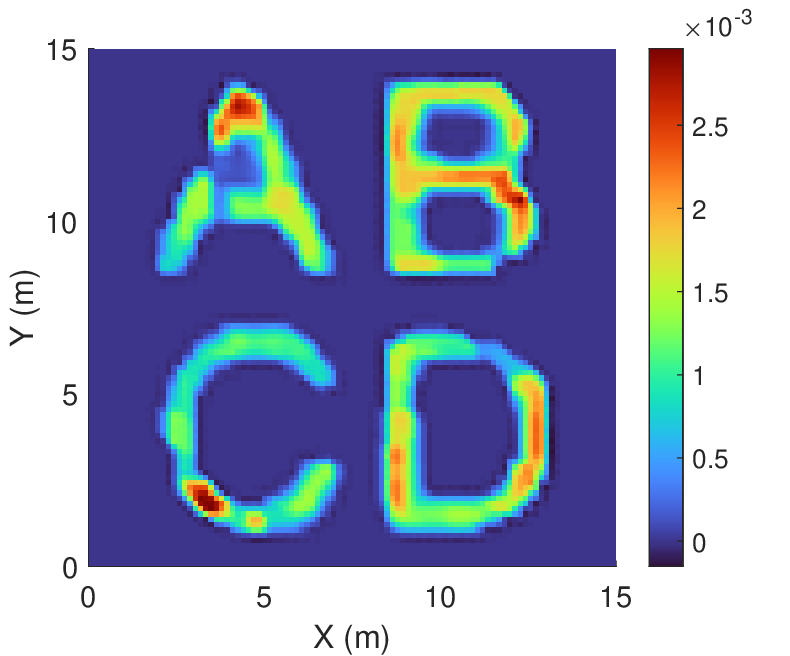}  }  \hspace{-3.5mm} 
    \subfigure[]{\includegraphics[width=0.165\linewidth]{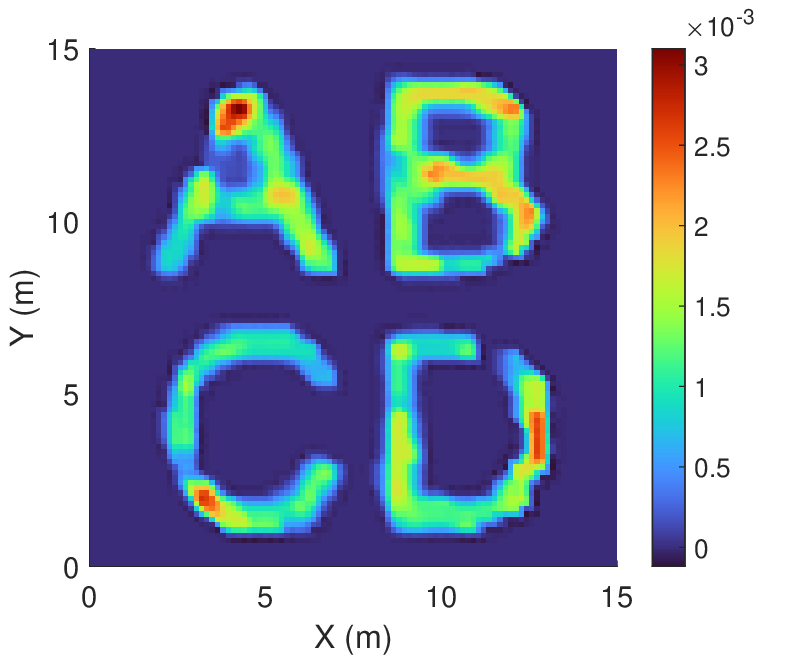}  } \hspace{-3.5mm}
    \subfigure[]{\includegraphics[width=0.165\linewidth]{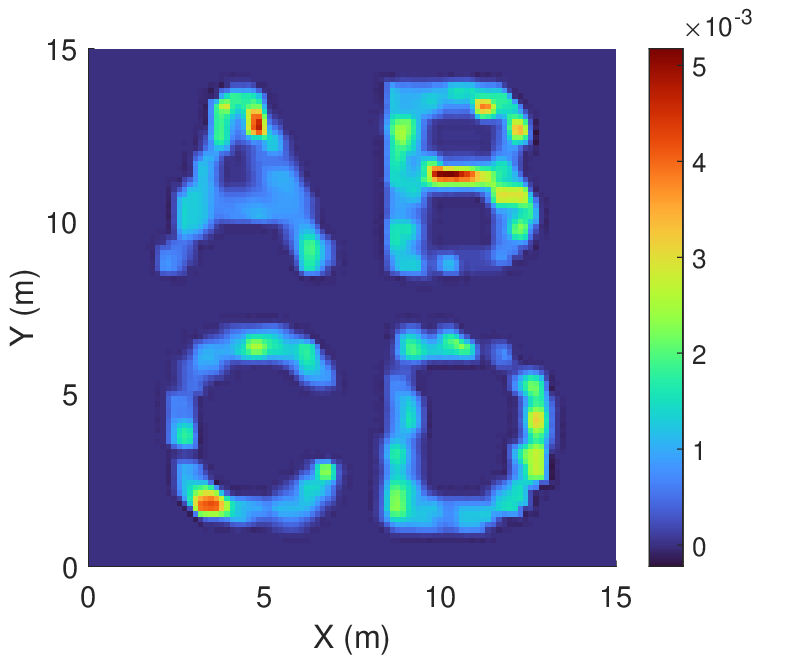}  }  
    \vspace{-1.5mm}
	\caption{Imaging performance comparison between different schemes with $K = 3, Q = 900$, and $Q' = 60 \times 60$. 
The first, the second, and the third rows present the images of benchmark scheme I, benchmark scheme II, and our proposed scheme, respectively. 
The first column is obtained assuming $N_{\rm tx} = N_{\rm rx} = L = 24, M = 20$, and $P = 10$ dBm. 
The second column is obtained assuming $N_{\rm tx} = N_{\rm rx} = L = 12, M = 20$, and $P = 10$ dBm.
The third column is obtained assuming $N_{\rm tx} = N_{\rm rx} = L = 8, M = 20$, and $P = 10$ dBm. 
The fourth column is obtained assuming $N_{\rm tx} = N_{\rm rx} = 24, L = 12, M = 20$, and $P = 10$ dBm.
The fifth column is obtained assuming $N_{\rm tx} = N_{\rm rx} = L = 24, M = 20$, and $P = -10$ dBm. 
The sixth column is obtained assuming $N_{\rm tx} = N_{\rm rx} = L = 24, M = 5$, and $P = 10$ dBm. 
} \label{fig:benchmark}  
\end{figure*} 

  \begin{table*}[!t] 
    \caption{Imaging Performance Comparison} \label{table:compare}
    \vspace{-1.5mm}
    \renewcommand{\arraystretch}{1}
    \centering
    {\scriptsize
    \begin{threeparttable}    
    \begin{tabular}{c|c|c|c|c|c|c|c}
      \hline
      \multicolumn{2}{c|}{~}  & Column 1 & Column 2 & Column 3 & Column 4 & Column 5  & Column 6
      \\  \hline  
      \multirow{2}*{Benchmark Scheme I} 
      & P-ISLR (dB) &  -3.61 &  0.86 & 2.82 & -1.62 &  -2.49 &  -3.60  \\   \cline{2-8}
      & IoU       &  0.59  &  0.37 & 0.29 & 0.44  &  0.51  &  0.56  \\   \hline
      \multirow{2}*{Benchmark Scheme II} 
      & P-ISLR (dB) &  -9.05 &  0.79 & 2.32  & -1.97  &  -2.54 &  -7.65 \\   \cline{2-8}
      & IoU       &  0.81  &  0.39 & 0.34 & 0.41 &  0.46  &  0.70  \\   \hline 
      \multirow{2}*{Proposed Scheme}  
      & P-ISLR (dB) & -11.37  &  -7.77 & -3.45 & -11.20  & -7.65  &  -9.81  \\   \cline{2-8}
      & IoU       &  0.89  &   0.82 & 0.64 &  0.88   &  0.82  &  0.84  \\   \hline 
    \end{tabular} 
    \end{threeparttable} }   
  \end{table*}  
  
Fig.~\ref{fig:benchmark} compares the multi-view imaging performance of the proposed scheme with the following benchmark schemes: 
\begin{itemize}
    \item Benchmark Scheme I: 
    We first multiply $\mathbf{Y}_{k,m}$ by $\mathbf{X}^H$ to obtain virtual array signal, and then apply IFFT to transform this signal into the wavenumber domain. The single-view image is formed based on the angular information, followed by multi-frame accumulation to enhance the signal-to-noise ratio. The final image is obtained by applying the fusing algorithm proposed in Section~\ref{Sec:fusion2}. 
       
    \item Benchmark Scheme II: 
    Based on the model in \eqref{eq:receive_Y_Nc_approx_vec}, we solve the CS problem applying the successive linear approximation variational Bayesian inference (SLA-VBI) algorithm proposed in \cite{ref:VBI}, and obtain the estimation of both the attenuation coefficients $\mathbf{c}_{k,m}$'s and grid positions $\mathbf{P}_k$'s. The final image is formed by accumulating $|\mathbf{c}_{k,q,m}|^2$ among $M$ frames and applying Algorithm~\ref{alg:fusion}. 
\end{itemize} 
We choose four capital letters as extended targets. We assume that each RBS has a limited field of view, with a $\pi/5$ obstacle-induced angular blind sector centered on the line-of-sight direction between the TBS and RBS. The three rows of Fig.~\ref{fig:benchmark} present the multi-view images of benchmark I, benchmark II, and our proposed scheme, respectively. The first column of Fig.~\ref{fig:benchmark} is obtained assuming $N_{\rm tx} = N_{\rm rx} = L = 24, M = 20$, and $P = 10$ dBm. In the second and third columns, we reduce $N_{\rm tx}, N_{\rm rx},$ and $L$ to $12$ and $8$, respectively. In the fourth column, we assume $N_{\rm tx} = N_{\rm rx} = 24$ and $L = 12$, where the non-orthogonal pilots are randomly selected from the sphere. The fifth column is obtained with lower transmit power $P = -10$~dBm. The sixth column is obtained using a smaller number of frames with $M = 5$. To quantitatively compare different schemes, Table \ref{table:compare} presents pseudo integrated sidelobe ratio (P-ISLR) and intersection over union (IoU) of the images in Fig.~\ref{fig:benchmark}. P-ISLR is defined as \cite{ref:manzoni2024wavefield}
\begin{equation}\label{ISLR}
  {\text {P-ISLR}} = 10 \log_{10} \tfrac{\sum_{ i\in {\text {sidelobe}}   } {\gamma}'_{r,i} }{ \sum_{ i\in {\text {mainlobe}}}  {\gamma}'_{r,i}  }, 
\end{equation}
where all the grids within the true target are regarded as main lobe and those outside are regarded as sidelobes. IoU measures how well the target shape is reconstructed, defined as \cite{ref:TV1}
\begin{equation}\label{IoU}
  {\text {IoU}} = \tfrac{ {\text {Area}}(\mathcal{S} \cap \hat{\mathcal{S}}) }{ {\text {Area}}(\mathcal{S} \cup \hat{\mathcal{S}}) }, 
\end{equation}
where $\mathcal{S}$ denotes the spatial region occupied by extended targets, and $\hat{\mathcal{S}}$ denotes its estimate obtained by sorting all detected grids according to corresponding intensities and retaining the highest-intensity ones until their cumulative intensity reaches $95\%$ of the total recovered intensity. As we can see, the letters are well-reconstructed by the proposed scheme, whereas benchmark schemes perform poorly, especially in the scenario with limited resources, such as a small number of transmit/receive antennas, low transmit power, or limited pilot length. Specifically, as illustrated in the first three columns, assuming that the orthogonal pilots are utilized, when the number of antennas is reduced, the benchmark schemes suffer from more pronounced sidelobes due to the decreased spatial resolution. In contrast, the proposed method still yields clear reconstructions of the letters. Quantitatively, in the case with $N_{\rm tx} = N_{\rm rx} = 12$, our scheme achieves at least $8$ dB reduction in P-ISLR and twofold increase in IoU compared to benchmarks. Columns 2 and 4 show that with fixed pilot length, the imaging performance improves as the number of antennas increases, and the proposed scheme substantially outperforms the benchmarks in both cases. The comparison between columns 1 and 4 shows that when the pilot length is reduced from $L=24$ to $L=12$, the benchmark methods exhibit blurred boundaries, but our proposed imaging scheme still presents sharp target contours and enhanced sidelobe suppression. In the case with $L = 12$, compared with benchmark methods, our approach reduces P-ISLR by nearly $10$ dB and doubles the IoU at the same time. When the transmit power is significantly reduced, compared with benchmark methods which show more spurious artifacts, our scheme maintains superior performance, as shown in columns 1 and 5. The comparison between columns 1 and 6 indicates that our scheme maintains low P-ISLR and high IoU even in the scenario with limited frames. The imaging performance comparison shown in Fig.~\ref{fig:benchmark} and Table \ref{table:compare} demonstrates the substantial performance improvement of the proposed scheme over benchmark schemes. The performance gain can be interpreted as the ability of our scheme to efficiently reduce the number of parameters to be estimated by exploiting channel statistical properties.

\begin{figure}
    \centering
    \includegraphics[width=0.93\linewidth]{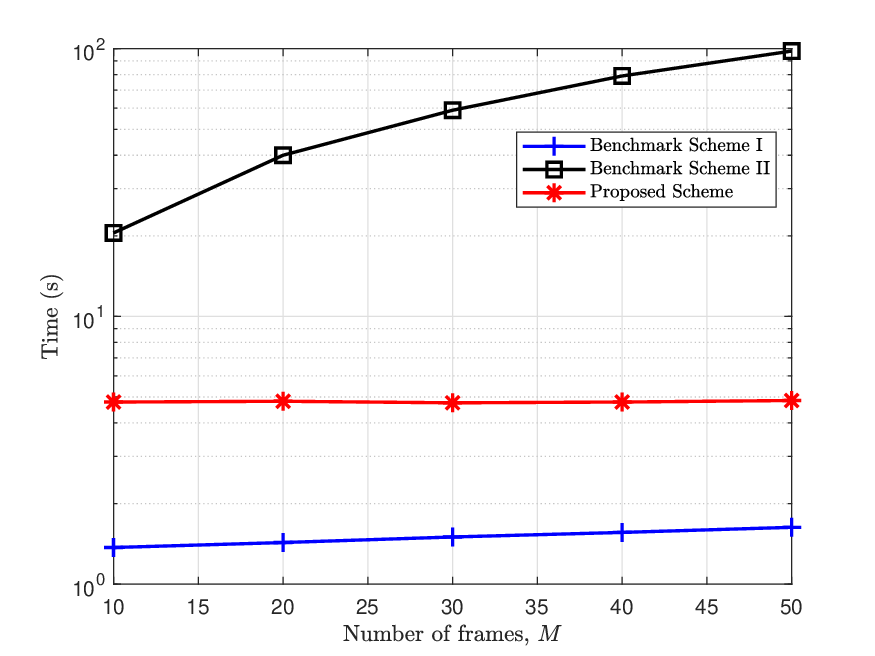}\\
    \caption{CPU time versus the number $M$ of frames with $N_{\rm tx} = N_{\rm rx} = L = 8$ and $P = 10$ dBm. 
} \label{fig:time}  
\end{figure}

In Fig.~\ref{fig:time}, we compare the CPU time of different schemes versus the number $M$ of frames on a desktop with Intel Core i9-11900K, assuming that $N_{\rm tx} = N_{\rm rx} = L = 8$ and $P = 10$ dBm. It can be observed that the proposed scheme achieves significantly lower CPU computation time than Benchmark Scheme II. Although the proposed scheme requires more time than Benchmark Scheme I, the latter suffers from poor imaging quality as previously discussed. As the number $M$ of frames increases, the CPU time of both the benchmark schemes increases linearly, since they require RBS to perform single-view imaging in each frame followed by multi-frame accumulation as aforementioned. In contrast, for the proposed scheme, increasing $M$ only affects the complexity of calculating the sample covariance matrix. Since its overhead is marginal, the complexity of the overall algorithm is mainly determined by other steps, i.e., the AO of the scattering intensity and grid position, and the multi-view fusion. Thus, the CPU time of the proposed scheme remains nearly constant as $M$ increases. The provided simulation results demonstrate the superior imaging quality and time-efficiency of the proposed framework, thereby indicating that it is well-suited for real-world imaging applications in stationary scenarios.

\section{Conclusion} \label{Sec:conclusion}

This work investigated the multi-view imaging problem of extended targets. An advanced two-phase framework was developed. In Phase I, to recover the single-view image at each RBS, we formulated a penalized ML problem to alternatively estimate the effective scattering intensity for each grid point and update grid shapes to conform to target geometries. Compared to FFT and CS based schemes that aim to estimate instantaneous channels, our formulation dramatically reduced the number of variables to be estimated. In Phase II, we developed a multi-view fusion strategy to aggregate individual images. Specifically, we designed an EP-NNI method to map each heterogeneous single-view image onto a common set of grids, and formulated an optimization problem to estimate the fused scattering intensity and ``informative'' subset of RBSs. Extensive numerical results were provided to verify the effectiveness of the proposed scheme, demonstrating its potential to support high-quality imaging in future 6G networks. In the future, it is interesting to explore more advanced methods for determining the number of grids in our framework and consider more practical issues, e.g., moving targets, clusters, etc.

\bibliographystyle{IEEEtran}  
\bibliography{IEEEabrv,reference}

\begin{IEEEbiography}[{\includegraphics[width=1in,height=1.25in,clip,keepaspectratio]{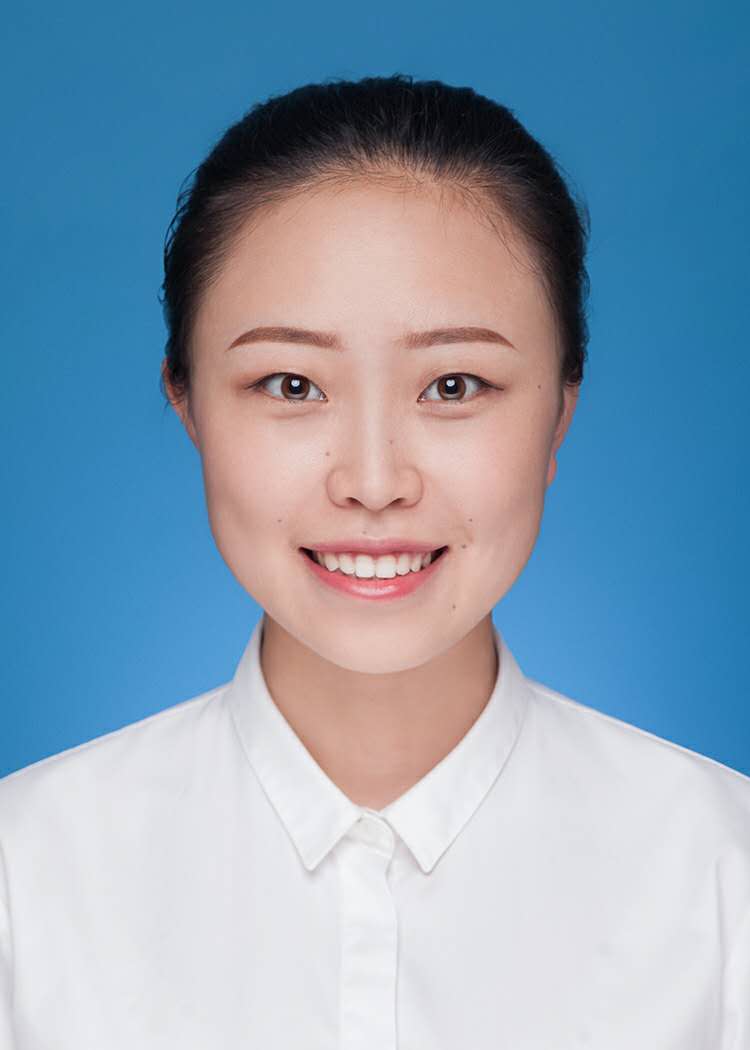}}]{Junyuan Gao}
received the B.S. degree in communication engineering from Chongqing University, Chongqing, China, in 2018, and the Ph.D. degree in information and communication engineering from Shanghai Jiao Tong University, Shanghai, China, in 2023. She is currently a postdoctoral fellow in The Hong Kong Polytechnic University, Hong Kong. Her research interests include massive random access, massive MIMO, finite-blocklength information theory, and integrated sensing and communication. 
\end{IEEEbiography}

\begin{IEEEbiography}[{\includegraphics[width=1in,height=1.25in,clip,keepaspectratio]{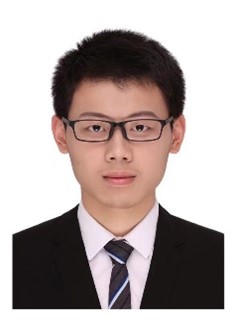}}]{Weifeng Zhu} 
(Member, IEEE) received the B.S. degree in communication engineering from Tianjin University, Tianjin, China, in 2017, and the Ph.D. degree in information and communication engineering from Shanghai Jiao Tong University, Shanghai, China, in 2023. Since 2023, he is a Postdoctoral Fellow in The Hong Kong Polytechnic University, Hong Kong. His research interests include statistical signal processing, high-dimensional statistics, machine learning, massive random access, neural radio-frequency radiance field, and integrated communication-sensing-computing. 
\end{IEEEbiography}

\begin{IEEEbiography}[{\includegraphics[width=1in,height=1.25in,clip,keepaspectratio]{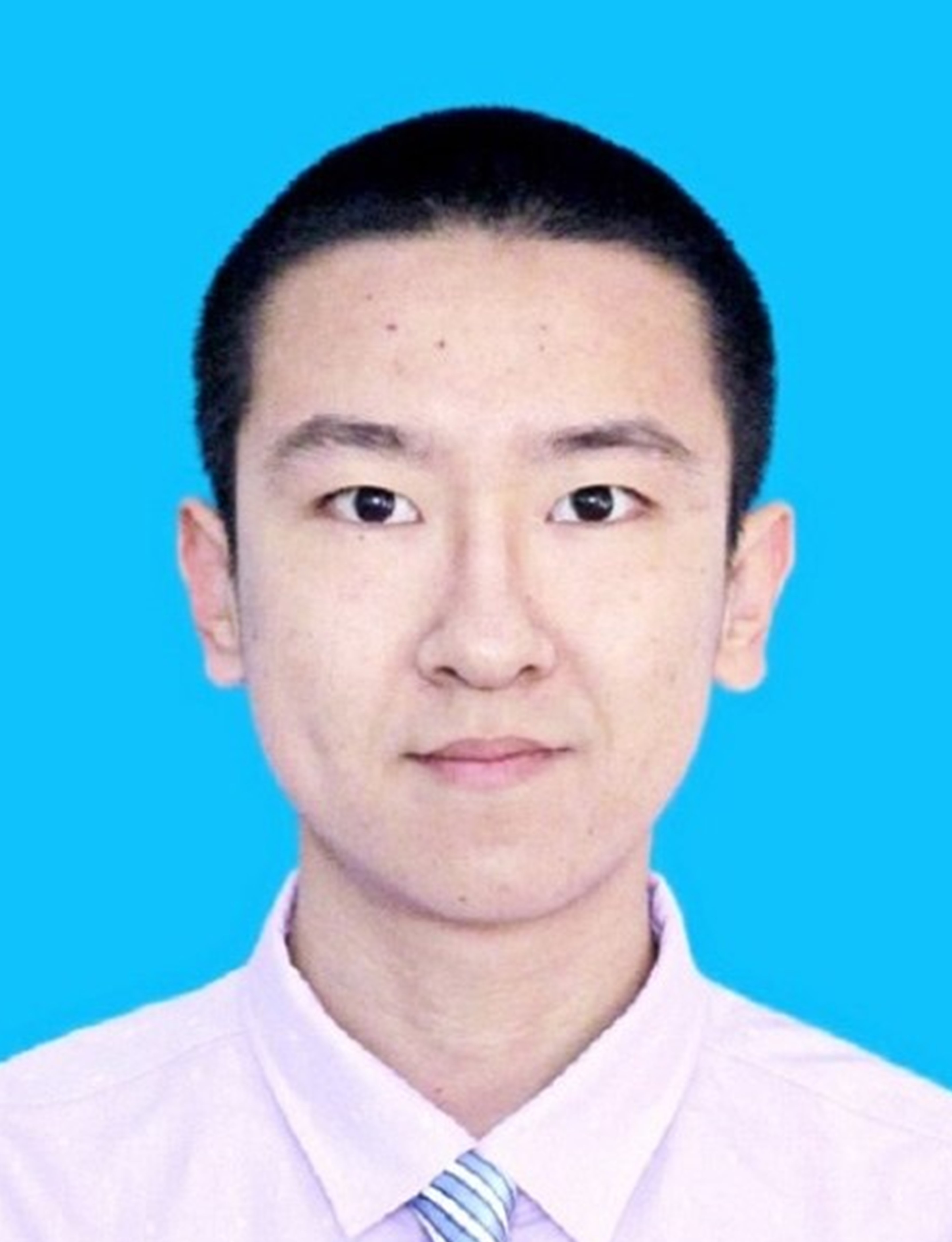}}]{Yanmo Hu}
(Member, IEEE) received the B.S. and Ph.D. degrees in information and communication engineering from Harbin Institute of Technology, Harbin, China, in 2019 and 2024, respectively. He is currently a Post-Doctoral Fellow with The Hong Kong Polytechnic University, Hong Kong. His research interests include target localization/tracking/imaging, high frequency surface wave radar, mmWave radar sensing, terahertz radar localization, and integrated sensing and communications. 
\end{IEEEbiography}

\begin{IEEEbiography}[{\includegraphics[width=1in,height=1.25in,clip,keepaspectratio]{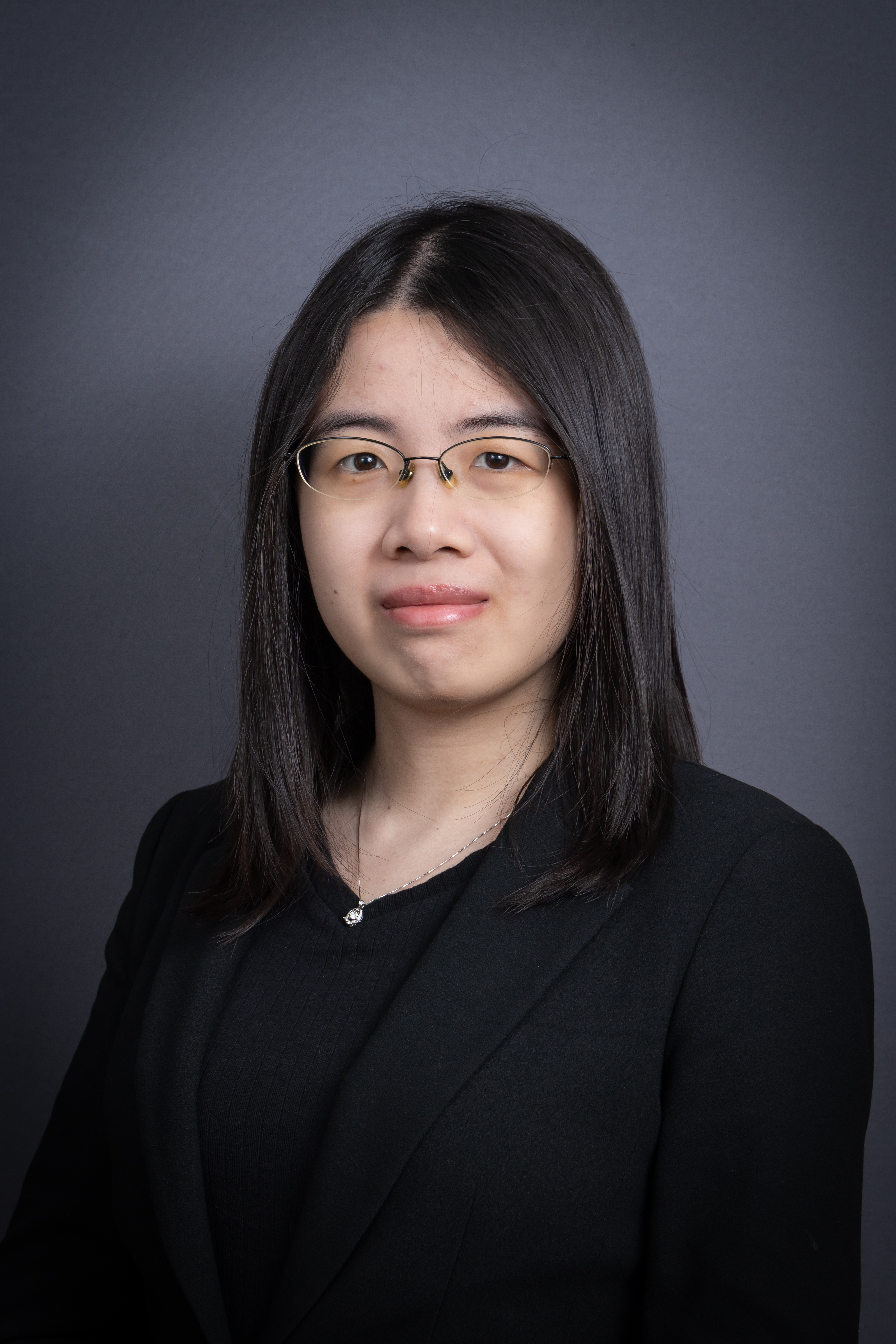}}]{Shuowen Zhang}(Senior Member, IEEE) 
received the B.Eng. degree in information engineering from the Chien-Shiung Wu Honors College, Southeast University, Nanjing, China, in June 2013, and the Ph.D. degree from NUS Graduate School for Integrative Sciences and Engineering (NGS), National University of Singapore, in January 2018 under the NGS scholarship. After completing her Ph.D. studies, she served as a Research Fellow at the Department of Electrical and Computer Engineering, National University of Singapore, and then joined The Hong Kong Polytechnic University where she is currently an Assistant Professor with the Department of Electrical and Electronic Engineering. Her research interests include integrated sensing and communication, intelligent surface aided communication, unmanned aerial vehicles, multiple-input multiple-output (MIMO), communication theory, and optimization methods. Dr. Zhang is currently serving as an Editor for IEEE Transactions on Wireless Communications and an Associate Editor for IEEE Transactions on Mobile Computing. She has served as a Guest Editor for various journals such as the IEEE Journal on Selected Areas in Communications. Dr. Zhang is the sole recipient of the 2021 Marconi Society Paul Baran Young Scholar Award, as well as a recipient of the 2022 IEEE Communications Society Young Author Best Paper Award, the 2023 IEEE Communications Society Best Tutorial Paper Award, the 2023 PolyU Young Innovative Researcher Award, and the 2024 IEEE Communications Society Asia-Pacific Outstanding Young Researcher Award.
\end{IEEEbiography}

\begin{IEEEbiography}[{\includegraphics[width=1in,height=1.25in,clip,keepaspectratio]{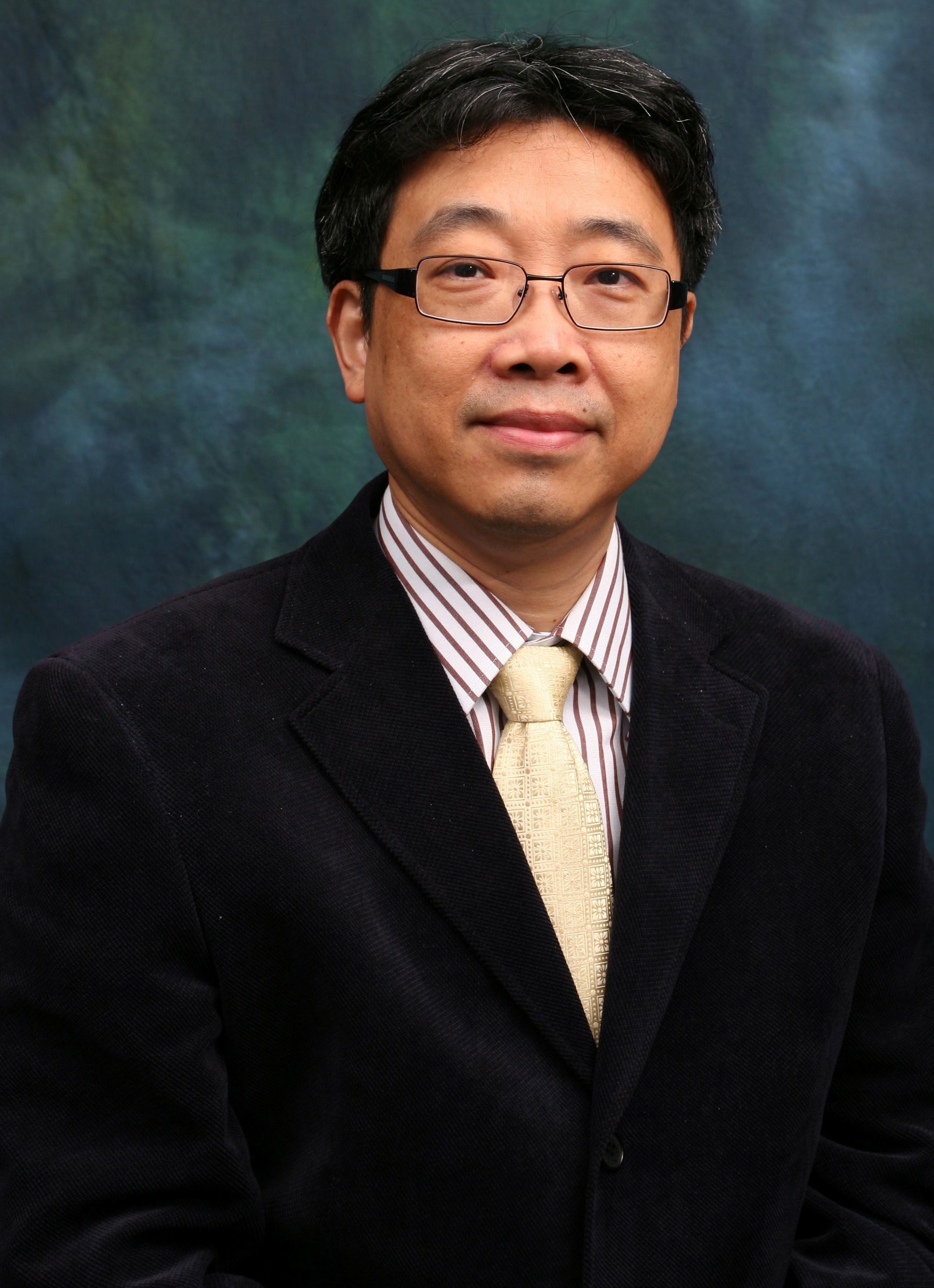}}]{Jiannong Cao}(Fellow, IEEE) is currently the Otto Poon Charitable Foundation Professor in Data Science and the Chair Professor of Distributed and Mobile Computing in the Department of Computing at The Hong Kong Polytechnic University (PolyU), Hong Kong. Prof. Cao is a member of Academia Europaea, a fellow of IEEE, a fellow of CCF, and an ACM distinguished member. Prof. Cao's research interests include distributed systems and blockchain, wireless sensing and networking, big data and machine learning, and mobile cloud and edge computing.
\end{IEEEbiography}

\begin{IEEEbiography}[{\includegraphics[width=1in,height=1.25in,clip,keepaspectratio]{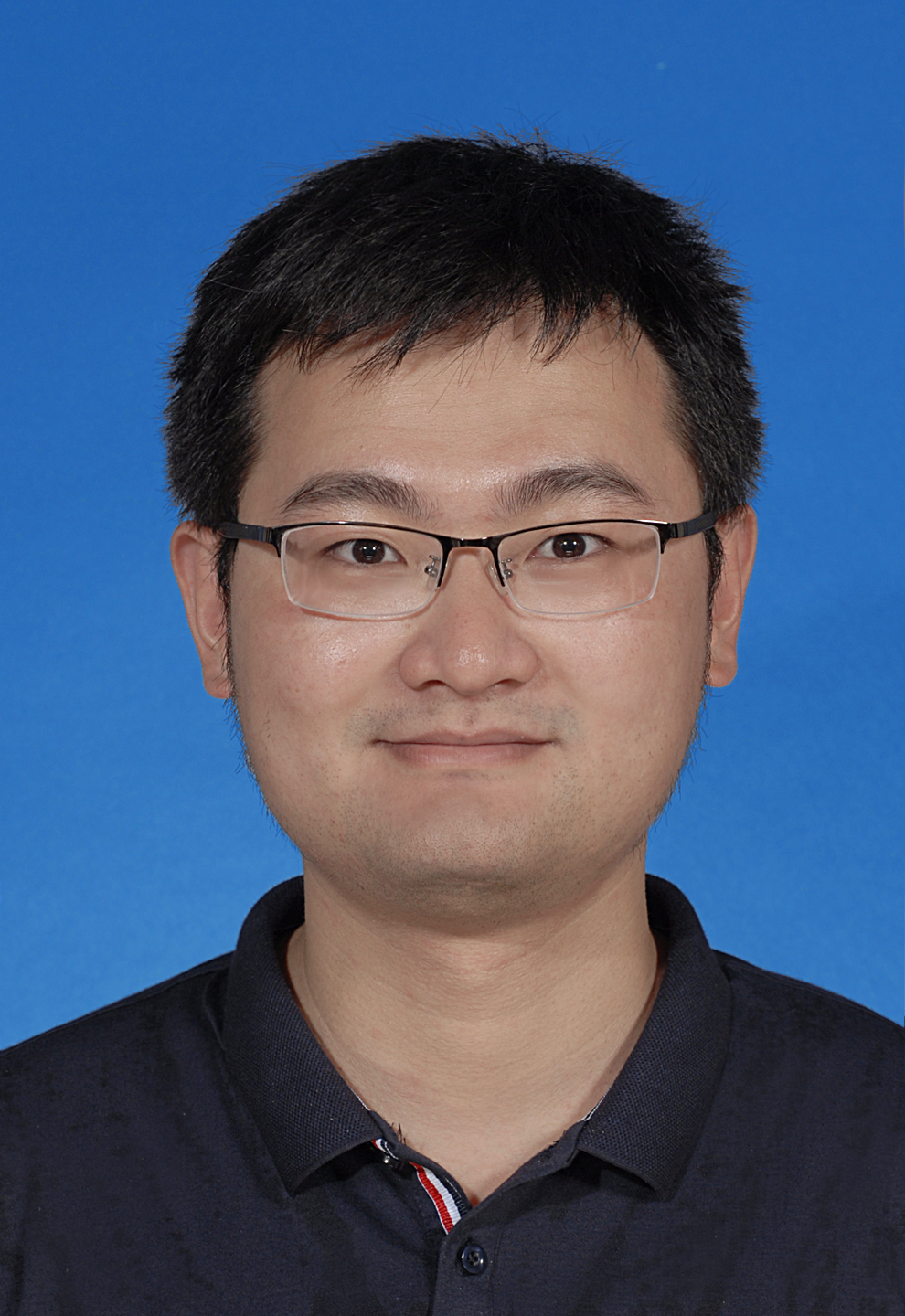}}]{Yongpeng Wu}(S'08--M'13--SM'17)
received the B.S. degree in telecommunication engineering from Wuhan University, Wuhan, China, in July 2007, the Ph.D. degree in communication and signal processing with the National Mobile Communications Research Laboratory, Southeast University, Nanjing, China, in November 2013.

Dr. Wu is currently a Professor with the Department of Electronic Engineering, Shanghai Jiao Tong University, China. Previously, he was senior research fellow with Institute for Communications Engineering, Technical University of Munich, Germany  and the Humboldt research fellow and the senior research fellow with Institute for Digital Communications, University Erlangen-N$\ddot{u}$rnberg,  Germany. During his doctoral studies, he conducted cooperative research at the Department of Electrical Engineering, Missouri University of Science and Technology, USA. His research interests include massive MIMO/MIMO systems, massive machine type communication,  physical layer security, and signal processing for wireless communications.

Dr. Wu was awarded the IEEE Student Travel Grants for IEEE International Conference on Communications (ICC) 2010,  the Alexander von Humboldt Fellowship in 2014, the Travel Grants for IEEE Communication Theory Workshop 2016, the Excellent Doctoral Thesis Awards of China Communications Society 2016, the Exemplary Editor Award of  IEEE Communication Letters 2017,  Young Elite Scientist Sponsorship Program by CAST 2017, and Excellent Youth Science Fund Project  of National Natural Science Foundation of China 2021.

He has been the lead guest editor of IEEE Journal of Selected Topics in Signal Processing, IEEE Journal on Selected Areas in Communications and IEEE Wireless Communications. He is currently an editor of IEEE Transactions on Information Theory/Wireless Communications, and was an Editor of IEEE Transactions on Communications and IEEE Communications Letters. He has been symposium chairs of various conferences, including Globecom, ICC, VTC, and PIMRC, etc.
\end{IEEEbiography}

\begin{IEEEbiography}[{\includegraphics[width=1in,height=1.25in,clip,keepaspectratio]{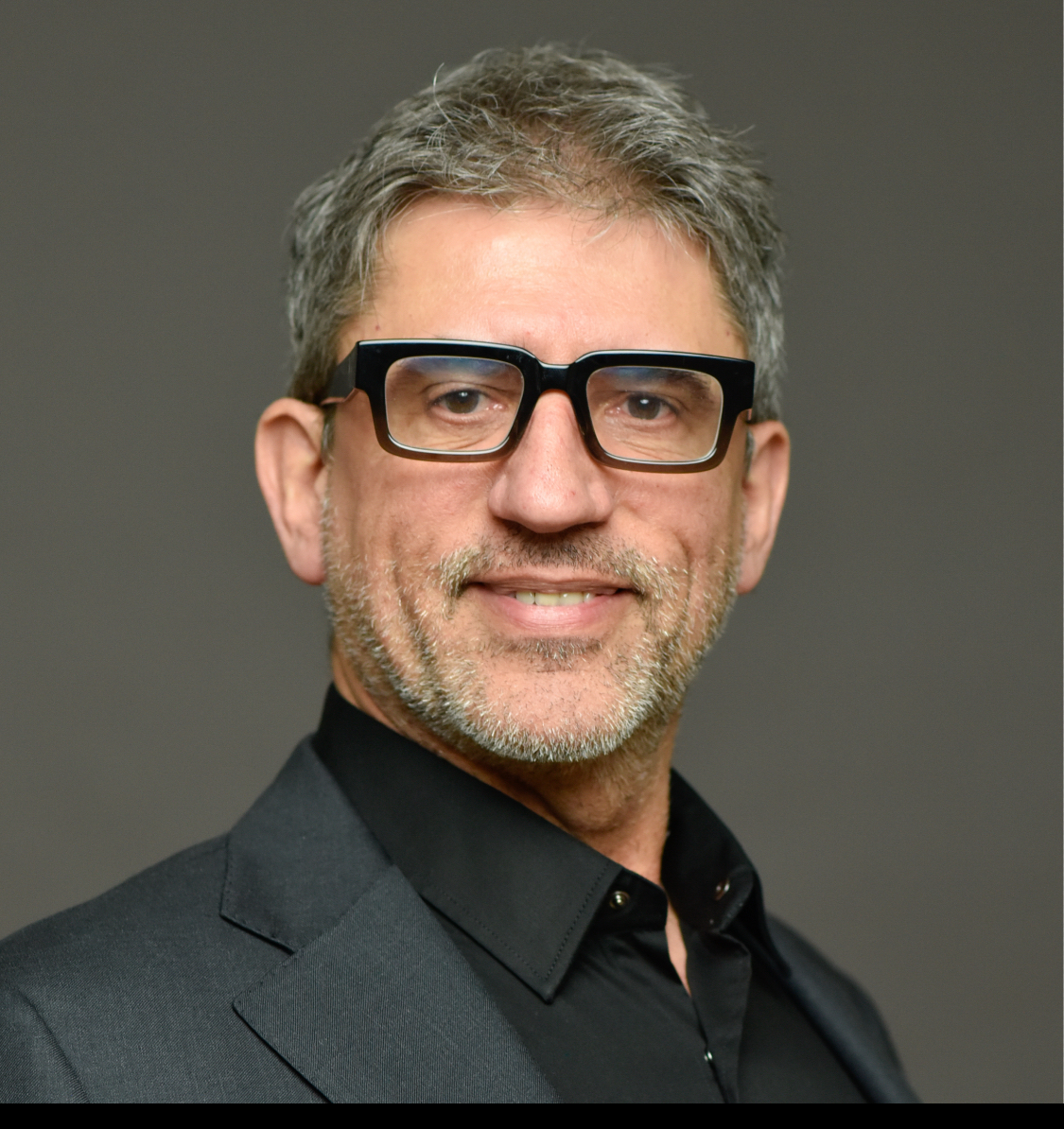}}]{Giuseppe Caire}(S'92 -- M'94 -- SM'03 -- F'05) 
was born in Torino in 1965. He received a B.Sc. in Electrical Engineering  from Politecnico di Torino in 1990, an M.Sc. in Electrical Engineering from Princeton University in 1992, and a Ph.D. from Politecnico di Torino in 1994. He has been a post-doctoral research fellow with the European Space Agency (ESTEC, Noordwijk, The Netherlands) in 1994-1995, Assistant Professor in Telecommunications at the Politecnico di Torino, Associate Professor at the University of Parma, Italy, Professor with the Department of Mobile Communications at the Eurecom Institute,  Sophia-Antipolis, France, a Professor of Electrical Engineering with the Viterbi School of Engineering, University of Southern California, Los Angeles, and he is currently an Alexander von Humboldt Professor with the Faculty of Electrical Engineering and Computer Science at the Technical University of Berlin, Germany. He is a member of the German National Academy of Sciences (Leopoldina) since 2024, and a member of the American National Academy of Engineering (NAE) since 2026. 

He received the Jack Neubauer Best System Paper Award from the IEEE Vehicular Technology Society in 2003, the IEEE Communications Society and Information Theory Society Joint Paper Award in 2004, in 2011, and in 2025, the Okawa Research Award in 2006,   the Alexander von Humboldt Professorship in 2014, the Vodafone Innovation Prize in 2015, an ERC Advanced Grant in 2018,  the Leonard G. Abraham Prize for best IEEE JSAC paper in 2019, the IEEE Communications Society Edwin Howard Armstrong Achievement Award in 2020, the 2021 Leibniz Prize  of the German National Science Foundation (DFG), and the  CTTC Technical Achievement Award of the IEEE Communications Society in 2023.  Giuseppe Caire is a Fellow of IEEE since 2005.  He has served in the Board of Governors of the IEEE Information Theory Society from 2004 to 2007, and as officer from 2008 to 2013. He was President of the IEEE Information Theory Society in 2011. His main research interests are in the field of communications theory, information theory, channel and source coding with particular focus on wireless communications.   
\end{IEEEbiography}

\begin{IEEEbiography}[{\includegraphics[width=1in,height=1.25in,clip,keepaspectratio]{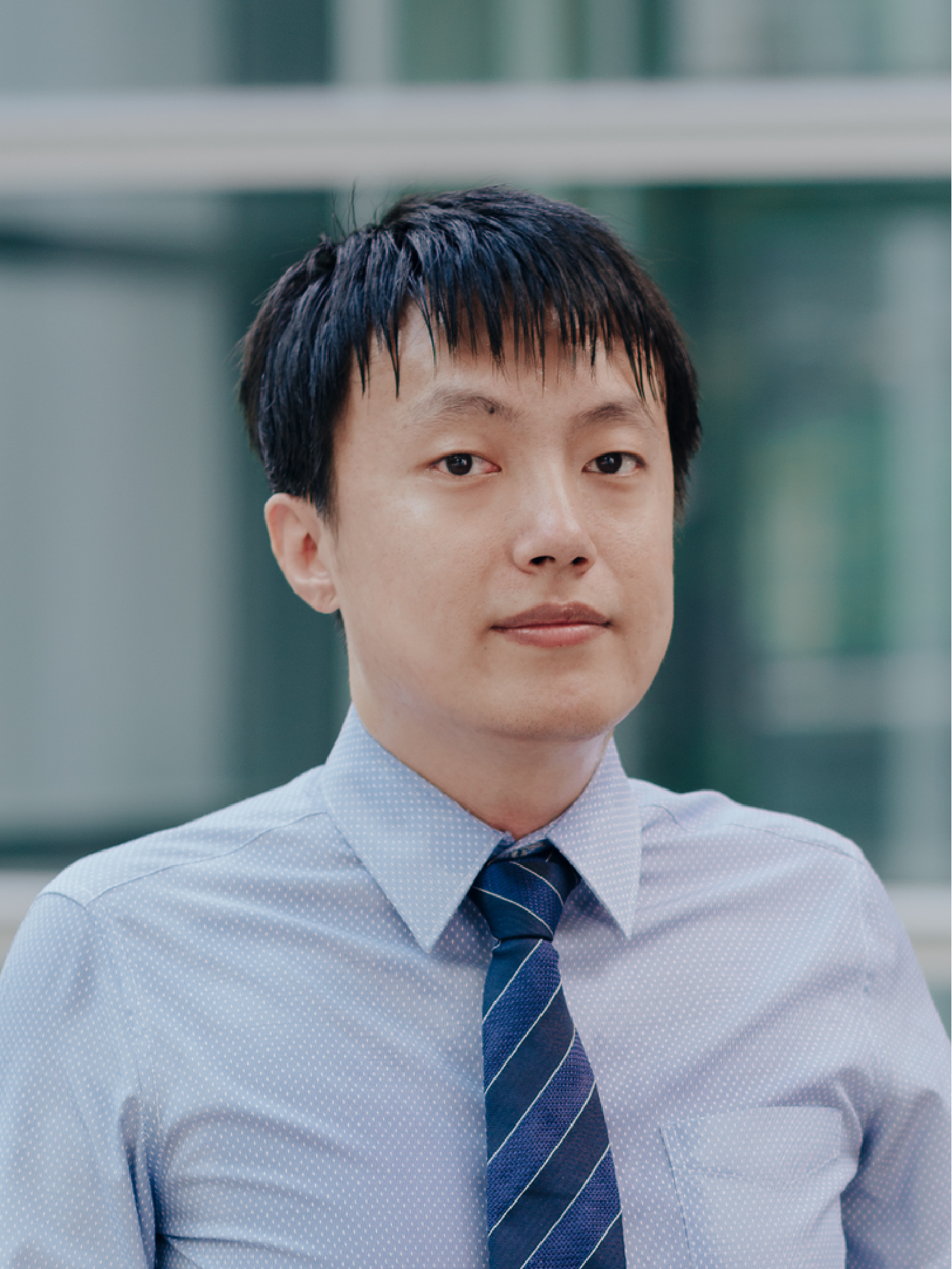}}]{Liang Liu}(Fellow, IEEE) is currently an Associate Professor with the Department of Electrical and Electronic Engineering, The Hong Kong Polytechnic University. He received the B.Eng. degree from Tianjin University, China, in 2010, and the Ph.D. degree from the National University of Singapore, Singapore, in 2014. He was a Post-Doctoral Fellow with the Department of Electrical and Computer Engineering, University of Toronto, from 2015-2017, a Research Fellow with the Department of Electrical and Computer Engineering, National University of Singapore, from 2017-2019, and an Assistant Professor with the Department of Electrical and Electronic Engineering, The Hong Kong Polytechnic University, from 2019-2024. His research interests lie in the next generation cellular technologies such as machine-type communications for the Internet of Things, integrated sensing and communication, etc. He is an IEEE ComSoc Distinguished Lecturer for the class of 2025-2026. He was a recipient of the 2021 IEEE Signal Processing Society Best Paper Award, the 2017 IEEE Signal Processing Society Young Author Best Paper Award, the Best Student Award of 2022 IEEE International Conference on Acoustics, Speech, and Signal Processing (ICASSP), and the Best Paper Award of the 2011 International Conference on Wireless Communications and Signal Processing. He was recognized by Clarivate Analytics as a Highly Cited Researcher in 2018. He is an Editor of IEEE TRANSACTIONS ON WIRELESS COMMUNICATIONS and an Associate Editor of IEEE TRANSACTIONS ON SIGNAL PROCESSING. He was a Leading Guest Editor of IEEE WIRELESS COMMUNICATIONS Special Issue on Massive Machine-Type Communications for IoT. He is a co-author of the book ``Next Generation Multiple Access'' published at Wiley-IEEE Press. He is a Fellow of IEEE.
\end{IEEEbiography}

\end{document}